\documentclass[useAMS, usenatbib]{mn2e}
\usepackage{amsmath}
\usepackage{amssymb}
\usepackage{graphicx}
\usepackage{subfigure}

\newcommand{\be}{\begin{equation}} 
\newcommand{\ee}{\end{equation}}

\newcommand{\ba}{\begin{eqnarray}} 
\newcommand{\ea}{\end{eqnarray}}

\newcommand{\msun}{M_{\odot}}

\newcommand{\hatLp}{\hat{{\bf L }}}
\newcommand{\Lp}{{\bf L }}

\newcommand{\hatLb}{\hat{{\bf L }}_b}
\newcommand{\Lb}{{\bf L }_b}
\newcommand{\Ss}{{\bf S }_\star}
\newcommand{\Sp}{{\bf S }_p}

\newcommand{\ep}{{\bf e }}

\newcommand{\hatS}{\hat{{\bf S }}_\star}
\newcommand{\hatSp}{\hat{{\bf S }}_p}
\newcommand{\hatep}{\hat{{\bf e }}}
\newcommand{\hateb}{\hat{{\bf e }}_b}
\newcommand{\SL}{{\rm sl}}
\newcommand{\PS}{\rm{ps}}

\newcommand{\LB}{\rm{lb}}
\newcommand{\GR}{\rm{GR}}
\newcommand{\ST}{\rm{Tide}}
\newcommand{\OP}{\rm{Rot}}
\newcommand{\mx}{\rm{max}}
\newcommand{\oct}{\mathrm{oct}}
\newcommand{\jvec}{\mathbf{j}}
\newcommand{\emax}{e_{\rm max}}
\newcommand{\emin}{e_{\rm min}}
\newcommand{\Mtot}{M_{\rm tot}}
\newcommand{\thetaSL}{\theta_{\rm sl}}
\newcommand{\thetaLB}{\theta_{\rm lb}}

\newcommand{\gr}{\mathrm{GR}}
\newcommand{\tide}{\mathrm{Tide}}
\newcommand{\rot}{\mathrm{Rot}}
\newcommand{\m}{\mathrm{max}}

\newcommand{\li}{\mathrm{lim}}

\newcommand{\abeff}{a_{b,\rm eff}}

\newcommand{\Mtunit}{\bar{M}_{\mathrm{tot}}}
\newcommand{\Mbunit}{\bar{M}_{b}}
\newcommand{\Mpunit}{\bar{M}_{p}}
\newcommand{\Msunit}{\bar{M}_{\star}}

\newcommand{\abunit}{\bar{a}_{b,\rm eff}}
\newcommand{\aunit}{\bar{a}}
\newcommand{\afunit}{\bar{a}_F}

\newcommand{\Rpunit}{\bar{R}_{p}}
\newcommand{\Rsunit}{\bar{R}_{\star}}
\newcommand{\Psunit}{\bar{P}_{\star}}
\newcommand{\Ppunit}{\bar{P}_{p}}

\def\go{\mathrel{\raise.3ex\hbox{$>$}\mkern-14mu
             \lower0.6ex\hbox{$\sim$}}}

\def\lo{\mathrel{\raise.3ex\hbox{$<$}\mkern-14mu
             \lower0.6ex\hbox{$\sim$}}}

\begin{document}
\title[Formation and Stellar Spin-Orbit Misalignment of Hot Jupiters]{\bf Formation and Stellar Spin-Orbit Misalignment of Hot Jupiters from Lidov-Kozai Oscillations in Stellar Binaries} 

\label{firstpage}

\author[Kassandra R. Anderson, Natalia I. Storch, \& Dong
Lai]{Kassandra R. Anderson$^{1}$\thanks{kra46@cornell.edu}, Natalia
  I. Storch$^{1}$\thanks{Current address: TAPIR, Walter Burke Institute for Theoretical Physics, Mailcode 350-17, Caltech, Pasadena, CA 91125, USA}, \& Dong
  Lai$^{1}$ \\ \\ $^{1}$Cornell Center for Astrophysics and Planetary
  Science, Department of Astronomy, Cornell University, Ithaca, NY
  14853, USA.}
\maketitle
\begin{abstract}
Observed hot Jupiter (HJ) systems exhibit a wide range of stellar
spin-orbit misalignment angles. The origin of these HJs remains unclear.
This paper investigates the inward migration of giant planets due to
Lidov-Kozai (LK) oscillations of orbital eccentricity/inclination
induced by a distant (100-1000 AU) stellar companion and orbital
circularization from dissipative tides.  We conduct a large population
synthesis study, including
octupole gravitational potential from the stellar companion, mutual precession
of the host stellar spin axis and planet orbital axis, pericenter
advances due to short-range-forces, tidal dissipation in the planet,
and stellar spin-down in the host star due to magnetic braking. We examine a range of
planet masses ($0.3-5\,M_J$) and initial semi-major axes ($1-5$~AU),
different properties for the host star, and varying tidal dissipation
strengths.  The fraction ($f_{\rm HJ}$) of systems that result in HJs
is a function of planet mass and stellar type, with $f_{\rm HJ}$ in
the range of $1-4\%$ (depending on tidal dissipation strength) for
$M_p=1\,M_J$, and larger (up to $8\%$) for more massive planets.  The
production efficiency of ``hot Saturns'' ($M_p=0.3M_J$) is much lower,
because most of the inward-migrating planets are tidally disrupted. We
find that the fraction of systems that result in either HJ formation
or tidal disruption, $f_{\rm mig} \simeq 11-14\%$ is roughly constant,
having little variation with planet mass, stellar type and tidal
dissipation strength.  This ``universal'' migration fraction can be
understood qualitatively from analytical migration criteria based on
the properties of octupole LK oscillations.  The distribution of final
stellar obliquities for the HJ systems formed in our calculations
exhibits a complex dependence on the planet mass and stellar type. For
$M_p = (1-3)M_J$, the distribution is always bimodal, with peaks
around $\sim 30^{\circ}$ and $\sim 130^{\circ}$.  The obliquity
distribution for massive planets ($M_p = 5 M_J$) depends on the host
stellar type, with a preference for low obliquities for solar-type
stars, and higher obliquities for more massive ($1.4M_{\odot}$) F-type
stars.
\end{abstract}  

\begin{keywords}
planets and satellites: dynamical evolution and stability --
planet-–star interactions  -- binaries: general
\end{keywords}

\begin{table*}
 \centering
 \begin{minipage}{120mm}
  \caption{Definitions of variables, along with the canonical value used in this paper (if applicable), and dimensionless form.
  }
  \begin{tabular}{@{}ll@{}}
  \hline
  \hline
  Quantity & Dimensionless/Normalized Form\\
 \hline
\\
{\bf Vector Quantities} & \\
Planet orbital angular momentum $\Lp $  & ..... \\
Planet eccentricity vector $\ep$ & .....  \\
Binary orbital angular momentum $\Lb$  & .....  \\
Binary eccentricity vector ${\bf e}_b$  & .....  \\
Stellar spin angular momentum $\Ss$ & ..... \\
Planetary spin angular momentum $\Sp$ & ..... \\
\hline
{\bf Physical Properties} & \\
Stellar mass $M_\star$ & $\Msunit = M_\star/ M_{\odot}$ \\
Stellar radius $R_\star$ & $\Rsunit = R_\star/ R_{\odot}$ \\
Planet mass $M_p$ & $\Mpunit = M_p/ M_{J}$ \\
Planet radius $R_p$ & $\Rpunit = R_p/ R_{J}$ \\
Binary companion mass $M_b$ & $\Mbunit = M_b/ M_{\odot}$ \\
Inner binary total mass $M_{\rm tot} \equiv M_\star + M_p$  & $\Mtunit = M_{\rm tot}/ M_{\odot}$ \\
\hline
{\bf Spin \& Structure Properties} & \\
Spin-orbit angle $\theta_{\SL}$ (defined by $\cos \theta_{\SL} = \hatLp \cdot \hatS $) & ..... \\
Stellar moment of inertia constant $k_\star$ ($I_\star = k_\star M_\star R_\star^2$) & $\bar{k}_\star = k_\star / 0.1$ \\
Planet moment of inertia constant $k_p$ ($I_p = k_p M_p R_p^2$) & $\bar{k}_p = k_p / 0.25$ \\
\\
Stellar rotational distortion coefficient $k_{q\star}$ (see Sec.~\ref{sec:spin}) & $\bar{k}_{q\star} = k_{q\star} / 0.05$ \\
Planet rotational distortion coefficient $k_{qp}$ (see Sec.~\ref{sec:spin}) & $\bar{k}_{qp} = k_{qp} / 0.17$ \\
Stellar spin period $P_\star = 2 \pi/\Omega_\star$ & $\bar{P}_\star = P_\star/ \rm day$ \\
Planet spin period $P_p = 2 \pi/\Omega_p$ & $\bar{P}_p = P_p/ \rm day$ \\
\hline
{\bf Tidal Properties} & \\
Planet tidal Love number $k_{2p}$ & $\bar{k}_{2p} = k_{2p}/0.37$ \\
Tidal lag time $\Delta t_L$ & ..... \\
Tidal enhancement factor $\chi$ ($\Delta t_L = 0.1 \chi \rm sec$) & ..... \\
\hline
{\bf Orbital Properties} & \\
Planet semi-major axis $a$ & $\aunit = a / \rm AU$ \\
Planet eccentricity $e$ & ..... \\
Planet inclination $\theta_{\LB}$  (relative to outer binary, defined by $\cos \theta_{\LB} = \hatLp \cdot \hatLb $) & ..... \\
Outer binary semi-major axis $a_b$ & $\bar{a}_b = a_b / 100 \rm AU$ \\
Outer binary eccentricity $e_b$ & ..... \\
Effective outer binary semi-major axis $\abeff \equiv a_b \sqrt{1 - e_b^2}$ & $\abunit = \abeff / 100 \rm AU$ \\
Orbital mean motion $n = \sqrt{G M_{\rm tot}/a^3}$ & ..... \\

\hline
\hline
\end{tabular}
\label{table:variables}
\end{minipage}
\end{table*}

\section{Introduction}

The growing sample of close-in giant planets (hot Jupiters, hereafter HJs) continues
to yield surprises.  
These planets (with orbital periods of $\sim 3$~days) could not have formed in
situ, given the large stellar tidal gravity and radiation fields close
to their host stars, and must have formed beyond a few AUs
and migrated inward. The recent discoveries of many HJs
with orbital angular momentum axes that are misaligned with respect to their
host star's spin axis
\citep[e.g.][]{hebrard2008,narita2009,winn2009,triaud2010,albrecht2012a,moutou2011}
has stimulated new studies on the dynamical causes behind such
configurations.  The presence (or lack) of such misalignment in an 
HJ system serves as a probe of the planet's dynamical history,
and can potentially constrain the planet's migration channel.
Therefore, understanding the dynamics behind spin-orbit misalignments
is an important endeavor.

HJ systems with low spin-orbit misalignments are commonly thought to have arisen from smooth disk-driven migration, thereby preserving
an initially low stellar obliquity.\footnote{Throughout this paper we use the
  terms ``spin-orbit misalignment'' and ``stellar obliquity''
  interchangeably.} In contrast, systems with high misalignments must
have undergone a more disruptive high-eccentricity migration, in which
the eccentricity becomes excited to a large value, with subsequent
orbital decay due to dissipative tides raised on the planet by the
host star. This assumption has been challenged recently with the
suggestion of a ``primordial misalignment'' \citep{bate2010, foucart2011, lai2011,thies2011, batygin2012, batygin2013, lai2014,spalding2014,fielding2015}, in which the protoplanetary disk itself becomes tilted with respect to the stellar spin and planets subsequently form and smoothly
migrate within the misaligned disk, resulting in close-in planets with
large stellar obliquities. Collectively, these works show that much remains to
be done in disentangling the various possible dynamical histories of
HJs.

High-eccentricity migration requires either one or more additional
planets in the system, or the presence of a stellar binary companion.
In the former case, the eccentricity excitation can be caused by
strong planet-planet scatterings 
\citep{rasio1996,chatterjee2008,ford2008, juric2008}, and various forms of secular interactions, such as secular chaos with at least three giant planets 
\citep{wu2011} and interactions between two modestly eccentric coplanar planets \citep{petrovich2015a}, or, most likely, a combination of both \citep{nagasawa2008,beauge2012}.
In the case of a stellar companion, high eccentricity is achieved from
``Lidov-Kozai'' (LK) oscillations \citep{lidov1962, kozai1962}, in which an
inclined stellar companion pumps up the planet's eccentricity to values close
to unity; during the brief high eccentricity phases, dissipative tides
within the planet cause orbital decay and inward migration, eventually
resulting in a planet with an orbital period of a few days
\citep[e.g.][]{wu2003,fabrycky2007,naoz2012,petrovich2015b}.  Note that LK oscillations with tidal
dissipation from stellar companions have also been invoked to explain
the existence of tight inner binaries in stellar triple systems
\citep[e.g.][]{mazeh1979,eggleton2001,fabrycky2007,naoz2014}.

To assess the feasibility of HJ formation from the dynamical
effects of distant perturbers, searches for both planetary and stellar
companions in HJ systems have been
conducted. \cite{knutson2014} searched for radial velocity signatures
from distant companions in systems known to host HJs, and estimated a
companion occurrence rate of $\sim 50 \%$ for HJ systems (corrected for
sample incompleteness), for companion masses in the range $\sim 1 - 13
M_J$ and separations $\sim 1 - 20$ AU.  By direct imaging,
\cite{ngo2015} performed a similar survey for stellar mass companions, 
and found an occurrence rate of $48\pm 9\%$ for companions at separations $\sim 50 - 2000$ AU; this is larger than $24\%$,
the fraction of binaries (of the same separation range)
among solar-type field stars \citep{raghavan2010}, suggesting that the presence of a stellar companion increases the likelihood of 
HJ formation. Taken together, \cite{ngo2015}
suggested a total companion fraction (including stars and planets) of
$\sim 70 \%$ for systems hosting HJs.  Using a combination of adaptive
optics imaging and radial velocity, \cite{wang2015} searched for
stellar companions in systems containing {\it Kepler} Objects of
Interest, focusing on gas giant planets with orbital periods ranging
from a few days to hundreds of days.  They found that the stellar
multiplicity fraction of companions with separations 
between 20 and 200~AU is a factor of $\sim 2$ higher for stars hosting
a giant planet, compared to a control sample with no planet detections.
Since many of the objects in their sample are HJs, this highlights the
potential role of companion stars in the formation of close-in giant
planets.

Despite these optimistic companion fractions, some aspects of HJ
formation via LK oscillations remain problematic.  Assuming
steady-state formation of HJs, high-eccentricity migration implies the
presence of giant planets at wide orbital separations and large
eccentricities, with $a \sim$ several AU and $e \gtrsim 0.9$
\citep[``super-eccentric migrating Jupiters,''][]{socrates2012}.
However, this class of planets is not observed \citep{dawson2015}.
Whether this apparent lack of ultra-eccentric giant planets is due to
the majority of HJs being formed from disk-driven migration, or whether our
understanding of high-eccentricity migration needs to be revised
remains to be determined. In addition, the discovery that a significant fraction of 
HJs have giant planet companions at a few AU's \citep{knutson2014},
including a number of systems with full orbit solutions for the companions \citep[e.g.][]{feng2015,becker2015,neveu2015},
and the observed stellar-metallicity trend of giant planet
eccentricities \citep{dawson2013}, suggest that LK oscillations driven by stellar companions may not
account for the majority of the observed HJ population.  Regardless,
these issues clearly highlight the need for a better understanding of
all channels of HJ formation.

In this paper, we focus on HJ formation 
in stellar binaries through LK oscillations with tidal
dissipation, and present the results of a large-scale population
synthesis.  Initial population studies of HJ formation by the
LK mechanism included the leading order (quadrupole) gravitational
potential of the binary companion on the planet's orbit
\citep{fabrycky2007, wu2007, correia2011}.  \cite{naoz2012}
incorporated the octupole potential of the binary \citep{ford2000},
and showed that the octupole terms could alter the outcome of the
population synthesis (e.g., they claimed that the efficiency of HJ
production can be significantly increased due to increases in the
maximum eccentricity).  Taking a slightly different approach,
\cite{petrovich2015b} conducted a thorough octupole-level population
synthesis study, focusing on the steady-state distributions of the planet's orbital
elements. He showed that the octupole potential leads to a significant 
increase in the fraction of tidally disrupted planets. Both \cite{naoz2012} and \cite{petrovich2015b} have presented results for
the distribution of the stellar obliquities of HJs formed in this scenario,
showing a broad spread in the spin-orbit misalignment angles 
(from $\sim 20^\circ$ to $\sim 140^\circ$).
Thus far, all population studies have focused on a single planet mass ($1 M_J$) and limited stellar
spin properties.  However, in a recent paper \citep{storch2014}, we showed 
that gravitational interaction between the planet and its oblate host
star can lead to chaotic evolution of the stellar spin axis during LK
cycles, and this evolution depends sensitively on the planet mass and
stellar rotation period. The chaotic spin dynamics arises from
secular spin-orbit resonances and related resonance overlaps 
(Storch \& Lai 2015). In the presence of tidal dissipation, the
complex spin evolution can leave an imprint on the final spin-orbit
misalignment angles. Thus, the result of Storch et al.~(2014) 
shows that the stellar spin properties 
and the planet mass can have a strong effect on the distribution of 
stellar obliquities in HJ systems produced by the LK mechanism.  The
goal of the present paper is to expand upon this previous work by
running a large ensemble of numerical simulations with varying planet
masses and stellar mass and spin properties.  We perform a thorough
survey of the parameter space and examine a range of planetary
semi-major axes, binary separations, inclinations, and eccentricities.
We show that, not only the spin-orbit misalignments are affected by
stellar types and planet masses, but also the various outcomes of the
planets (HJ formation and tidal disruption) are strongly influenced by
the properties of the planets and host stars.  We also present a
number of new analytical calculations and estimates to help understand
our numerical population synthesis results.

This paper is organized as follows.  In Section 2, we describe the
problem setup and present the secular equations of motion that govern
the evolution of the system.  Section 3 presents several analytical
results for understanding the dynamics of the planet's orbit and
stellar spin evolution -- these results will be useful for
interpreting the numerical calculations of later sections.  In Section
4, we investigate the properties of the stellar spin evolution, and
illustrate the various possible paths of generating spin-orbit misalignments.
Section 5 presents our population synthesis calculations. We first
discuss results (with and without octupole effects)
for a given value of binary separation and initial planet
semi-major axis (Sections 5.2-5.3; Table 2). The most general
population synthesis results are presented in Sections 5.4-5.5 (Table 3).
 We conclude in Section 6 with a summary of results
and discussion of their implications.

\section{Formulation}
We consider a hierarchical triple system, consisting of an inner
binary (host star and planet) of masses $M_\star$ and $M_p$, with
a distant, inclined outer (stellar) companion $M_b$.  The planet and binary companion have semi-major axes $a$ and $a_b$ respectively, with $a/a_b \ll 1$.  We include the secular gravitational perturbations
on the planet from the outer companion
to octupole order in the disturbing potential, along with spin-orbit coupling between the oblate host star and planet, tidal dissipation within the planet, and periastron precession due to various short-range forces (General Relativity, and rotational and tidal distortions of the planet).  We ignore the perturbations from the inner binary ($M_\star$ and $M_p$) on the outer binary ($M_\star$ and $M_b$). The planetary orbit is characterized by the unit vectors ($\hatLp$, $\hatep$), where $\hatLp$ is normal to the orbital plane (in the direction of the angular momentum vector $\Lp$) and $\hatep$ is in the direction of the eccentricity vector $\ep$.  Similarly, the orbit of the outer binary is characterized by the unit vectors ($\hatLb$, $\hateb$).  The invariant plane is determined by the outer binary angular momentum axis $\hatLb$.  The secular equations of motion for the planetary orbit take the forms
\be
\frac{d \Lp}{d t} =  \left . \frac{d \Lp}{d t} \right |_{{\rm LK}}
+ \left . \frac{d \Lp}{d t} \right |_{{\rm SL}} + \left
  . \frac{d \Lp}{d t} \right |_{{\rm Tide}}, 
\ee
and
\be
\frac{d \ep}{d t} =  \left . \frac{d \ep}{d t} \right |_{{\rm LK}}
+ \left . \frac{d \ep}{d t} \right |_{{\rm SL}}
+ \left . \frac{d \ep}{d t} \right |_{{\rm SRF}} 
+ \left . \frac{d \ep}{d t} \right |_{{\rm Tide}},
\ee
where we are including contributions from the binary companion that give rise to Lidov-Kozai (LK) oscillations, spin-orbit coupling between the host star spin $S_\star$ and $L$ (SL), dissipative tides (Tide) within the planet, and periastron precession due to short-range forces (SRFs).  Explicit forms for each term are given in Appendix \ref{sec:eqns}.  

Note that the ``LK'' term from the binary companion consists of two pieces: a quadrupole term, and an octupole term.  The quadrupole has a characteristic timescale for LK oscillations $t_k$, given by
\be
\frac{1}{t_k} = \frac{M_{b}}{M_{\rm tot}} \frac{a^3}{\abeff^3} n = \left( \frac{2 \pi}{10^6 \rm yr} \right) \frac{\Mbunit \aunit^{3/2}}{\Mtunit^{1/2} \abunit^3},
\label{tk}
\ee
where $\abeff \equiv a_b \sqrt{1 - e_b^2}$, and $n = \sqrt{G M_{\rm tot} / a^3}$ is the planetary mean motion.  The octupole term has a relative ``strength'' $\varepsilon_{\oct}$ (compared to the quadrupole contribution), given by 
\be
\varepsilon_{\oct} = \frac{M_\star-M_p}{M_\star+M_p}\frac{a}{a_b}\frac{e_b}{1-e_b^2}.
\label{eq:epsilonOct}
 \ee 
(See Table \ref{table:variables} for a summary of various physical quantities and their normalized forms used throughout the paper.)
In terms of the unit vector $\hatLp$, the effect of the binary companion is to induce precession of $\hatLp$ around $\hatLb$, with simultaneous nutation.  The rate of change of $\hatLp$ due to the quadrupole potential of the binary companion is given by
\be
\begin{split}
\Omega_{\rm L}  & = \left | \frac{d \hatLp}{d t} \right |_{{\rm LK, Quad}} \\
& = \big[(\Omega_{\rm pl} \sin \theta_{\LB})^2 + \dot{\theta}_{\LB}^2 \big]^{1/2},
\end{split}
\label{OmegaPL}
\ee
where $\Omega_{\rm pl} = \dot{\Omega}$, the precession rate of the classical orbital node $\Omega$, and $\theta_{\LB}$ (defined as $\cos \theta_{\LB} = \hatLp \cdot \hatLb$) is the angle between the planet orbital axis $\hatLp$ and the binary axis $\hatLb$.  The first term in Eq.~(\ref{OmegaPL}) represents precession of $\hatLp$ around the binary axis $\hatLb$, and the second term represents nutation of $\hatLp$.  An approximate expression for $\Omega_L$ as a function of $e$ and $\theta_{\LB}$ is (see Appendix)
\be
\Omega_L \simeq \frac{3 (1 + 4 e^2)}{8 t_k \sqrt{1 - e^2}} |\sin 2 \theta_{\LB}|.
\label{OmegaL_approx}
\ee  
(Note that Eq.~(\ref{OmegaL_approx}) is exact at $e = 0$ and the maximum eccentricity.)  At zero eccentricity the expression becomes
\ba
\left . \Omega_{\rm L} \right |_{e = 0} & = & \frac{3}{4 t_k} \cos \theta_{\LB} \sin \theta_{\LB} \nonumber\\
& \simeq & 4.71 \times 10^{-6} {\rm yr^{-1}} \frac{\Mbunit \aunit^{3/2}}{\Mtunit^{1/2} \abunit^3} \cos \theta_{\LB} \sin \theta_{\LB}.
\ea

\subsection{Spin Evolution due to Stellar Quadrupole}
\label{sec:spin}
The oblate host star has angular momentum $\Ss = I_\star \Omega_\star \hatS$, where $I_\star = k_\star M_{\star} R_\star^2$ is the moment of inertia, with $k_\star \simeq 0.1$ for a solar-type star \citep{claret1992}, $\Omega_\star$ is the stellar spin frequency (with period $P_\star = 2 \pi / \Omega_\star$), and $\hatS = \Ss / S_\star$ is the unit vector along the spin axis.  The stellar rotational distortion generates a quadrupole moment, thus introducing a torque between the star and planet.  This results in mutual precession of $\Ss$ and $\Lp$ around the total angular momentum ${\bf J} = \Lp + \Ss$ (we ignore the small contribution to ${\bf J}$ due to the planet spin, see Section \ref{sec:planetspin}).  The star also spins down via magnetic braking: we adopt the Skumanich law \citep{skumanich1972}, with $d \Omega_\star / dt \propto -\Omega_\star^3$. The stellar spin evolution thus has two contributions, and is given by
\be
\begin{split}
\frac{d \Ss}{d t} & = \left . \frac{d \Ss}{d t} \right |_{{\rm SL}} + \left . \frac{d \Ss}{d t} \right |_{{\rm MB}} \\
& = \Omega_{\PS} \hatLp \times \Ss - \alpha_{\rm MB} I_\star \Omega_\star^3 \hatS,
\end{split}
\ee
where the first term describes the precession of $\Ss$ around $\Lp$ (SL), and the second term describes the spin-down due to magnetic braking (MB), with the efficiency parameter $\alpha_{\rm MB}$.  In this paper we set $\alpha_{\rm MB} = 1.5 \times 10^{-14} $ yr to model solar-mass (type G) stars, and $\alpha_{\rm MB} = 1.5 \times 10^{-15} $ yr to model more massive ($1.4 \msun$, type F) stars, as in \cite{barker2009}.  This is consistent with observed stellar rotation periods, with massive stars spinning more rapidly on average \citep{mcquillan2014}, and more sophisticated stellar spin-down models (see \citealt{bouvier2013} for a review).

The precession frequency of $\Ss$ around $\Lp$, $\Omega_{\PS}$, is given by
\ba
\Omega_{\PS} & = & - \frac{3 G M_p (I_3 - I_1) \cos \theta_{\SL}}{2 a^3 j^3 S_\star} = - \frac{3}{2} \frac{k_{q \star}}{k_\star} \frac{M_p}{M_\star} \frac{R_\star^3}{a^3} \frac{\Omega_\star}{j^3} \cos{\theta_{\SL}} \nonumber\\
& \simeq & - 1.64 \times 10^{-7} {\rm yr^{-1}} \frac{\bar{k}_{q \star} \Mpunit \Rsunit^3}{\bar{k}_\star \Psunit \Msunit \aunit^3} \frac{\cos \theta_{\SL}}{j^3},
\label{OmegaPS}
\ea
where the stellar spin-orbit angle $\theta_{\SL}$ is defined by $\cos \theta_{\SL} = \hatLp \cdot \hatS$, $j = \sqrt{1 - e^2}$, and the stellar quadrupole moment ($I_3 - I_1$) is related to the spin frequency via $(I_3 - I_1) = k_{q\star} M_\star R_\star^2 \hat{\Omega}_\star^2$.  Here $\hat{\Omega}_\star = \Omega_\star (G M_\star/R_\star^3)^{-1/2}$ is the stellar rotation rate in units of the breakup frequency, and $k_{q\star}$ is a ``rotational distortion coefficient'' (we adopt the canonical value $k_{q\star} = 0.05$ in this paper; \citealt{claret1992}).\footnote{Note that $k_{q\star}$ is related to the apsidal motion constant $\kappa$, the Love number $k_2$, and the $J_2$ parameter by $k_{q\star} = 2 \kappa /3 = k_2/3$ and $J_2 = k_{q\star} \hat{\Omega}_\star^2$.} The stellar quadrupole also affects the planet's orbit through a backreaction torque, and precession of the pericenter (see Section \ref{sec:feedback} and Appendix \ref{sec:eqns}).

As discussed in \cite{storch2014}, qualitatively distinct types of behavior for the stellar spin axis arise, depending on the ratio of the stellar spin precession rate $|\Omega_{\PS} |$ to the nodal precession rate due to the binary companion $| \Omega_{\rm L} |$ (see Eqs.~[\ref{OmegaPS}] and [\ref{OmegaPL}]):

If $|\Omega_{\PS}| \ll |\Omega_{L}|$ throughout the LK cycle, the stellar spin axis effectively precesses around the binary axis $\hatLb$, so that the angle between $\hatS$ and $\hatLb$ is nearly constant.  We refer to this as the ``non-adiabatic'' regime.

On the other hand, if $|\Omega_{\PS}| \gtrsim |\Omega_{\rm L}|$ throughout the LK cycle, the stellar spin axis is strongly coupled to the evolution of the orbital axis.  Two different types of behavior can occur in this ``adiabatic regime'':
(i) The stellar spin axis $\hatS$ essentially follows the orbital axis $\hatLp$, with $\theta_{\SL} \sim$ constant.  For systems that begin with $\hatS$ and $\hatLp$ aligned ($\theta_{\SL,0} = 0^\circ$), the spin-orbit angle remains relatively small $(\theta_{\SL} \lesssim 30^{\circ})$ throughout the evolution. (ii) The spin-orbit angle is initially small, but gradually increases towards the end of the evolution when the planet semi-major axis has decayed appreciably due to tidal dissipation.  In this situation, the final misalignment angle settles to a final value $\theta_{\SL,\rm f} < 90^{\circ}$.  We term this behavior ``adiabatic advection'' and will discuss it in Section \ref{sec:path} (see also \citealt[submitted]{SLA15}). 

Finally, if during the LK cycle, $|\Omega_{\PS}| \sim |\Omega_{L}|$, secular resonances develop, and overlapping resonances can lead to complex, and often chaotic behavior of the stellar spin axis.  The spin-orbit angle $\theta_{\SL}$ may cross $90^{\circ}$, and a wide distribution of final misalignment angles is possible.  Note that $\theta_{\SL}$ can also cross $90^{\circ}$ in the non-adiabatic regime, but the addition of secular resonances in the trans-adiabatic regime leads to much more complex evolution than the non-adiabatic regime.  

To help characterize the dynamics, we introduce an ``adiabaticity parameter'' $\mathcal{A}$:
\be
\mathcal{A} \equiv \left | \frac{\Omega_{\PS}}{\Omega_{L}} \right|. 
\label{eq:Aparameter}
\ee
This parameter will be used throughout the paper to help characterize the spin-orbit dynamics. In general, $\mathcal{A}$ is a strong function of eccentricity and time.  At the start of the evolution (so that $e \approx 0$)
\be
\mathcal{A}_0 \equiv \left | \frac{\Omega_{\PS}}{\Omega_L} \right|_{e = 0} = 0.07 \frac{\bar{k}_{q \star} \Mpunit \Mtunit^{1/2} \Rsunit^3 \abunit^3}{\bar{k}_\star \Msunit \Mbunit \aunit^{9/2} \Psunit} \left | \frac{\cos \theta_{\SL,0}}{\sin 2 \theta_{\LB,0}} \right |.
\label{adiabatic}
\ee
     
\section{LK Migration and Stellar Spin Evolution: Analytical Results}
\label{sec:analytic}

Before presenting our detailed population synthesis calculations, we discuss some general properties of LK migration and stellar spin evolution. These will be useful
for understanding the results of later sections.  Readers interested in the full population synthesis and observational implications are referred to Section \ref{sec:popsynth}.

\subsection{LK Oscillations: Range of Eccentricity and Freezing of 
Oscillations}
\label{sec:efreeze}

\begin{figure*}
\includegraphics[width=\textwidth]{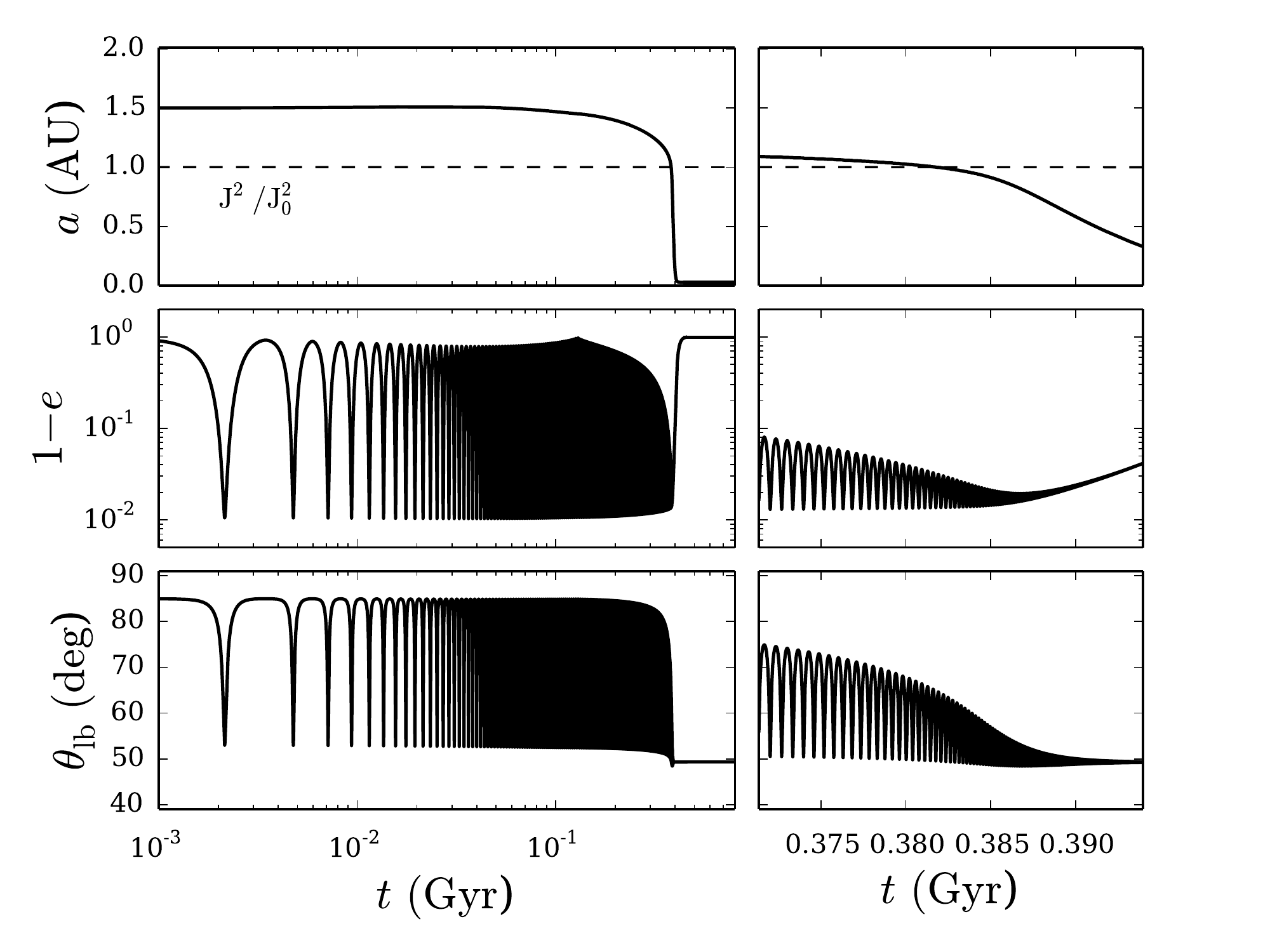}
\caption{Semi-major axis $a$ (top), eccentricity (middle), and
  inclination $\theta_{\LB}$ (bottom) as functions of time, showing the
  evolution until the planetary orbit has decayed and circularized
  (left panels, with logarithmic scale on the x-axis), as well as a zoomed-in version showing the suppression of LK oscillations and tidal
  decay (right panels, with linear scale on the x-axis).  As the orbit decays, the maximum eccentricity of each
  LK cycle is approximately constant, while the minimum eccentricity
  steadily increases, until eventually the LK cycles are completely
  suppressed due to the effects of short-range forces.  The dashed line shows that the angular momentum projected along the binary axis $\hatLb$ (defined by Eq.~[\ref{kozai_constant_diss}]) is conserved throughout the evolution. Parameters are
  $M_p = 5 M_J$, $a_0 = 1.5$ AU, $a_b = 200$ AU, $e_b = 0$, $\theta_{\LB,0} =
  85^{\circ}$.  The other parameters assume their canonical values, as defined in Table \ref{table:variables}.}
\label{fig:ecc_example}
\end{figure*}

Figure \ref{fig:ecc_example} gives a ``canonical'' example of the
formation of an HJ due to LK oscillations with tidal dissipation.  For
simplicity, this example neglects the feedback of the
stellar spin on the orbit.  Here we set the binary eccentricity $e_b = 0$, so that the octupole-level
perturbation from the binary companion vanishes.  The planet starts with initial semi-major
axis $a_0 = 1.5$ AU, and eccentricity $e_0 = 0.01$, and then undergoes
cyclic excursions to maximum eccentricity $\emax$, with
accompanying oscillations in the inclination $\theta_{\LB}$ (recall that $\cos \theta_{\LB} = \hatLp \cdot \hatLb$), between the initial
(maximum) $\theta_{\LB,0} = 85^{\circ}$ and minimum (occurring at $e =
\emax$) $\theta_{\LB,\rm max} \approx 53^{\circ}$.  Note that short-range forces (SRFs) cause $\theta_{\LB,\rm max} > 40^{\circ}$ here, in contrast to planets subject only to LK oscillations (without SRFs).  As the planetary
orbit decays, the range of eccentricity oscillations
becomes smaller.  The example shows that before the oscillations freeze, $\emax$ is approximately
constant in time, while the minimum eccentricity 
$\emin$ steadily increases toward $\emax$.  Eventually, when $a$ is sufficiently small,
the LK oscillations freeze, and the planet undergoes ``pure'' orbital decay/circularization
governed by tidal dissipation, at nearly constant angular momentum.

As is well recognized in previous work \citep[e.g.][]{holman1997,
  wu2003,fabrycky2007,liu2015}, SRFs
play an important role in determining the maximum eccentricity $e_{\rm
  max}$ in LK cycles.  The range of eccentricity oscillations during
the LK migration can also be understood from the effects of SRFs, as we
discuss below.  As in the example depicted in Fig.~\ref{fig:ecc_example}, we ignore the stellar spin feedback on the planetary
orbit, as well as octupole-level perturbations from the binary
companion.

In the absence of tidal dissipation, the evolution of the planetary
orbit is governed by two conservation laws.  The first, which is
related to the component of the angular momentum along the binary
axis, is the well-known ``Kozai constant'', given by
\be
K = j \cos \theta_{\LB}, \quad {\rm where} 
\quad
j = \sqrt{1 - e^2}.
\label{kozai_constant}
\ee
The second conserved quantity is the energy per unit mass, which in secular form is given by \citep[e.g.][]{fabrycky2007,liu2015}
\be
\Phi = \Phi_{\rm Quad} + \Phi_{\GR} + \Phi_{\ST} + \Phi_{\OP},
\ee
where the subscripts ``${\rm Quad}$'', ``${\GR}$'', ``${\ST}$'', and ``${\OP}$'' denote contributions from the binary companion (to
quadrupole order), General Relativity, static tidal deformation of the
planet, and the rotational deformation of the planet.  In terms of the
planet's eccentricity ($e$), inclination ($\theta_{\LB}$), and
argument of pericenter ($\omega$), the energy (per unit mass) from the
binary companion takes the form
\be
\Phi_{\rm Quad} = \frac{\Phi_0}{8} \big(1 - 6 e^2 - 3 K^2 + 15e^2
  \sin^2 \theta_{\LB} \sin^2 \omega \big),
\label{energy_constant}
\ee  
where
\be
\Phi_0 = \frac{G M_b a^2}{\abeff^3}.
\ee
The remaining energy terms due to SRFs can be written as
\ba
&& \Phi_{\GR} = - \varepsilon_{\GR} \frac{\Phi_0}{j}, \nonumber\\
&& \Phi_{\ST} = - \varepsilon_{\ST} \frac{\Phi_0}{15} \frac{1 + 3 e^2 + 3 e^4/8}{j^{9}}, \nonumber\\
&& \Phi_{\OP} = - \varepsilon_{\OP} \frac{\Phi_0}{3 j^3},
\ea
where we have defined dimensionless parameters $\varepsilon_{\GR}$, $\varepsilon_{\ST}$ and $\varepsilon_{\OP}$ that quantify the 
``strengths'' of the SRFs:
\ba 
\varepsilon_{\gr} & \equiv & \frac{3G \Mtot^2 \abeff^3}{M_b a^4c^2} 
\simeq 0.03 \frac{ \Mtunit^2 \abunit^3 }{ \Mbunit \aunit^4},\label{eq:epsigr}
\ea
\ba 
\varepsilon_{\ST} & \equiv & \frac{15 k_{2p} M_\star M_{\rm tot} \abeff^3
R_p^5}{M_bM_p a^8} \nonumber\\
& \simeq & 1.47\times 10^{-7} \frac{ \Msunit \Mtunit \abunit^3 \Rpunit^5} {
\Mbunit \Mpunit \aunit^8},\label{eq:epsitide}
\ea
\ba 
\varepsilon_{{\OP}} & \equiv & \frac{3 k_{qp}}{2} \hat{\Omega}_p^2
\frac{M_{\rm tot}}{M_b} \left( \frac{R_p}{a} \right)^2 \left(
\frac{\abeff}{a} \right)^3 \quad \nonumber\\ 
& \simeq & 8.48 \times 10^{-4}
\bar{k}_{qp} \left ( \frac{P_p}{1 \rm day}
\right)^{-2} \frac{\Mtunit \Rpunit^5 \abunit^3}{\Mpunit \Mbunit \aunit^5}. 
\ea
(see Table \ref{table:variables} for definitions of $k_{2p}$ and $k_{qp}$).

With tidal dissipation included, the semi-major axis is no longer
constant.  We expect that the first conservation law, Eq.~(\ref{kozai_constant}) is replaced by
\be
J = \sqrt{a(1 - e^2)} \cos \theta_{\LB} = \sqrt{a} j \cos \theta_{\LB}.
\label{kozai_constant_diss}
\ee
Figure \ref{fig:ecc_example} shows that $J$ is indeed conserved to
high precision throughout the LK migration.  With $a \neq $ constant,
the energy expression, Eq.~(\ref{energy_constant}) is no longer
conserved.  However, since the timescale for tidal dissipation
(see Section \ref{sec:migrationrate}, Eq.~[\ref{eq:maxrate}]) is much longer than the timescale for LK
oscillations (Eq.~[\ref{tk}]), the energy is very nearly constant
over a single LK cycle.

As seen from Fig.~\ref{fig:ecc_example}, during the oscillatory
phase of the LK migration, the maximum eccentricity of each LK cycle
$e_{\mx} \approx$ constant, while the minimum eccentricity steadily
increases, so that the range of eccentricity variation narrows (see
right panels of Fig.~\ref{fig:ecc_example}). The inclination at maximum eccentricity, $\theta_{\LB,\max}$, is also nearly constant.  For given $\emax$ and $\theta_{\LB,\max}$, the minimum eccentricity $\emin$ can
be determined using the two (approximate) conservation laws, giving
\ba
&&\frac{3}{4} \emin^2  = \frac{3}{8} \emax^2 \big(2 - 5 \sin^2 
\theta_{\LB,\rm max}\big)\nonumber\\
&& \quad +\left[\frac{\varepsilon_{\GR}}{j}
  + \frac{\varepsilon_{\ST}}{15 j^9}\Bigl(1 + 3e^2 + {3e^4\over 8}\Bigr) 
  + \frac{\varepsilon_{\rm Rot}}{3j^3}\right]\Biggr|_{\emin}^{\emax}.
\ea
Here we have used the fact that the maximum eccentricity occurs when
$\omega = \pi/2$ or $3 \pi / 2$, while the minimum eccentricity occurs
at $\omega = 0$ or $\pi$ (provided that $\omega$ is in the
circulating, rather than librating regime).
For reasonable values of the planetary rotation rate (see Section \ref{sec:planetspin}), the SRF effect
due to the rotational bulge can be neglected compared to the tidal effect.

We can now determine the condition for the suppression (freezing) of
LK oscillations.  Since the freezing occurs at $\emax$ close to 1, 
it is more appropriate to consider the freezing
of $j$.  For $\Delta j \equiv j_{\rm min} - j_{\rm max} = \sqrt{1 -
  \emin^2} - \sqrt{1 - \emax^2} \ll j_{\rm max}$, we find that
\be 
\frac{\Delta j}{j_{\rm
      max}} \approx \frac{15}{8}  
\sin^2\!\theta_{\LB,\rm max}\left( \frac{\varepsilon_{\GR}}{j_{\rm max}} + \frac{21}{8}
\frac{\varepsilon_{\ST}}{j_{\rm max}^9} \right)^{-1}.
\label{eq:djmax}
\ee
(Note that the subscript ``max'' indicates the value at maximum eccentricity.) As $a$ decreases, both $\varepsilon_{\GR}$ and $\varepsilon_{\ST}$ 
increase rapidly, which leads to the decrease of $\Delta j$.  The fact that $\theta_{\LB,\max}$ is nearly constant (see Fig.~\ref{fig:ecc_example}), along with conservation of $J$ (see Eq.~[\ref{kozai_constant_diss}]), together imply that $j_{\max} \propto a^{-1/2}$.
For $\varepsilon_{\GR}/j_{\rm max}\go (21/8)\varepsilon_{\ST}/j_{\rm max}^9$, 
or
\be
j_{\rm max}\go \left({21\varepsilon_{\rm Tide}\over 8\varepsilon_{\rm GR}}
\right)^{1/8}=0.245{\Rpunit^{5/8}\over \Mpunit^{1/8} \aunit^{1/2}},
\label{eq:jmaxgr}\ee
the GR term dominates, and we have
\be
\frac{\Delta j}{j_{\rm max}} \simeq 0.1 \frac{\Mbunit}{\Msunit^2 \abunit^2} \left(\frac{a}{0.3 \, {\rm AU}} \right)^4 \left( \frac{j_{\rm max}}{0.2} \right)\sin^2\!\theta_{\rm lb,max}.
\ee
When equation (\ref{eq:jmaxgr}) is not satisfied, the tidal term dominates,
and we have
\be
{\Delta j\over j_{\rm max}}\simeq 0.01 {\Mbunit \Mpunit \over \Msunit^2
\abunit^3}\left({a\over 0.5\,{\rm AU}}\right)^8
\left({j_{\rm max}\over 0.2}\right)^9\sin^2\!\theta_{\rm lb,max}.
\ee

\begin{figure}
\centering 
\includegraphics[scale=0.5]{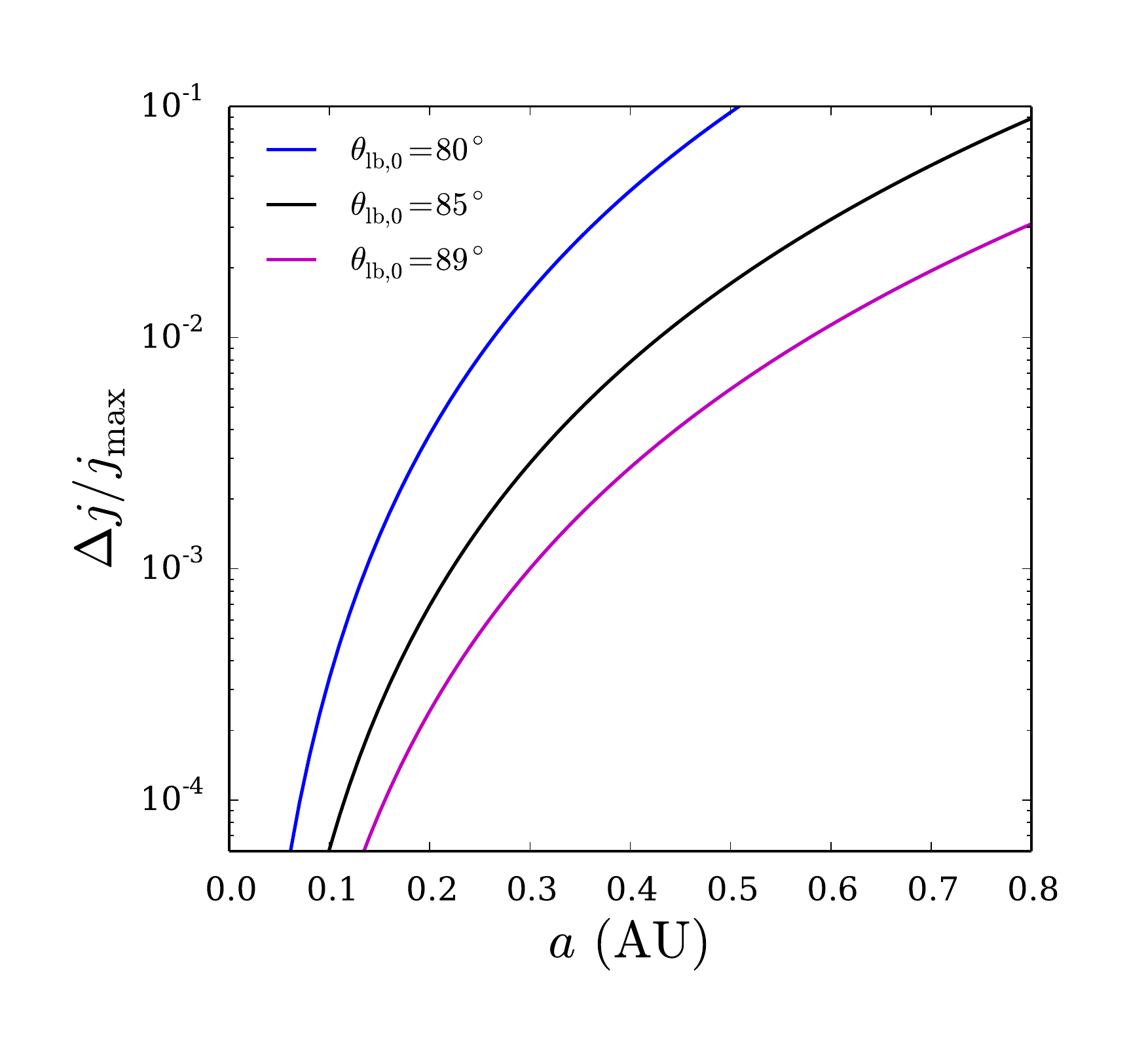}
\caption{Condition for freezing of LK oscillations, $\Delta j / j_{\rm max}$ as a function of $a$ using Eq.~(\ref{eq:djmax}) (where we assumed $\Delta j / j_{\rm max} \ll 1$), where $j_{\max}  = \sqrt{1 - \emax^2}$ has been calculated from Eq.~(\ref{kozai_constant_diss}), with the assumption that $\theta_{\LB,\max} \sim \theta_{\LB,0}$.  We have chosen three values of $\theta_{\LB,0}$, as labeled, and all other parameters the same as in Fig.~\ref{fig:ecc_example}.  As $a$ decreases (so that $\varepsilon_{\gr}$ and $\varepsilon_{\tide}$ increase), SRFs limit the eccentricity variation, causing $\Delta j$ to decrease.}
\label{fig:djmax}
\end{figure}

Figure \ref{fig:djmax} shows $\Delta j / j_{\rm max}$ as a function of $a$ using Eq.~(\ref{eq:djmax}) (where $j_{\max}$ has been calculated from Eq.~[\ref{kozai_constant_diss}]), for the same system parameters as depicted in Fig.~\ref{fig:ecc_example}, and three values of $\theta_{\LB,0}$.  We see that $\Delta j / j_{\rm max}$ decreases with decreasing $a$, as SRFs increasingly suppress the LK oscillations.  

\subsection{Migration Rate: Upper Limit and Estimate}
\label{sec:migrationrate}
For a given $a$ and $e$, the orbital decay rate (using weak friction tidal theory) takes the form \citep{alexander1973,hut1981}
\be
\left( \frac{1}{a} \frac{da}{dt} \right)_{\rm Tide} = - \frac{1}{t_a} \frac{1}{j^{15}} \bigg[f_1 (e) - j^3 f_2 (e) \frac{\Omega_p}{n}\bigg], 
\ee
where the dimensionless functions of eccentricity $f_1$ and $f_2$ are defined in Eqs.~(\ref{eq:F1}) and (\ref{eq:F2}).
The timescale $t_a$ is given by
\ba
\frac{1}{t_a} & = & 6 k_{2p} \Delta t_{\rm L} 
\frac{M_*}{M_p} \left( \frac{R_p}{a}\right)^5 n^2 \nonumber\\
& \approx & \frac{7.3 \times 10^{-21}}{\rm yr} \chi \bar{k}_{2p} \frac{\Msunit \Mtunit}{\Mpunit} \frac{\Rpunit^5}{\aunit^8},
\ea
where $\Delta t_L$ is the lag time, $k_{2p}$ is the tidal Love number, and we have introduced a tidal enhancement factor $\chi$ (relative to Jupiter), defined such that $\Delta t_L = 0.1 \chi$ sec.  Our canonical value is $\chi = 10$.  It is convenient to introduce the quantity
\be
a_F \equiv a (1 - \emax^2),
\ee
because $a_F$ varies by at most $\sim 20 \%$ during the inward migration of a planet undergoing LK cycles.  Note that $a_F$ is approximately equivalent to the final (``circularized'') semi-major axis of the planet.  To produce HJs, we require $a_F \lesssim 0.05$ AU (i.e. orbital periods less than $\sim 4$ days).

For a given value of the planetary spin rate $\Omega_p$, the maximum decay rate occurs for $e = \emax$ (see Section \ref{sec:planetspin} for a discussion of our treatment of the planetary spin).  Setting $\Omega_p \simeq 0$ for simplicity, the maximum decay rate is
\be
\begin{split}
\left|  \frac{1}{a} \frac{da}{dt} \right|_{\rm Tide, max} & = \frac{1}{t_a} \frac{f_1 (\emax)}{j_{\rm max}^{15}} \\ & \approx \frac{2.52 \times 10^{-9}}{\rm yr}  \chi \bar{k}_{2p} \frac{\Mtunit \Msunit \Rpunit^5}{\Mpunit \aunit^{1/2}} \left( \frac{\bar{a}_F}{0.05} \right)^{-15/2}. 
\end{split}
\label{eq:max_decay_rate}
\ee
Non-zero values of the planetary spin rate $\Omega_p$ would slightly modify the numerical coefficient in Eq.~(\ref{eq:max_decay_rate}).

Eq.~(\ref{eq:max_decay_rate}) overestimates the actual LK migration rate, since the planet spends only a small fraction of time near high eccentricity during an LK cycle.  We can estimate the time spent in the vicinity of $\emax$ as follows.  Neglecting SRFs, the planet's argument of pericenter $\omega$ evolves according to
\be
\frac{d \omega}{d t} = \frac{3}{4 t_k \sqrt{1 - e^2}} \big[2(1 - e^2) + 5 \sin^2 \omega (e^2 - \sin^2 \theta_{\LB})\big].
\label{omegadot}
\ee
Near maximum eccentricity, $\omega$ centers around $\pi / 2$ or $3 \pi
/ 2$, with width of $\Delta \omega \sim 1$ radian \cite[see,
e.g.][Fig.~3]{holman1997}.  Thus, the second term in
Eq.~(\ref{omegadot}) is of order unity and the first term is
negligible, so that the time spent near $\emax$ can be approximated by
\be
\Delta t (\emax) \sim t_k \sqrt{1 - \emax^2}.
\label{eq:t_emax}
\ee
Thus, the actual orbital decay rate during LK migration is roughly
\be
\begin{split}
\left| \frac{1}{a} \frac{da}{dt} \right|_{\rm Tide, LK} & \sim  \left| \frac{1}{a} \frac{da}{dt} \right|_{\rm Tide, max} (1 - \emax^2)^{1/2} \\
 & \simeq \frac{5.6 \times 10^{-10}}{\rm yr}  \chi \bar{k}_{2p} \frac{\Mtunit \Msunit \Rpunit^5}{\Mpunit \aunit} \left(\frac{\afunit}{0.05} \right)^{-7}
\end{split}
\label{eq:maxrate}
\ee
(see also \citealt{petrovich2015b} for a more detailed exploration of
the LK migration rate).  Since the main-sequence lifetime of a solar-type star is $\sim 10^{10}$ yr, inward migration resulting in HJ formation requires that $a_F \lesssim 0.05$ AU.

\subsection{Evolution of Planet Spin During LK Cycles with Tidal Friction}
\label{sec:planetspin}
Similar to the spin axis of the host star, the spin axis of the oblate planet $\hatSp$ (where the spin angular momentum is $\Sp = S_p \hatSp$) precesses around the orbital axis $\hatLp$ according to
\be
\frac{d \hatSp}{dt} = \Omega_{{\rm prec},p} \hatLp \times \hatSp, 
\ee
where the precession rate $\Omega_{{\rm prec},p}$ is given by
\ba
\Omega_{{\rm prec},p} & = & - \frac{3}{2} \frac{k_{qp}}{k_p} \frac{M_\star}{M_p} \frac{R_p^3}{a^3} \frac{\Omega_p}{j^3} \cos{\theta_{p}}\nonumber\\ 
& \simeq & - 2.69 \times 10^{-4} {\rm yr^{-1}} \frac{\bar{k}_{qp} \Msunit \Rpunit^3}{\bar{k}_p \Mpunit \aunit^3} \frac{\cos{\theta_{p}}}{j^3},
\ea
with $\cos \theta_{p} = \hatSp \cdot \hatLp$ (see Table 1 for definitions and canonical values of all other quantities). 
We can define a planetary ``adiabaticity parameter'' $\mathcal{A}_{\rm p,0}$ (analogous to the stellar adiabaticity parameter $\mathcal{A}_0$, see Eq.~[\ref{adiabatic}]), where
\be
\mathcal{A}_{p,0} \equiv \left | \frac{\Omega_{{\rm prec},p}}{\Omega_L} \right|_{e = 0} \simeq 57.1 \frac{\bar{k}_{qp} \Msunit \Mtunit^{1/2} \Rpunit^3 \abunit^3}{\bar{k}_p \Mpunit \Mbunit \aunit^{9/2} \Ppunit} \left | \frac{\cos \theta_{p}}{ \cos \theta_{\LB} \sin \theta_{\LB}} \right |.
\ee
Clearly, for all plausible parameters, $\mathcal{A}_{p,0} \gg 1$, provided that the planetary obliquity $\theta_p$ is not too close to $90^{\circ}$.  The planetary spin axis is thus always in the adiabatic regime (see Section \ref{sec:spin}), with the planetary spin orbit angle $\theta_{p} \approx$ constant.  

We thus treat the direction of the planetary spin axis as always being aligned with the orbital axis $\hatLp$, and the spin magnitude $S_p = k_p M_p R_p^2 \Omega_p$ evolves only due to tidal dissipation.  After averaging over the periastron precession \citep[e.g.][]{alexander1973,hut1981,correia2011}, the evolution of $S_p$ is governed by the expression
\be
\left ( \frac{1}{S_p} \frac{d S_p}{dt} \right)_{\rm Tide} = -\frac{1}{2 t_a} \frac{L}{S_p} \frac{1}{j^{13}} \left[ j^3 f_5(e) \frac{\Omega_p}{n} - f_2(e) \right],
\label{eq:dSpdt}
\ee
where $f_2$ and $f_5$ are functions of eccentricity, defined in Eqs.~(\ref{eq:F2}) and (\ref{eq:F5}). The magnitude of the orbital angular momentum evolves according to $(d L/dt)_{\rm Tide} = - (d S_p / dt)_{\rm Tide}$.

\begin{figure}
\centering 
\includegraphics[scale=0.5]{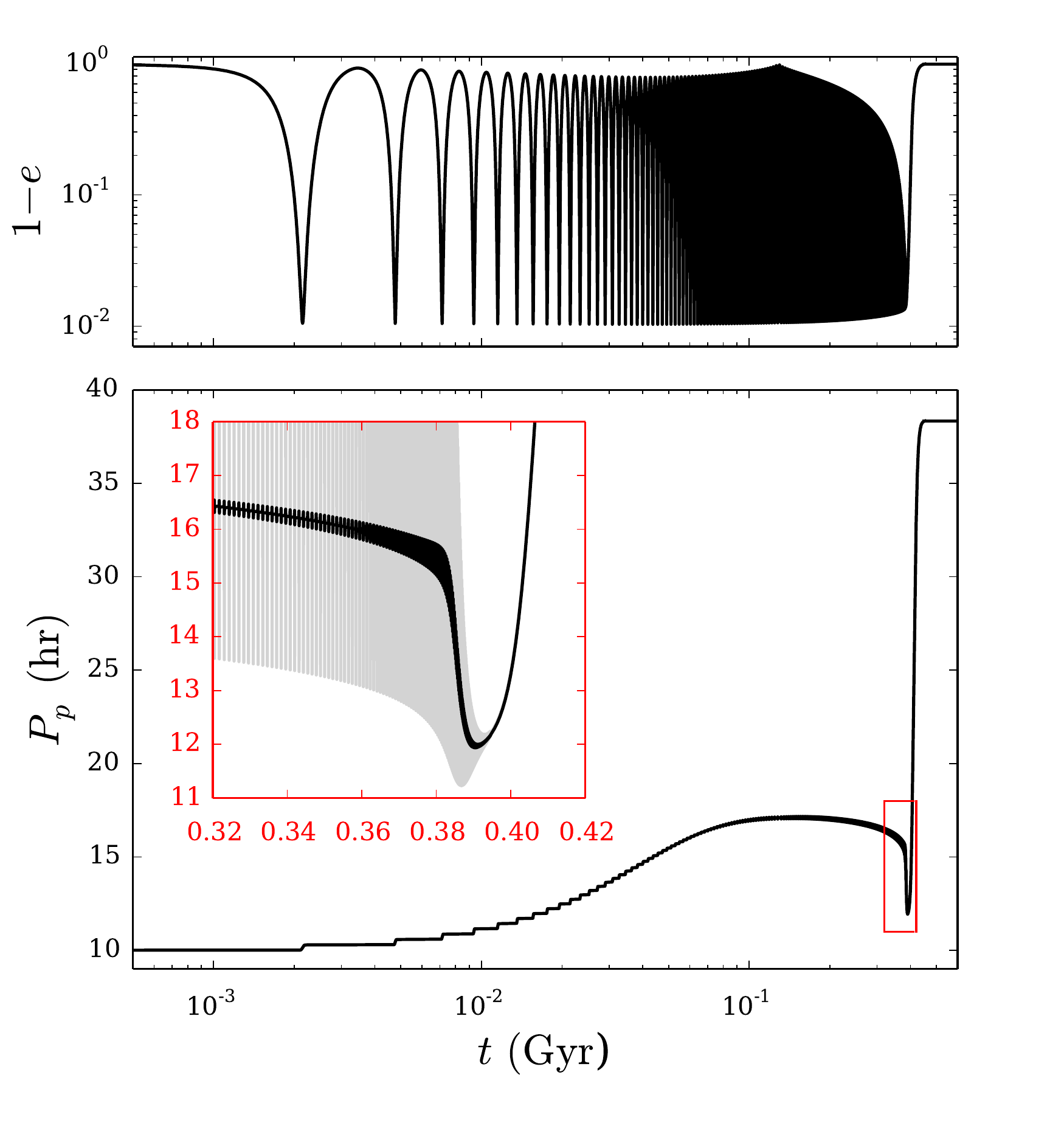}
\caption{Planet spin period as a function of time, for the same parameters shown in Fig.~\ref{fig:ecc_example}.  For reference, we also show the variation of the orbital eccentricity $1 - e$ (top panel).  The planet spin remains constant during the low-eccentricity phase of each LK cycle, and undergoes a rapid ``jump'' during the brief high-eccentricity phase.  The bottom panel shows $P_p$ over the entire evolution (until the LK cycles are suppressed and the semi-major axis decays to the final value), and the inset shows a zoomed-in portion of the spin evolution, as indicated by the red-boxed region ($0.32$ Gyr $\lesssim t \lesssim 0.42$ Gyr).  On timescales much longer than $t_k$, but shorter than the orbital decay time, the spin period reaches ``Kozai spin equilibrium'' (see text).  As the LK oscillations are suppressed (see Section \ref{sec:efreeze}), the equilibrium spin period approaches the pseudo-synchronized value (Eq.~\ref{pseudo}), drawn in light-grey in the inset panel.}
\label{fig:Pspin}
\end{figure}

\begin{figure}
\centering 
\includegraphics[scale=0.5]{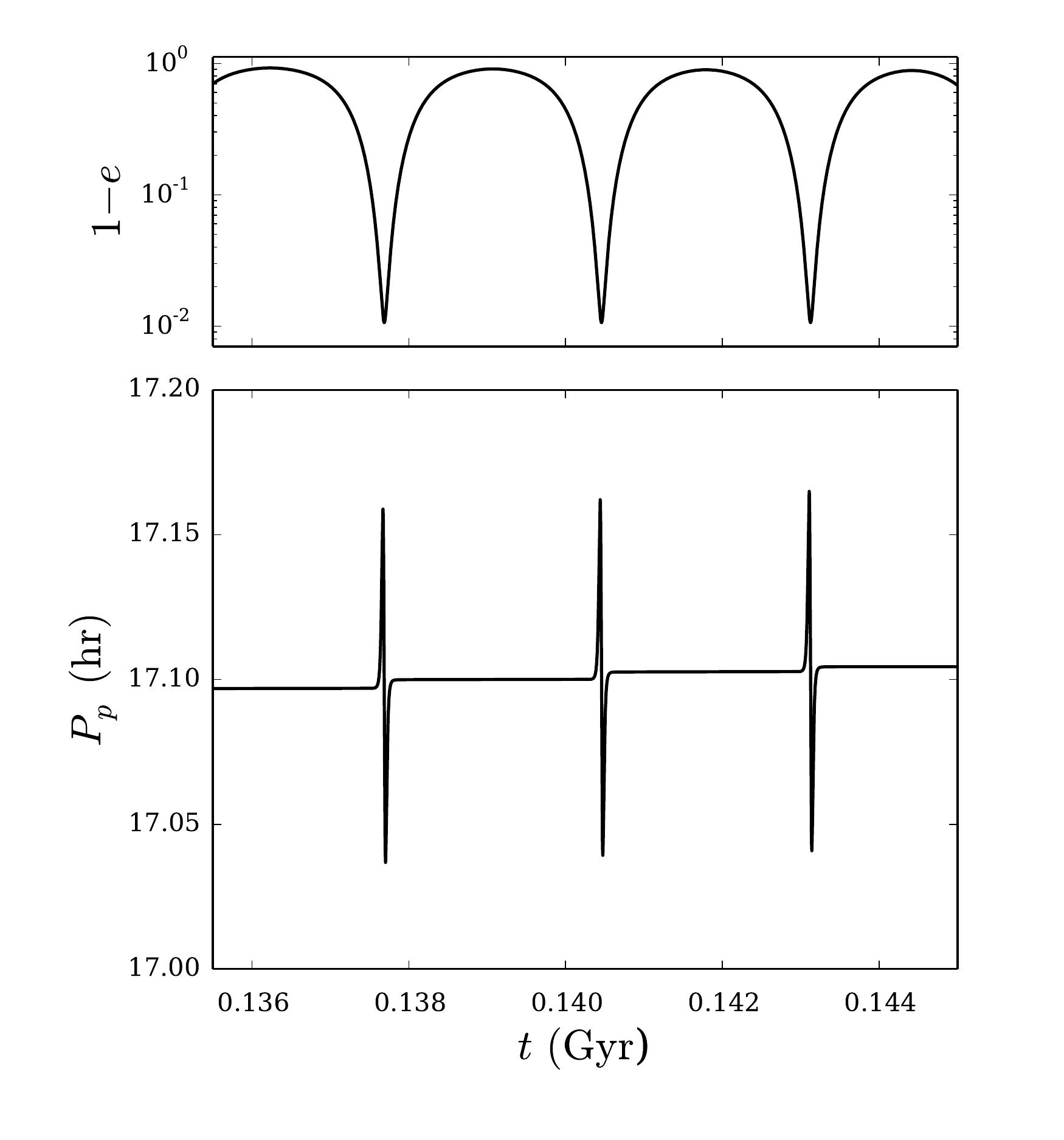}
\caption{Same as Figure \ref{fig:Pspin}, but showing only three LK cycles, once the planet spin has achieved the ``Kozai spin equilibrium'' (see text).}
\label{fig:Pspin_zoom}
\end{figure}

A fiducial example of the planetary spin behavior is shown in Fig.~\ref{fig:Pspin}, for the same parameters as in Fig.~\ref{fig:ecc_example}.  The planet spin period is initialized to $P_p = 10$ hours, and exhibits complex behavior, as it tidally evolves while under the external forcing of the binary companion.  During the low-$e$ phase of each LK cycle, the planet spin magnitude remains nearly constant, and then undergoes a rapid ``jump'' (with $|\Delta P_p|/P_p \ll 1$) during the high-$e$ phases.  After many LK cycles, a state of near equilibrium is reached, so that the spin period at low eccentricity returns to nearly the same value after the high-$e$ ``jump'' (see Fig.~\ref{fig:Pspin_zoom}).  As the LK cycles begin to be suppressed due to orbital decay, the range of eccentricity narrows (see Section \ref{sec:efreeze}), and the spin period gradually decreases.  Once the LK cycles are completely suppressed, the spin period increases and eventually settles to a final value $P_p \simeq 38$ hours, synchronized with the final orbital period of the planet.     

We can understand the behavior of the planetary spin under the influence of LK cycles as follows.  The timescale for planetary spin variation due to tidal dissipation is (see Eq.~[\ref{eq:dSpdt}])
\be
\begin{split}
t_{\rm spin} & =  \left | \frac{S_p}{\dot{S}_p} \right | \sim \frac{S_p}{L} t_a j^{13} \\ 
& \simeq 2.9 \times 10^{3} {\rm yr} \frac{\bar{k}_p}{\bar{k}_{2p} \chi} \frac{\Mpunit \aunit^{15/2}}{\Msunit^2 \Mtunit^{1/2} \Rpunit^3} \left( \frac{P_p}{1 \rm day} \right)^{-1} \left( \frac{j}{0.1}\right)^{12}.
\end{split}
\label{eq:tspin}
\ee
This is much less than the orbital decay circularization timescale due to tides, $t_{\rm circ} \sim t_a j^{13}$, or the orbital decay time ($\sim t_a j^{15}$) for all values of $a$ and $e$.  Therefore, in the absence of an external perturber (i.e. when the system is governed purely by tidal dissipation), the planetary spin reaches a state of pseudo-synchronization, with
\be
\Omega_{p, \rm eq} = \Omega_{p, \rm pseudo} = \frac{f_2(e)}{j^3 f_5(e)} n.
\label{pseudo}
\ee

\begin{figure}
\centering 
\includegraphics[scale=0.5]{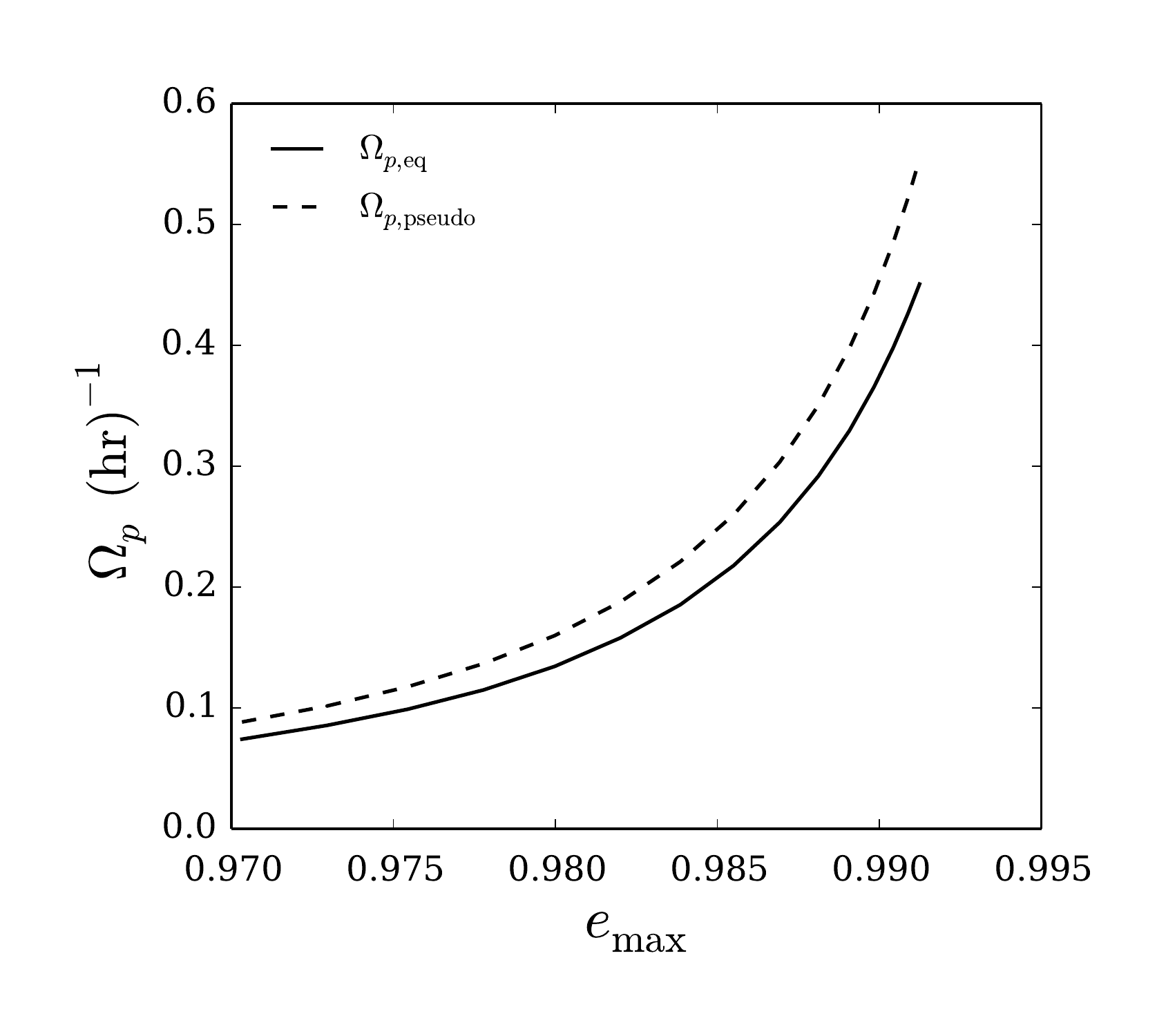}
\caption{``Kozai spin equilibrium rate'' rate ($\Omega_{p,\rm eq}$,
  solid curve), as a function of $\emax$, the maximum eccentricity
  attained in an LK cycle.  For comparison, we also plot the
  pseudo-synchronized rate at $\emax$ ($\Omega_{p,\rm pseudo}$, dashed
  curve).  We vary the maximum eccentricity by varying the initial
  inclination $\theta_{\LB,0}$, and integrate a set of simplified
  equations for a single LK cycle (accounting for pericenter
  precession due to GR and static tides, but neglecting the precession
  due to planetary rotation).  We further ignore orbital decay.
  Parameters are $M_p = 1 M_J$, $a = 1.5 \rm AU$, $a_b = 200 \rm AU$,
  $e_b = 0$.}
\label{fig:eqspin}
\end{figure}

The situation is very different when the planet undergoes LK oscillations driven by an external perturber.  The time the planet spends around eccentricity $e$ in each LK cycle is of order $\Delta t_k \sim t_k \sqrt{1 - e^2}$ (see Eqs.~[\ref{tk}] and [\ref{eq:t_emax}]).  Note that the spin evolution timescale $t_{\rm spin}$ (see Eq.~[\ref{eq:tspin}]) depends strongly on eccentricity.
During the low-eccentricity phase of the LK cycle, $t_{\rm spin} \gg \Delta t_{k}$, so that the spin magnitude remains constant.  However, during the brief high-eccentricity phase, $t_{\rm spin} $ can be comparable to $\Delta t_{k}$, and the planetary spin magnitude undergoes a small ``jump'' $\Delta \Omega_p$. Assuming $| \Delta \Omega_p | / \Omega_p \ll 1$, this jump can be calculated from
\be
\frac{\Delta \Omega_p}{\Omega_p} \simeq - \int_{-t_k / 2}^{t_k / 2} \frac{1}{2 t_a j^{13}} \frac{L}{S_p} \left[ j^3 f_5(e) \frac{\Omega_p}{n} - f_2(e) \right] dt,
\label{deltaOmega}
\ee
where $e = e(t)$, and the time integration covers a single LK cycle centered around the eccentricity maximum.  On timescales much longer than $t_k$ but shorter than the orbital decay time, the spin rate approaches a constant value $\Omega_{p, \rm eq}$, the ``Kozai spin equilibrium,'' such that $\Delta \Omega_p = 0$.  For ``canonical'' system parameters ($M_p = 1 M_J$, $a_0 = 1.5 \rm AU$, $a_b = 200 \rm AU$), and varying initial inclination (corresponding to varying $\emax$), we determine $\Omega_{p, \rm eq}$ by adjusting the initial planetary spin rate, integrating for a single LK cycle, and iterating until $\Delta \Omega_p = 0$ in Eq.~(\ref{deltaOmega}).  The results are depicted in Figure \ref{fig:eqspin}.  We see that the Kozai spin equilibrium differs from the pseudo-synchronized value at $\emax$, with the ratio $\Omega_{p,\rm eq} / \Omega_{p, \rm pseudo}(\emax) \approx 0.8$.

\subsection{Limiting Eccentricity and Necessary Conditions for Planet Migration and Disruption}
\label{sec:disrupt}

When the octupole potential from the binary companion is neglected,
the maximum eccentricity $e_\m$ attained by the planet in LK cycles can 
be determined by the conservation laws discussed in Section 3.1.
If the initial eccentricity of the planet is close to zero and the initial
inclination is $\theta_{\rm lb,0}$, we find
(Liu et al.~2015):
\ba
&&\varepsilon_{\gr}\Big(\frac{1}{j_\m}-1\Big)
+\frac{\varepsilon_{\tide}}{15}\Big(\frac{1+3e^2_\m+\frac{3}{8}e^4_\m}
{j_\m^9}-1\Big)\nonumber\\
&& +\frac{\varepsilon_{\rot}}{3}\Big(\frac{1}{j_\m^3}-1\Big)
=\frac{9e_\m^2}{8j_\m^2}\Big(j_\m^2-\frac{5}{3}\cos^2\!
\theta_{\rm lb,0}\Big).
\label{eq:emax}
\ea
The limiting eccentricity $e_\li$ is achieved at $\theta_{\rm lb,0}=90^\circ$.
For $e_\m\simeq 1$, we have
\be
{\varepsilon_\gr\over j_\li}+{7\varepsilon_\tide \over 24 j_\li^9}\simeq 
{9\over 8},
\label{eq:jlim}\ee
where
\be
j_\li\equiv (1-e_\li^2)^{1/2},
\ee
and we have neglected the effect associated with the planetary rotational bulge
(since it is generally smaller than the tidal term).

When the octupole potential is included, the ``Kozai constant'' $K$
[Eq.~(\ref{kozai_constant})] is no longer a constant of motion, thus Eq.~(\ref{eq:emax}) is not valid. Nevertheless, \cite{liu2015}
show that the limiting eccentricity, as determined by Eq.~(\ref{eq:jlim}) still provides an upper limit to the achievable
eccentricity in the LK cycles in the presence of SRFs.
The effect of the octupole potential is to make the planet
undergo occasional excursion into $e_\li$ even when $\theta_{\rm lb,0}
\neq 90^\circ$. In general, $e_\li$ can be attained for a range of
$\theta_{\rm lb,0}$ centered around $90^\circ$, with the range
becoming wider as the octupole  
parameter $\varepsilon_{\oct}$ increases (see Eq.~[\ref{eq:epsilonOct}]).

For a given set of system parameters ($M_\star, M_b, M_p, R_p, a, a_b, e_b$), Eq.~(\ref{eq:jlim}) determines the limiting eccentricity (or limiting periastron distance $a_{p,\li}\equiv 
a[1-e_\li]$)
\be
0.021{\Msunit^2\abunit^3\over \Mbunit \bar{a}_{p,\li}^{1/2}\aunit^{3.5}}
+1.89\times 10^{-9}
{\Msunit^2\abunit^3 \Rpunit^5\over
\Mbunit \Mpunit \bar{a}_{p,\li}^{4.5}\aunit^{3.5}}={9\over 8},
\ee
where we have used Eqs.~(\ref{eq:epsigr}) and (\ref{eq:epsitide}).  For $j_\li\go j_{\li,c}$, where
\be
j_{\li,c}^2=\left({7\varepsilon_\tide\over 24\varepsilon_\gr}\right)^{1/4}
=3.46\times 10^{-2}{\Rpunit^{5/4}\over \Mpunit^{1/4}\aunit},
\ee
the GR effect dominates SRFs, and we have
\be
j_\li^2=1-e_\li^2=7.1\times 10^{-4}
\left({\Msunit^2 \abunit^3\over \Mbunit \aunit^4}\right)^2.
\ee
For $j_\li\lo j_{\li,c}$, tides dominate the SRF, and we have
\be
j_\li^2=1-e_\li^2=2.25\times 10^{-2}\left({\Msunit^2 \Rpunit^5 \abunit^3
\over \Mbunit \Mpunit \aunit^8}\right)^{2/9}.
\ee

As discussed in Section \ref{sec:migrationrate}, for a planet to migrate, its pericenter distance $a_p$ must be sufficiently small, so that tidal dissipation can damp and circularize the orbit within a few Gyrs.  We therefore require $a_{p,\li} \lesssim a_{p, \rm crit}$, where $a_{p,\rm crit}$ is the maximum pericenter distance needed to circularize the orbit within a specified time frame.  Note that $a_{p,\rm crit}$ depends on the tidal dissipation strength, and therefore is a fuzzy number.  However, for reasonable tidal dissipation strengths, and circularization times of a few Gyr or less, $a_{p,\rm crit} \simeq 0.025$ AU (so that $a_F \lesssim 0.05$ AU).  Setting $a_{p,\li} \lesssim a_{p,\rm crit}$, a necessary condition for LK migration is
\be 
\begin{split}
\abunit \lesssim & 2.03 \ \aunit^{7/6} \left(\frac{a_{p,\rm crit}}{0.025 {\rm AU}} \right)^{1/6}  \left( \frac{\Mbunit}{\Msunit^2} \right)^{1/3} \\
& \times \left[1 + 0.23 \frac{\Rpunit^5}{\Mpunit} \left(\frac{a_{p,\rm crit}}{0.025 {\rm AU}} \right)^{-4} \right]^{-1/3}.
\end{split}
\label{eq:migration_condition}
\ee  
Note that this is a necessary, but not sufficient condition, because as discussed above, the outer binary must be sufficiently inclined in order for a planet to achieve $e_{\li}$.

The planet is tidally disrupted if the planet's periastron distance is less than the tidal radius \citep[e.g.][]{guillochon2011}
\be
r_{\rm Tide} =2.7 f R_p\left({M_\star\over M_p}\right)^{1/3},
\label{Rtide}
\ee
where $f \sim 1$ (we set $f = 1$ for all calculations in this paper).  Setting $a_{p,\li} \leq r_{\rm Tide}$, we obtain a necessary condition for tidal disruption:
\ba \label{eq:disrupt}
\abunit & \leq & 1.81 \ \aunit^{7/6} (f \Rpunit)^{1/6} \left({\Msunit \over \Mpunit}\right)^{\!\!1/18} \left({\Mbunit \over \Msunit ^2}\right)^{\!\!1/3} \\ \nonumber
& & \times \left(1+{3.54\Rpunit \Mpunit^{1/3} \over f^4 \Msunit^{4/3}}\right)^{\!\!-1/3}.
\ea
Note that since the tidal disruption radius (Eq.~[\ref{Rtide}]) is not a precisely defined quantity (the coefficient $f$ has uncertainties, and it depends on the planetary mass-radius relation, which can vary widely for giant planets), there are associated uncertainties in the disruption condition in Eq.~(\ref{eq:disrupt}).

\begin{figure}
\centering 
\includegraphics[scale=0.6]{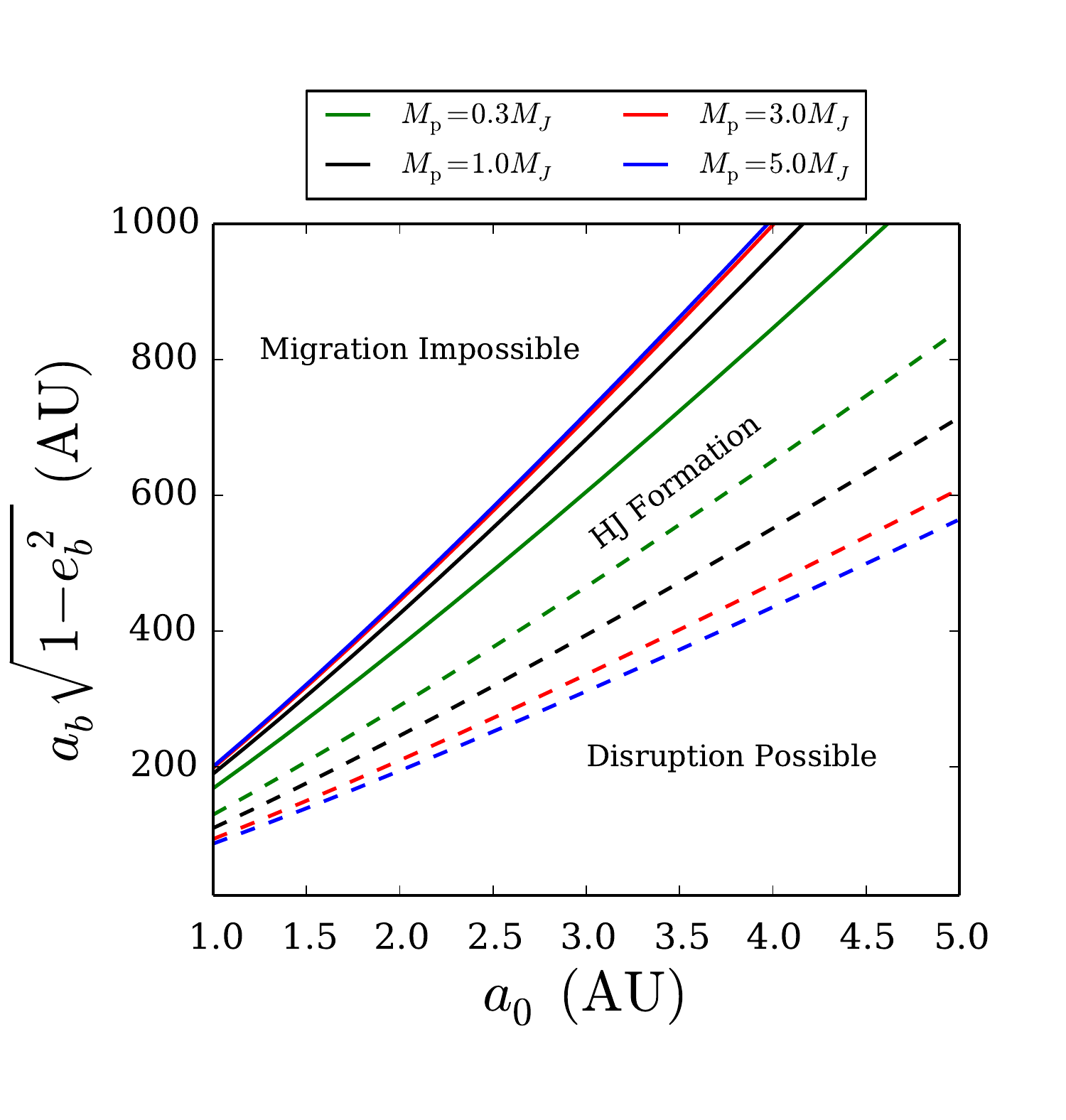}
\caption{Boundaries in $(a_0,a_{b,\rm eff})$ parameter space for migration (solid lines), and tidal disruption (dashed lines).  The migration and disruption boundaries are determined by Eq.~(\ref{eq:migration_condition}) (with $a_{p,\rm crit} = 0.025$ AU) and Eq.~(\ref{eq:disrupt}) (with $f = 1$) for several planet masses (as indicated by the color).  For each planet mass, migration is impossible (for all initial planet-outer binary inclinations) above the solid line, and tidal disruption is impossible above the dashed line.  Below the solid (dashed) line, migration (disruption) is possible (depending on the binary inclination), but not guaranteed.  HJ formation only occurs below the solid line, and is usually, but not always, confined to the region between the solid and dashed lines.}
\label{fig:migration_condition}
\end{figure}

Figure \ref{fig:migration_condition} delineates the parameter space in terms of the initial planet semi-major axis $a_0$ and effective binary separation $a_{b,\rm eff}$ for migration and disruption, as determined from Eqs.~(\ref{eq:migration_condition}) and (\ref{eq:disrupt}) for various planetary masses.  For a given planet mass, the parameter space can be divided into a ``Migration Impossible'' zone, a ``HJ Formation'' zone, and a ``Disruption Possible'' zone. Migration is possible below the solid line when the planet is sufficiently inclined relative to the binary, while below the dashed line, tidal disruption is possible.  The ``HJ Formation'' zone, the region between the solid and dashed lines, narrows substantially with decreasing planet mass, implying that HJ production efficiency should decline with decreasing planet mass.  Finally, note that while HJs are never able to form above the solid line, they do occassionally form below the dashed line, for systems where the mutual inclination is not high enough to result in tidal disruption.  Therefore, while the upper boundary (solid line) of the HJ formation zone is robust, the lower boundary is somewhat uncertain.  However, the vast majority of HJs will reside in the region between the solid and dashed lines.

Further discussion of the planet migration and disruption fractions can be found in Section \ref{sec:migrationfrac}.  

\subsection{Freezing of Spin-Orbit Angle}
\label{sec:thetafreeze}

The evolution of the spin-orbit angle $\theta_\SL$ is 
complex. Here we examine how $\theta_\SL$ is frozen into
its final value near the end of the LK migration.

As shown in Storch \& Lai (2015) (hereafter SL15), the dynamics of the stellar 
spin axis $\hatS$ relative to the planet's orbital axis $\hatLp$ depends on 
three dimensionless ratios
\ba
&&\epsilon\beta=-\frac{\Omega_{\rm pl}}{\alpha}\sin\theta_{\rm lb},
\label{beta} \\
&&\epsilon\gamma=\frac{\dot{\theta}_{\rm lb}}{\alpha}, 
\label{gamma}\\
&&\epsilon\psi=-\frac{\Omega_{\rm pl}}{\alpha}\cos\theta_{\rm lb},
\label{psi}
\ea
where we have defined the function $\alpha$ via
\be
\Omega_{\rm ps}=-\alpha\cos\theta_\SL, 
\ee
and the dimensionless parameter $\epsilon$ is defined by 
\be
\epsilon= \left | \frac{\Omega_{\rm pl}}{\alpha} \right |_{e = 0}.
\ee
The parameter $\epsilon$ is related to the ``adiabaticity parameter'' $\mathcal{A}_0$ [see Eq.~(\ref{adiabatic})] by $\epsilon = \mathcal{A}_0^{-1} |\cos \theta_{\SL,0}/ \sin \theta_{\LB,0}|$.
In general $\beta,\gamma,\psi$ are strong functions of time, with the period 
given by the LK period of the eccentricity variation (when neglecting the
feedback effect of the stellar spin on the orbit and the dissipative effect).
They can be decomposed into various Fourier components, each giving rise
to a resonance (see SL15). Near the end of LK migration, the amplitude of the
eccentricity oscillation becomes small (see Section \ref{sec:efreeze}). So
when $\theta_\SL$ begins to freeze, the dynamics of $\hatS$ is dominated
by the $N=0$ (time-independent) components ($\bar\beta$ and $\bar\psi$,
with $\bar\gamma=0$). Thus, the effective 
Hamiltonian for the stellar spin axis is (see Eq.~[53] of SL15) 
\be
H=-{1\over 2}p^2+\epsilon\,\bar\psi\, p-\epsilon\sqrt{1-p^2}
\,\bar\beta\cos\phi,
\ee
where $p=\cos\theta_\SL$ and $\phi$ (the phase of precession of $\hatS$ around
$\hatLp$) are the conjugate canonical variables.

Since $H$ is time-independent, the range of variation of $p$ can be
derived from energy conservation. Suppose $p=p_F$ at $\phi=\pi/2$.
For $\epsilon\ll 1$, we find 
\be
p\simeq p_F - {\epsilon\bar\beta\sqrt{1-p_F^2}\over p_F}\cos\phi.
\ee
Thus the spread (full width) of $\theta_\SL$ as $\phi$ circulates
between $0$ and $2\pi$ is
\be
\Delta\theta_\SL\simeq {2\epsilon\bar\beta\over|\cos\theta_{\SL,F}|}
={2\over \mathcal{A}_F},
\ee
where
\be
\mathcal{A}_F \equiv \frac{\langle |\Omega_{\rm pl}| \rangle}{\langle |\Omega_{\rm pl} \sin \theta_{\LB}| \rangle}.
\label{eq:AF}
\ee
The bracket $\langle ... \rangle$ in Eq.~(\ref{eq:AF}) indicates time averaging over the small ``residual'' LK oscillations.  If the eccentricity variation is ``frozen'' or has small amplitude, then the averaging is unnecessary and $\mathcal{A}_F$ is the same as $\mathcal{A}$ defined in Eq.~(\ref{eq:Aparameter}).
Thus, in order for the spin-orbit angle to freeze at $\theta_{\SL,F}$ to within $\Delta\theta_\SL$ (e.g., $2^\circ$) requires
\be
\mathcal{A}\go 60 \left({\Delta\theta_\SL\over 2^\circ}\right)^{-1}.
\ee

\section{Paths Toward Misalignment}
\label{sec:path}
In this section we present a series of numerical experiments to
illustrate various paths of spin-orbit evolution during LK migration.
These will be useful for understanding our population synthesis
results of the final spin-orbit angles for HJs in Section
\ref{sec:popsynth}. The theoretical basis for these different
evolutionary paths is presented in \citet[submitted]{SLA15}. 
\subsection{Effects of Varying Stellar Spin Rate}
\label{sec:path_Pstar}
To isolate the effects of the stellar spin dynamics, and highlight the importance of the stellar spin properties on the final
spin-orbit angle, we first ignore the feedback of the stellar spin on the planetary orbit (thus ignoring the mutual precession
of ${\bf S}_\star$ and $\Lp$). Possible types of evolution are illustrated in Figs.~\ref{fig:chaos} and \ref{fig:adiabatic}. In both figures, we vary the stellar spin period
while keeping all other system parameters constant. 
Figure \ref{fig:chaos} presents an example of chaotic spin evolution: 
three closely spaced values of the stellar spin period result 
in very different spin evolutions and final spin-orbit misalignments.  
Figure \ref{fig:adiabatic} presents three different types of non-chaotic spin evolution, 
only two of which are able to generate spin-orbit misalignment. 

The leftmost panel (with $P_\star = 30$ days) of Fig.~\ref{fig:adiabatic} (with $\theta_{\SL}$ in the middle row) shows an example of non-adiabatic spin behavior. Here, the spin-orbit misalignment angle $\thetaSL$ evolves slowly, with step-like changes corresponding to LK eccentricity maxima, during which the spin evolves the most rapidly. Since the planet orbit changes much faster than the spin can respond, the spin axis effectively precesses about the time average of the planet orbital angular momentum vector. 

On the opposite end of the spectrum, the middle panel of Fig.~\ref{fig:adiabatic} (with $P_\star = 7.07$ days) is an example of adiabatic spin behavior. Here, the stellar spin axis evolves quickly enough that it easily ``keeps up'' with the planet angular momentum vector, and hence $\thetaSL$ is approximately conserved, making it difficult to generate misalignment. 

The rightmost panel of Fig.~\ref{fig:adiabatic} (with $P_\star = 1.67$ days) shows a more complicated variation of the adiabatic evolution, which we term ``adiabatic advection''. As discussed in detail in SL15, the adiabatic regime of stellar spin evolution is governed by a set of resonances between the time-averaged spin precession rate and the mean LK oscillation rate. Under certain conditions, it is possible for a trajectory to become trapped inside one of the resonances. As tidal dissipation acts to make the system even more adiabatic, the resonance moves in phase space, dragging the trajectory with it and thus generating misalignment. We discuss and clarify the mechanism of this phenomenon in \cite[submitted]{SLA15}.  

Fig.~\ref{fig:typeI} presents final spin-orbit angles $\theta_{\rm sl,f}$ for many different values of the stellar spin period, for three different orbital evolutions (characterized by different initial inclinations $\theta_{\rm lb,0}$).  This illustrates the role of the adiabaticity parameter $\mathcal{A}_0$ (see Eq.~[\ref{adiabatic}]) in determining which of the four types of evolution the spin-orbit angle undergoes. For low values of $\mathcal{A}_0$, chaotic and regular non-adiabatic behaviors are prevalent. For intermediate values, e.g. $10 \lesssim \mathcal{A}_0 \lesssim 100$ in the rightmost panel, adiabatic advection dominates, with each of the striated lines corresponding to adiabatic advection by resonances of different orders \citep[see][submitted]{SLA15}. For $\mathcal{A}_0 \gtrsim 100$, stationary adiabatic behavior prevails. Thus, $\mathcal{A}_0$ can be used as an indicator for the behavior of a system with a particular set of initial conditions.

\subsection{Effects of Varying Inclination}
\label{sec:path_Inc}
\begin{figure*}
\centering 
\includegraphics[width=0.85\textwidth]{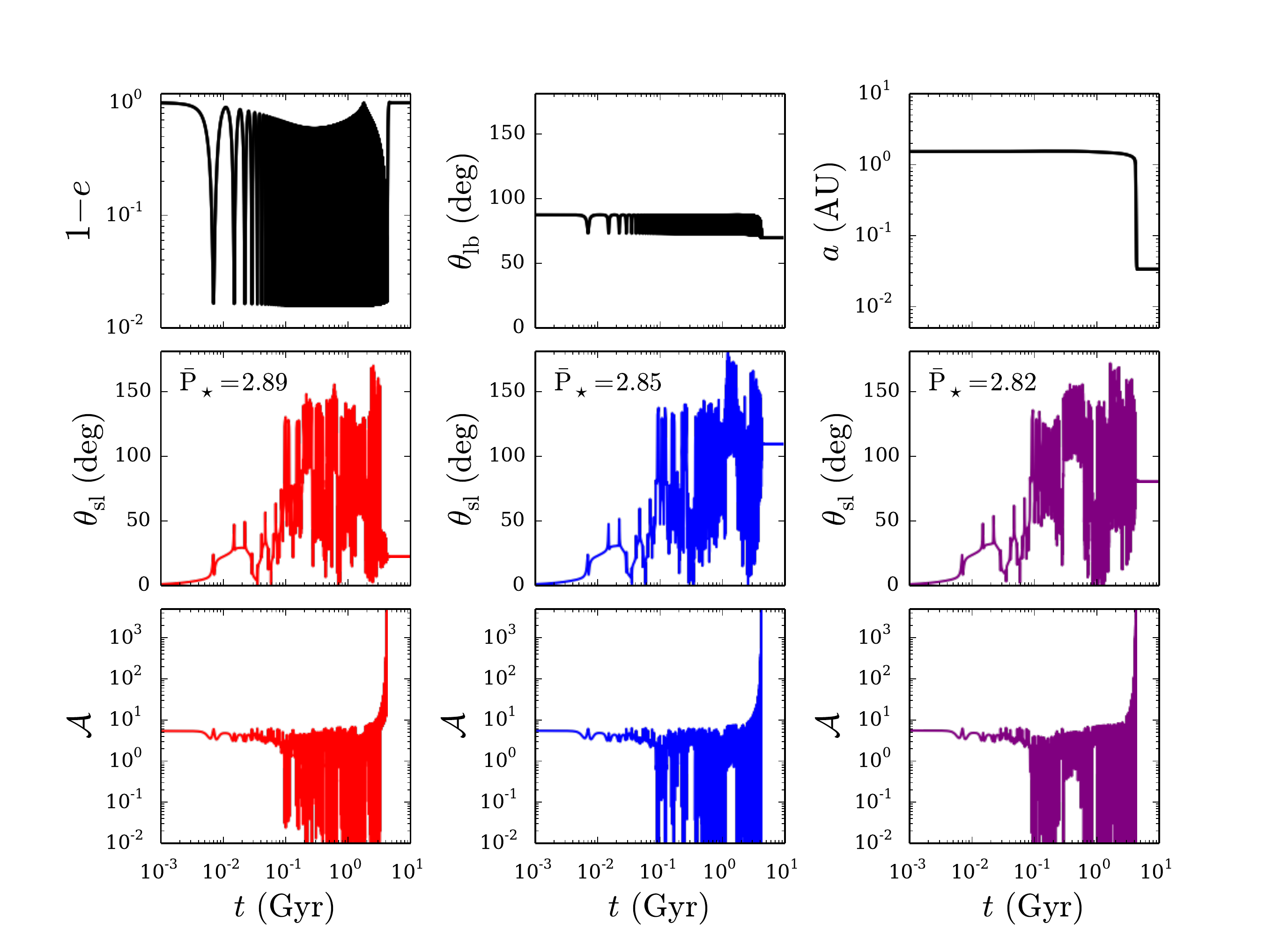}
\caption{Examples of chaotic evolution for three values of the stellar spin period (in days) as labeled, neglecting the feedback torque from the stellar quadrupole on the orbit.  Without feedback, the orbital evolution for each system is identical (shown in the top panels), while the spin-orbit angle settles to a final value that is highly sensitive to the initial conditions.  The adiabaticity parameter $\mathcal{A}$ is defined in Eq.~(\ref{eq:Aparameter}). Parameters are $M_p = 5 M_J$, $a_0 = 1.5$ AU, $a_b = 300$ AU, $e_b = 0$, $\theta_{\LB,0} = 87^{\circ}$.}
\label{fig:chaos}
\end{figure*}

\begin{figure*}
\centering 
\includegraphics[width=0.85\textwidth]{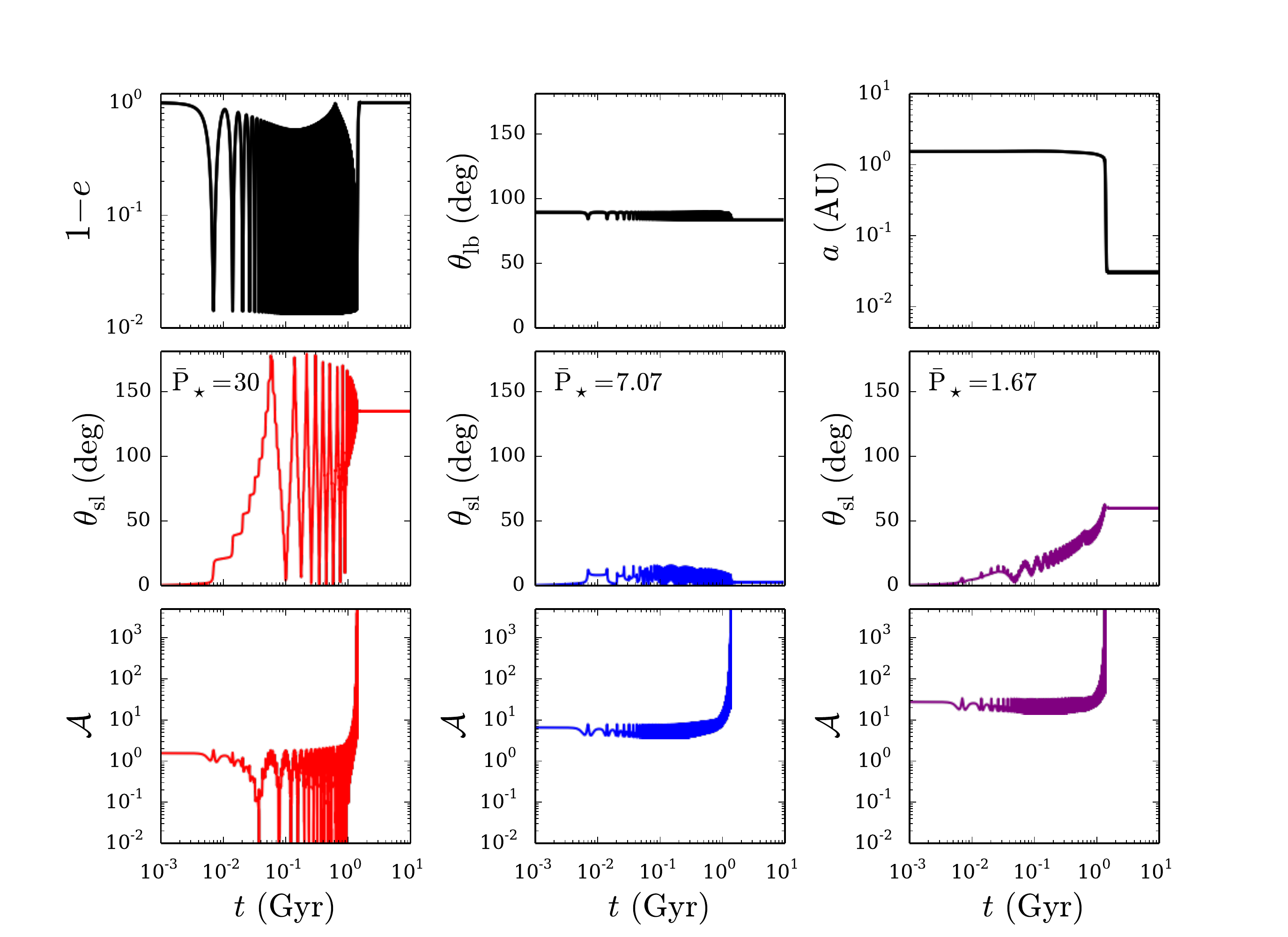}
\caption{Examples of possible non-chaotic evolution of the spin-orbit angle, depending on the stellar spin rate.  As in Fig.~\ref{fig:chaos}, feedback has been neglected, so that the orbital evolution, shown in the top row, is identical for all three examples: Non-adiabatic with $P_\star = 30$ days (left), stationary adiabatic with $P_\star = 7.07$ days (middle), and adiabatic advection with $P_\star = 1.67$ days (right).  Parameters are $M_p = 5 M_J$, $a_0 = 1.5$, $a_b = 300$ AU, $e_b = 0$, $\theta_{\LB,0} = 89^{\circ}$.}
\label{fig:adiabatic}
\end{figure*}

In this subsection we take a different tack and examine the effect of varying the initial planet orbit inclination $\thetaLB$, for different values of the stellar spin period and the planet mass. As before, we continue to ignore the back-reaction torque the star exerts on the planet orbit. 
Fig.~\ref{fig:time_ev_typeII} demonstrates that changing the initial inclination effectively changes $\mathcal{A}_0$, and thus systems with different initial inclinations can also exhibit the different behaviors shown in Figs.~\ref{fig:chaos} and \ref{fig:adiabatic} of Section \ref{sec:path_Pstar}. In particular, the three columns of Fig.~\ref{fig:time_ev_typeII} correspond to chaotic evolution (left panels), adiabatic advection (middle panels), and an extreme case of stationary adiabatic evolution (right panels). 

In Fig.~\ref{typeII} we show the dependence of the final spin-orbit misalignment angle on the initial inclination, for several combinations of planet mass and stellar spin period. As expected, chaotic behavior occurs mainly at lower initial inclinations (less adiabatic -- see the right two panels of Fig.~\ref{typeII}). We note, however, that despite spanning approximately the same range of $\mathcal{A}_0$, heavier planets are much more likely to produce chaotic behavior than lower-mass planets - this implies that $\mathcal{A}_0$ is not the only parameter governing the evolution of $\theta_{\SL}$ \citep[submitted]{SLA15}. Stationary adiabatic behavior manifests here as the ``tail'' of the distributions at higher initial inclinations, e.g. between $88.5^{\circ}$ and $90^{\circ}$ in the top left panel, and near $90^{\circ}$ in the bottom right panel. The long stretches of nearly-constant $\theta_{\rm sl,f}$ present in the higher-mass (more adiabatic) panels are due to adiabatic advection. 

The non-adiabatic behavior regime shown in Fig.~\ref{fig:adiabatic} (left panels) manifests here as a bimodal split in $\theta_{\rm sl,f}$ (see the left two panels of Fig.~\ref{typeII}). This bimodality is the result of a bifurcation phenomenon that occurs at the moment the system transitions from being non-adiabatic to being adiabatic (due to the orbital decay from tidal dissipation). Before the transition, the system undergoes wide $0-180^{\circ}$ degree oscillations in $\theta_{\SL}$; after the transition, the system must evolve adiabatically and be confined either above or below $\thetaSL=90^{\circ}$. The transition between these two states is akin to a bifurcation. We illustrate this in Fig.~\ref{fig:bimodal} by showing the time evolution of two trajectories with nearly identical initial conditions.  Unlike the previous chaotic examples shown (with positive Lyapunov exponents) the trajectories in Fig.~\ref{fig:bimodal} do not quickly diverge, but rather remain qualitatively similar while accumulating some phase difference. This phase difference, if pronounced enough, leads to a bifurcation in the final spin-orbit angle. We discuss this phenomenon in detail in \cite[submitteds]{SLA15}.

\begin{figure*}
\centering 
\includegraphics[width=\textwidth]{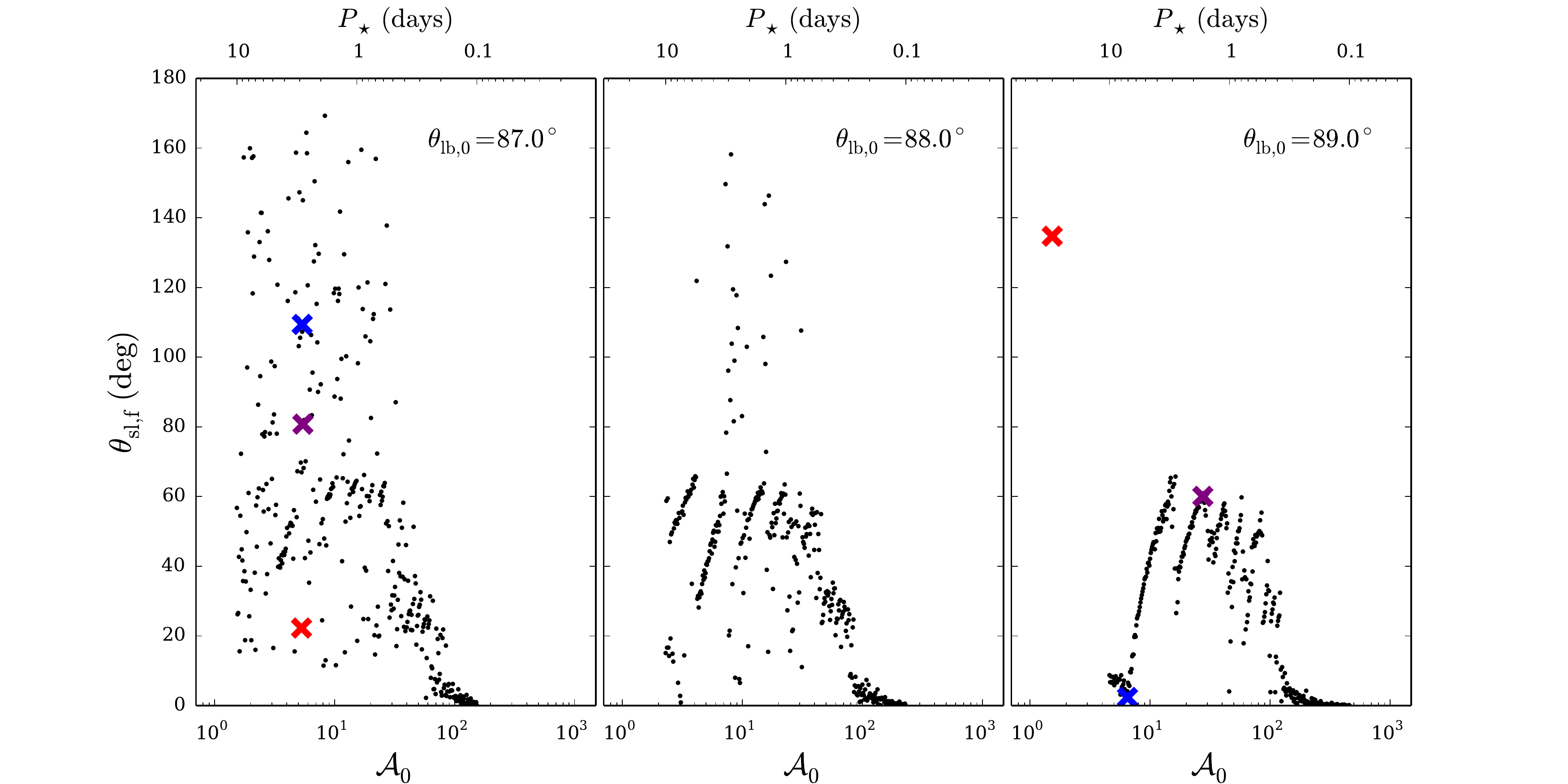}
\caption{The final spin-orbit angle $\theta_{\SL,\rm f}$ (for systems with planets that undergo inward migration to produce hot Jupiters) as a function of the adiabaticity parameter $\mathcal{A}_0$.  Here, we vary $\mathcal{A}_0$ by varying $P_\star = 0.1 - 10$ days (as depicted on the upper x-axis).  Results are shown for initial inclinations $\theta_{\LB,0} = 87^{\circ}$ (left), $88^{\circ}$ (middle), and $89^{\circ}$ (right).  The colored marks correspond to the time evolution presented in Fig.~\ref{fig:chaos} and \ref{fig:adiabatic}.  As the initial inclination increases, the adiabaticity parameter $\mathcal{A}_0$ increases, leading to systems with a smaller spread in $\theta_{\SL,\rm f}$. Parameters are $M_p = 5 M_J$, $a_0 = 1.5$, $a_b = 300$ AU, $e_b = 0$, no feedback.}
\label{fig:typeI}
\end{figure*}

\begin{figure*}
\centering 
\includegraphics[width=0.85\textwidth]{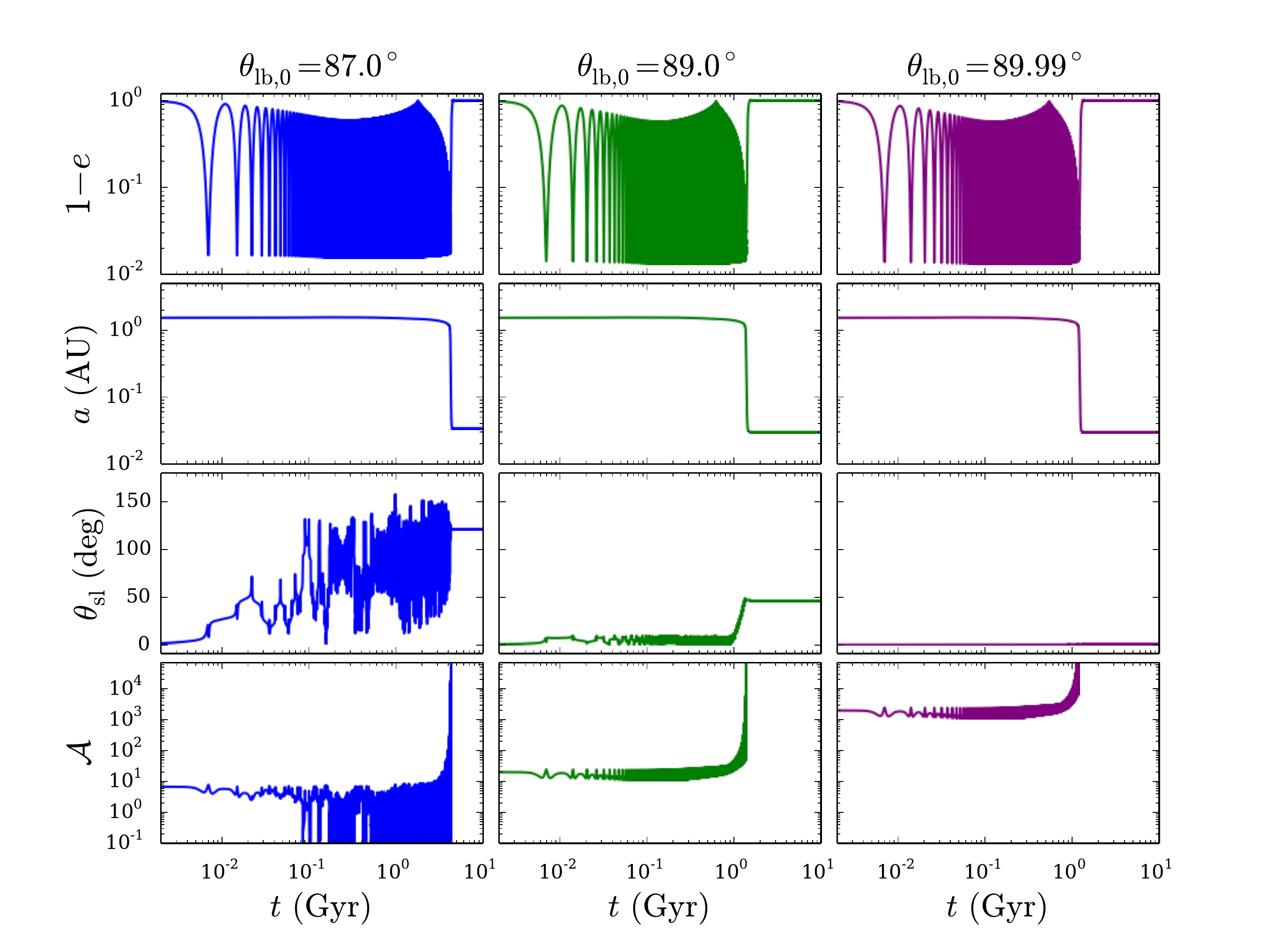}
\caption{Examples of possible evolution of the spin-orbit angle, depending on the initial inclination.  All examples have $M_p = 5 M_J$, $a_0 = 1.5$ AU, $a_b = 300$ AU, $P_\star = 2.3$ days, and the feedback torque from the stellar quadrupole has been neglected.  The system with $\theta_{\LB,0} = 87^\circ$ (left panels) has $\mathcal{A}_0 \lesssim 10$, sufficiently low to generate large spin-orbit misalignments.  The system with $\theta_{\LB,0} = 89^\circ$ (middle panels) has $\mathcal{A}_0 \gtrsim 10$, sufficiently high to preserve the initially low misalignment, but eventually undergoes adiabatic advection (see text).  The extreme example shown on the right with $\theta_{\LB,0} = 89.99^\circ$ has $\mathcal{A}_0 \gtrsim 10^3$, so that $\theta_{\SL}$ is very nearly constant for all time.}
\label{fig:time_ev_typeII}
\end{figure*}

\begin{figure*}
\centering 
\includegraphics[width=0.8\textwidth]{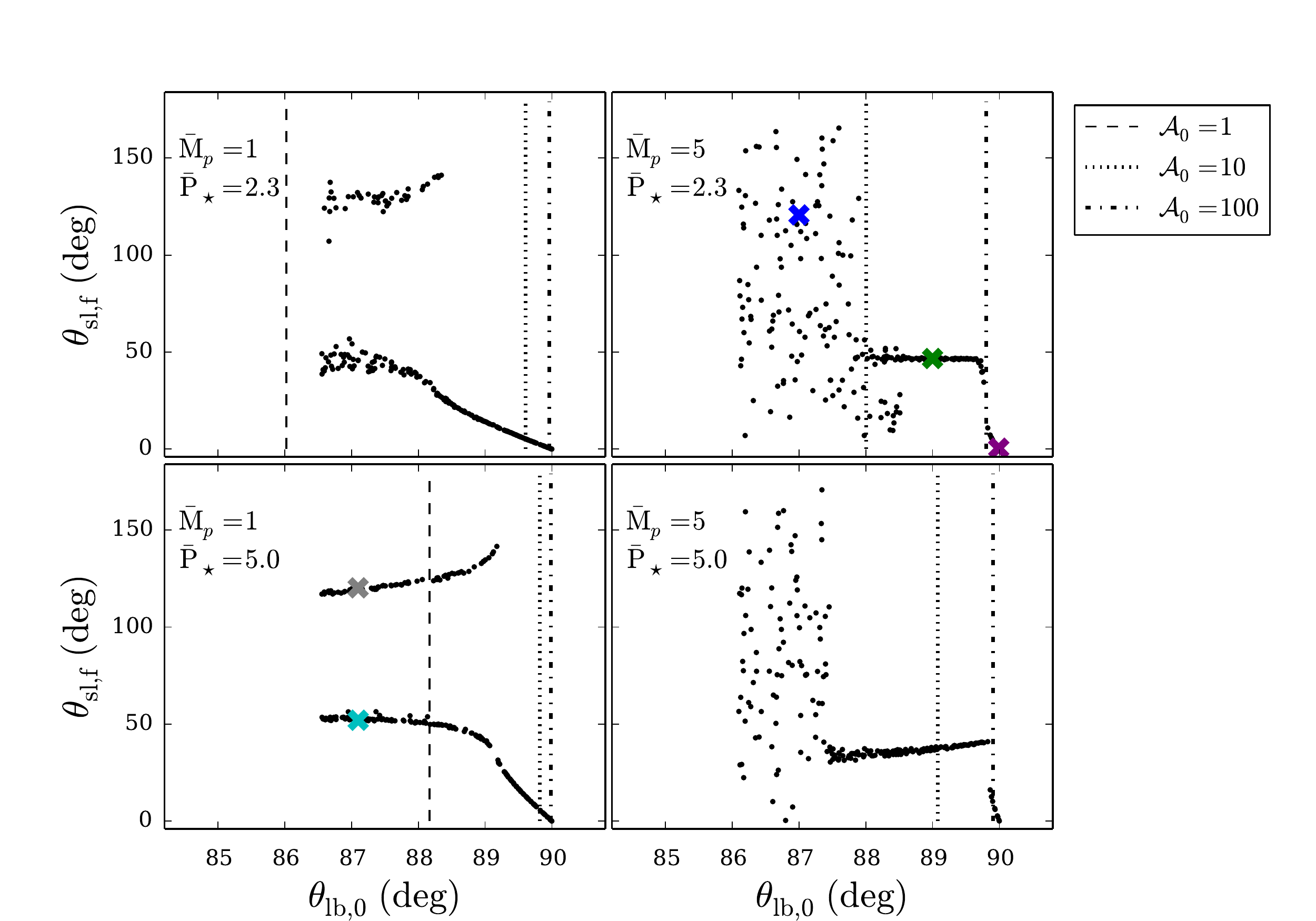}
\caption{Final spin orbit misalignments as a function of the initial inclination, for various combinations of planet mass and (constant) stellar spin period, as labeled.  In this example, we neglect the feedback torque from the stellar quadrupole on the planetary orbit.  We indicate various benchmark values of $\mathcal{A}_0$ by the vertical lines.  The colored crosses correspond to the time evolution presented in Fig.~\ref{fig:time_ev_typeII} (upper right panel), and Fig.~\ref{fig:bimodal} (lower left panel).  Parameters are $a_0 = 1.5$ AU, $a_b = 300$ AU, $e_b = 0$.}
\label{typeII}
\end{figure*}

\begin{figure*}
\centering 
\includegraphics[width=0.5\textwidth]{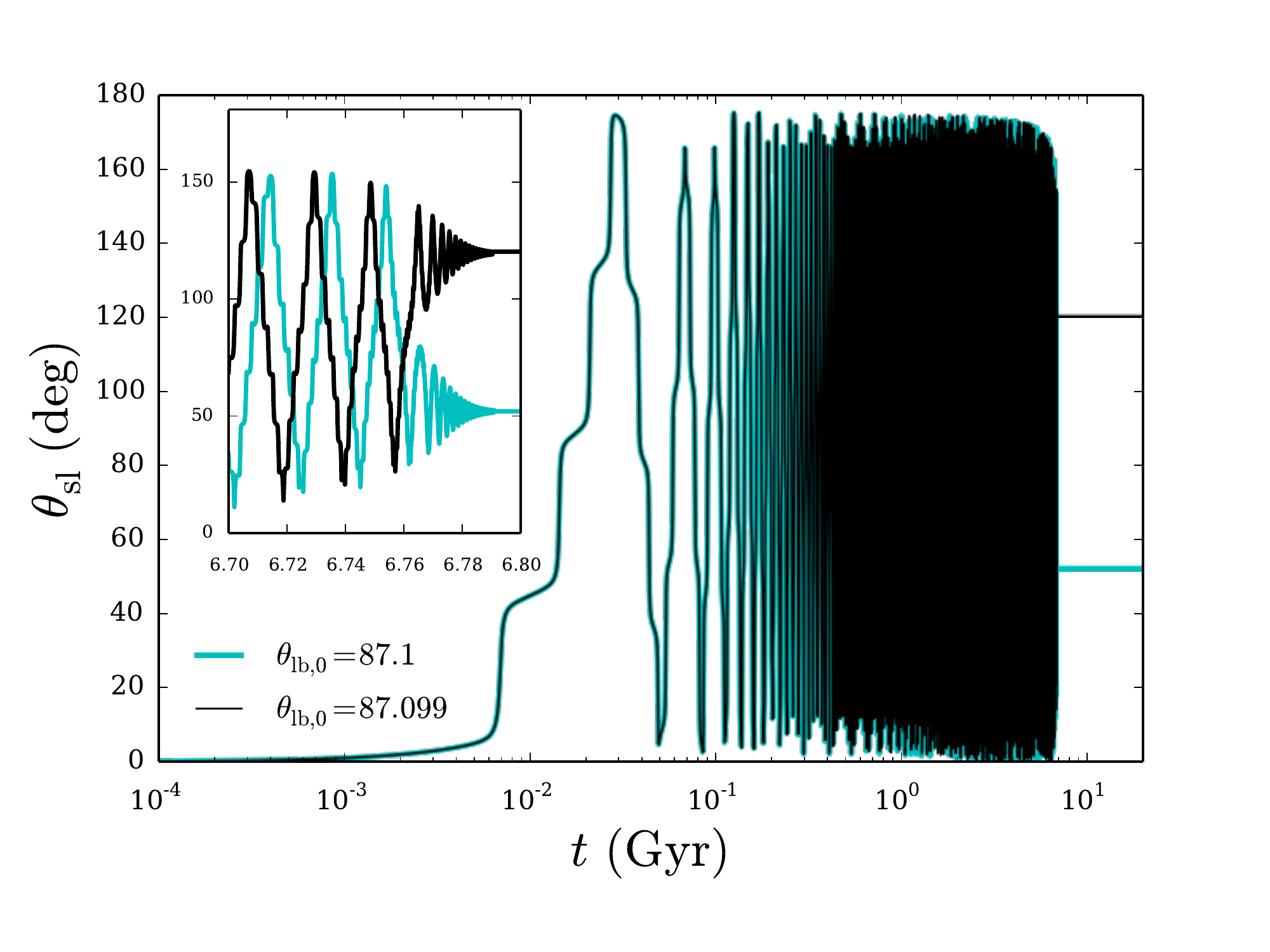}
\caption{Time evolution for two systems with very similar initial inclinations, illustrating the bimodality in the final misalignment angle, as depicted in the lower left panel of Figure \ref{typeII}. Parameters are $M_p = 1M_J$, $P_\star = 5$ days, $a_b = 300$ AU, no feedback.  Nearly identical initial inclinations accumulate some phase difference over the course of the evolution, which at the moment of transition to the adiabatic regime, give rise to different final angles, with $\theta_{\LB,\rm f} \approx 52^{\circ}$ and $120^{\circ}$.}
\label{fig:bimodal}
\end{figure*}

In summary, the evolution of the spin-orbit misalignment angle can proceed in four distinct ways. (i) {\it Chaotic}. Neighboring spin trajectories diverge exponentially and $\theta_{\rm sl,f}$ is very sensitive to initial conditions. (ii) {\it Regular non-adiabatic}. $\thetaSL$ initially undergoes wide, regular $0-180^{\circ}$ oscillations. After significant semi-major axis decay has occurred, the evolution of $\thetaSL$ undergoes a bifurcation and becomes confined either above or below $90^\circ$ degrees. This leads to the bimodality seen in Fig.~\ref{typeII} (left panels). (iii) {\it Stationary adiabatic}. $\thetaSL$ is approximately conserved and no misalignment can be generated. (iv) {\it Adiabatic advection}. The phase space trajectory becomes trapped in a resonance and advected to higher misalignments. $\theta_{\rm sl,f}$ depends sensitively on the stellar spin period (Fig.~\ref{fig:typeI}, right panel), but only weakly on the initial inclination (Fig.~\ref{typeII}, right panels). 


\subsection{Effects of the Backreaction Torque from the Stellar Quadrupole on the Orbit}
\label{sec:feedback}
All examples in Sections \ref{sec:path_Pstar} and \ref{sec:path_Inc} have neglected the backreaction torque from the stellar quadrupole on the planet's orbit, in order to simplify the analysis of the spin-orbit dynamics.  However, under some conditions, the backreaction torque can significantly affect the evolution of the spin-orbit misalignment.  In the following discussion, we show how including this torque affects (and complicates) the dynamics, and delineate the parameter space where this torque can compete with the torque from the binary companion in changing the orbital axis.

The stellar quadrupole has two effects on the planetary orbit.  First, it changes the direction of the angular momentum axis $\hatLp$ at the rate given by
\be
\left . \frac{d \hatLp}{d t} \right |_{{\rm SL}} = \Omega_{\PS} \frac{S_\star}{L} \hatS \times \hatLp \enspace \propto M_{\star}^{-1/2} R_\star^5 \Omega_\star^2.
\label{eq:feedback}
\ee
Second, it causes the eccentricity vector $\ep$ to precess around $\hatLp$,
\be
\left . \frac{d \ep}{d t} \right |_{{\rm SL,rot}} = \frac{\dot{\omega}_\star}{2} (5 \cos^2 \theta_{\SL} - 1) \hatLp \times \ep
\label{eq:eccfeedback}
\ee
where 
\be
\dot{\omega}_\star = - \frac{S_\star}{L} \frac{\Omega_{\PS}}{\cos \theta_{\SL}}.
\ee
The subscript ``rot'' in Eq.~(\ref{eq:eccfeedback}) implies that the time derivative is done in the frame rotating with the nodal precession of the orbit (at the rate $\Omega_{\PS} S_\star/L$), so that $\hatLp$ is fixed in space (compare Eq.~[\ref{eq:eccfeedback}] with Eq.~[\ref{eq:evecfeedback}]).  The effect of the stellar quadrupole on the eccentricity vector does not introduce any new features in the orbital evolution, but simply contributes to the rate of pericenter precession due to other SRFs (GR, tidal and rotational distortions of the planet). By contrast, the effect on the orbital axis $\hatLp$ does directly change $\theta_{\LB}$, thereby influencing the evolution of the spin-orbit angle.

Consider now the change in $\theta_{\LB}$ due to the backreaction torque of the stellar quadrupole (Eq.~[\ref{eq:feedback}]).  The maximum possible change is 
\be
\begin{split}
(\Delta \theta_{\LB})_{\rm max} & \sim \left ( \frac{S_\star}{L} \right )_{e_{\rm max}} \\ 
& \simeq 0.12 \frac{\bar{k}_\star \Mtunit^{1/2} \Rsunit^2}{\Mpunit} 
\left ( \frac{\afunit}{0.05} \right)^{-1/2} \left( \frac{P_\star}{30 \rm days} \right)^{-1},
\end{split}
\label{eq:delta_thetaLB_max}
\ee
assuming $L \gtrsim S_\star$.
The actual change of $\theta_{\LB}$ in an LK cycle can be obtained by integrating Eq.~(\ref{eq:feedback}) through time $t_k$ around the eccentricity maximum, yielding
\be
\begin{split}
(\Delta \theta_{\LB})_{\rm actual} & \sim \left( \left | \frac{d \hatLp}{d t} \right | \Delta t \right)_{e_{\rm max}} \\
& \sim \left( |\Omega_{\PS}| \frac{S_\star}{L} \right)_{e_{\rm max}} t_k \sqrt{1 - e_{\rm max}^2} \\
& \simeq 0.1 \frac{\bar{k}_q \Rsunit^5 \Mtunit \abunit^3}{\Mbunit \Msunit \aunit^{7/2}} \left( \frac{\afunit}{0.05} \right)^{-3/2} \left(\frac{P_\star}{6 \rm days} \right)^{-2}
\end{split}
\label{eq:delta_thetaLB_actual}
\ee
where we have used Eq.~(\ref{eq:t_emax}) for $\Delta t (e_{\rm max})$.  Note that $(\Delta \theta_{\LB})_{\rm actual}$ is also approximately equal to the ratio between $|d \hatLp / dt|_{\rm SL}$ and $|d \hatLp / dt|_{\rm LK}$.  Eq.~(\ref{eq:delta_thetaLB_actual}) assumes $\Delta \theta_{\LB,\rm actual} \lesssim \Delta \theta_{\LB,\rm max}$.  That is, the actual change in $\theta_{\LB}$ due to the backreaction torque is given by Eq.~(\ref{eq:delta_thetaLB_max}) or Eq.~(\ref{eq:delta_thetaLB_actual}), whichever is smaller.

\begin{figure*}
\centering 
\includegraphics[width=0.8\textwidth]{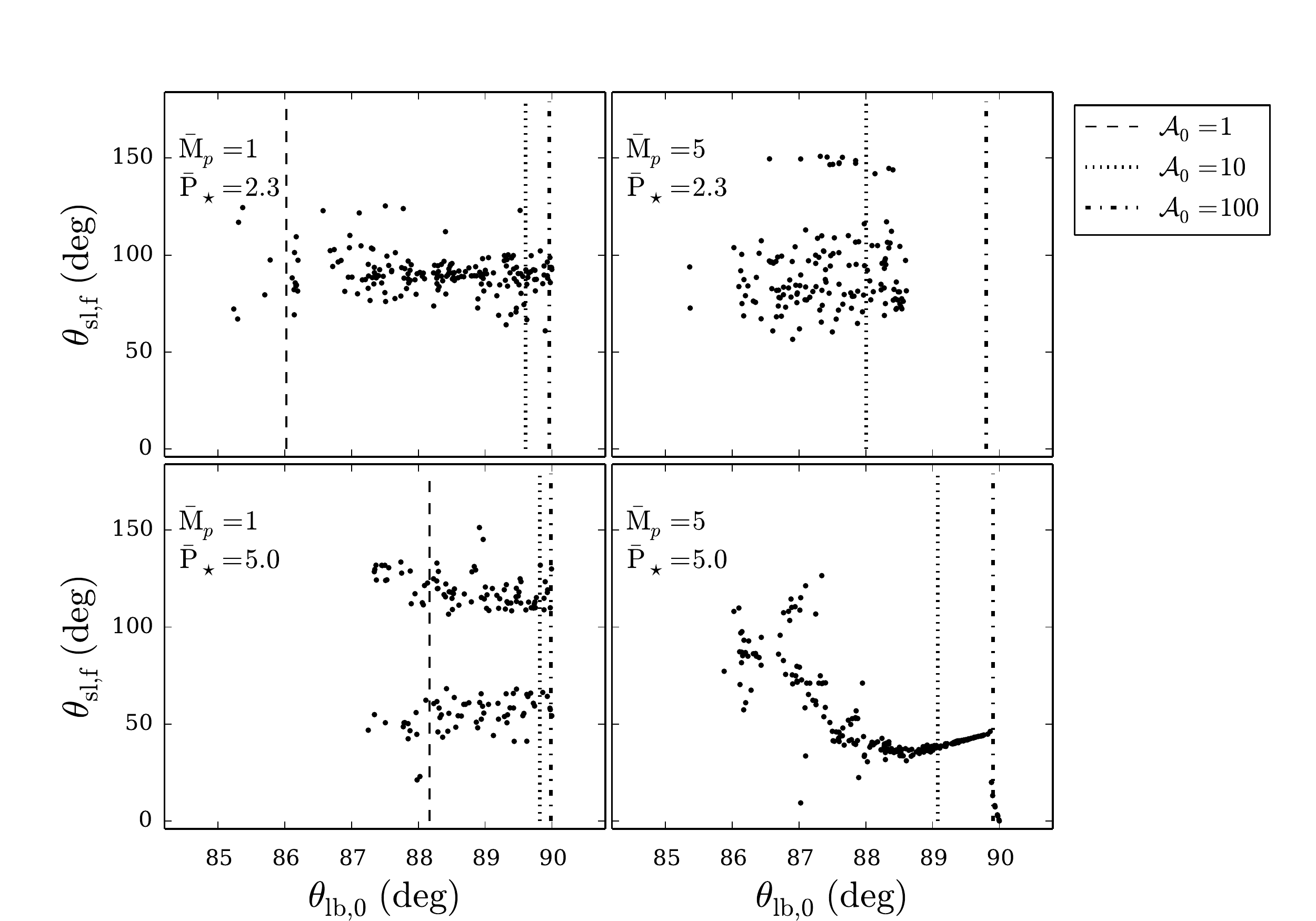}
\caption{Same as Fig.~\ref{typeII}, but including feedback from the stellar quadrupole on the orbit.}
\label{typeII_feedback}
\end{figure*}

We have already seen from Fig.~\ref{typeII} that the final spin-orbit misalignment can depend strongly on $\theta_{\LB,0}$.  We expect that the backreaction torque will significantly affect $\theta_{\SL,\rm f}$ when $(\Delta \theta_{\LB})_{\rm actual} \gtrsim 0.1$.  Eqs.~(\ref{eq:delta_thetaLB_max}) and (\ref{eq:delta_thetaLB_actual}) indicate that this condition is satisfied for $P_\star \lesssim$ a few days, depending on various parameters (such as $a_{b, \rm eff}$ and $M_p$).  Fig.~\ref{typeII_feedback} shows $\theta_{\SL, \rm f}$ as a function of $\theta_{\LB,0}$ for several values of $P_\star$ and $M_p$, with the backreaction torque included in the calculations (cf.~Fig.\ref{typeII}, which neglects the backreaction torque).

Comparing Figs.~\ref{typeII} and \ref{typeII_feedback} reveals the main effects of the backreaction torque on the final spin-orbit angle.  Systems with the lowest planet mass and shortest spin period ($M_p = 1 M_J$, $P_\star = 2.3$ days, top left) are most strongly affected by feedback, and the clean bimodality present in $\theta_{\SL,\rm f}$ in Fig.~\ref{typeII} is erased, and replaced by clustering near $\theta_{\SL,\rm f} \sim 90^{\circ}$.  The results for the large planet mass and short spin period ($M_p = 5 M_J$, $P_\star = 2.3$ days, top right) are also significantly affected, due to planets becoming tidally disrupted at high inclinations. The systems with longer stellar spin periods (bottom panels) are less affected by feedback, and the general structure found in Fig.~\ref{typeII} is partially preserved.

\section{Population Synthesis}
\label{sec:popsynth}
\subsection{Setup and Computational Procedure}
In this section we perform a detailed parameter space survey for giant
planets undergoing LK migration, exploring the dependence of the final
spin-orbit misalignment angle distribution on the planet mass and stellar spin
properties.  We focus on two types of host stars: a solar-mass
($M_\star = 1 M_\odot$, spectral type G) star, and a massive ($M_\star
= 1.4 M_\odot$, spectral type F) star.  The initial spin period of
both types of stars is set to $P_\star = 2.3$ days, corresponding to
$5 \%$ of breakup for the G star; both stars subsequently spin-down
according to the Skumanich law (see Section \ref{sec:spin}).  The G
(F) star is calibrated to reach a spin period of $28$ (9) days after 5
Gyr, to account for the fact that massive stars are observed to rotate
more rapidly at a given age \citep[e.g.][]{mcquillan2014}.  The stellar radius is set to $R_\star = 1 R_{\odot}$ for G-type stars, and $R_\star = 1.26 R_{\odot}$ for F-type stars.  We consider four planet masses ($M_p = 0.3, 1, 3$, and $5 M_J$), all having a radius $R_p = 1 R_J$.  Note that this is a simplification, as some observed close-in gas giant planets are found to be inflated in size, while others are more compact \citep[e.g.][]{laughlin2011}.  

We integrate the full equations of motion for the planetary orbit,
including the octupole terms from the stellar companion, feedback
torque from the host stellar spin, and all short-range forces,
together with evolution equations for the host stellar spin, and the
planetary spin rate (due to tidal dissipation).  As in previous population studies \citep{naoz2012,petrovich2015b}, systems that do not obey the stability condition \citep{mardling2001} 
\be
\frac{a_b}{a} > 2.8 \left(1 + \frac{M_b}{M_{\rm tot}} \right)^{2/5} \frac{(1 + e_b)^{2/5}}{(1 - e_b)^{6/5}} \left[1 - 0.3 \frac{\theta_{\LB,0}}{180^{\circ}} \right]
\ee  
are discarded.
To increase the efficiency of the parameter survey, for each integration we adopt the following stopping conditions:
\begin{enumerate}
\item If after 500 LK timescales (Eq. [\ref{tk}]) the pericenter distance has never reached $r_p = a(1 - e) < 0.07$ AU, we terminate the calculation to avoid unnecessary  integrations, and classify the planet as non-migrating. The time needed for such planets to undergo significant orbital decay is greater than $\sim 10^{11}$ years (see Section \ref{sec:migrationrate}, Eq.~[\ref{eq:maxrate}]).  This is far too long to allow significant migration within the lifetime of the host star.\footnote{Note that with the octupole terms from the binary companion included, the planet can achieve extreme values of eccentricity $e_{\lim}$ when $\theta_{\LB,0}$ is sufficiently large (see Section \ref{sec:disrupt}).  Although these octupole extreme eccentricities are nearly always achieved sooner than $500 t_k$ (depending on $\varepsilon_{\rm oct}$, see \citealt{liu2015}), the possibility of the planet achieving such a high eccentricity cannot be ruled out for $t > 500 t_k$.  We therefore run the risk of terminating systems that might later undergo orbital decay.  However, note that in such cases, the eccentricity usually becomes so high that the planet would be tidally disrupted, and removed from the sample of HJs.  We have tested this stopping criterion and found that the approximation causes a very small fraction of tidally disrupted planets to be misclassified as non-migrating, but the fraction of HJs is unaffected.}

\item If at any point the pericenter distance $r_p = a(1 - e) < r_{\rm Tide}$, where $r_{\rm Tide}$ is the tidal disruption radius, given in Eq.~(\ref{Rtide}), we terminate the integration, and classify the planet as tidally disrupted.
\item If the semi-major axis has decayed to $a < 0.1$ AU, we terminate the integration and classify the planet as a hot Jupiter.  In such cases, the spin-orbit angle has always safely reached the adiabatic regime (so that the adiabaticity parameter $\mathcal{A}$ has become sufficiently large), with $\hatS$ and $\hatLp$ undergoing mutual precession, and $\theta_{\SL}$ is nearly constant, varying by less than $1^\circ$.  At this point, LK oscillations from the binary companion are completely suppressed (see Section \ref{sec:efreeze}), and the planet will continue to undergo pure tidal evolution at nearly constant angular momentum, with final semimajor axis $a_{\rm f} \simeq a(1 - e^2)$, where $a$ and $e$ are evaluated at the point at which the integration is stopped.
\item If none of these conditions are satisfied during the integration, we terminate the integration at $t = 5$ Gyr and classify the planet as non-migrating.
\end{enumerate}

For each set of system parameters, we begin by integrating the full equations of motion. However, in situations where the planet experiences sufficient orbital decay, the LK oscillations become suppressed so that the range of eccentricity variation narrows, and the stellar spin axis enters the adiabatic regime where $\theta_{\SL} \approx$ constant (see Sections \ref{sec:efreeze} and \ref{sec:thetafreeze}).  In such cases, the eccentricity vector $\ep$ precesses much more rapidly compared to the tidal decay rate.  Resolving this rapid precession is computationally expensive, but does not influence the final result.  Therefore, once the LK eccentricity oscillations and spin-orbit angle have both ``frozen'' we stop following the eccentricity precession (i.e. by neglecting the SRF and LK terms in the planet's equations of motion), and allow the orbit to evolve purely under tidal dissipation.\footnote{In practice, we consider the $e$-oscillations to have frozen when $\varepsilon_{\gr} > 30$, and $\theta_{\SL}$ to have settled to its final value when the adiabaticity parameter satisfies $\mathcal{A}_0 \sin 2 \theta_{\LB} > 5$ (see Sections \ref{sec:efreeze} and \ref{sec:thetafreeze}). We have tested both conditions extensively and find they are extremely conservative estimates, so that the LK oscillations and variation in $\theta_{\SL}$ are always safely quenched at the point when the SRF and LK terms are neglected in the equations of motion.}

We assume that the initial planet orbital axis $\hatLp$ is isotropically distributed with respect to $\hatLb$. In principle, the initial inclination should be sampled over the entire range ($\theta_{\LB,0} = [0^{\circ}, 90^{\circ}]$).\footnote{Since $M_{\rm p} \ll M_\star, M_b$, the triple systems considered here exhibit symmetry around $\theta_{\LB,0} = 90^{\circ}$, so that $90^{\circ}\leq \theta_{\LB,0} \leq 180^{\circ}$ need not be considered \citep[e.g.][]{liu2015}.}  In practice however, we explore a limited range of $\theta_{\LB,0}$ to avoid unnecessary computation for planets that have no chance of migrating.  Note that systems with inclinations $\theta_{\LB,0} \lesssim 40^{\circ}$ (the critical ``Kozai angle'') can be safely excluded, because they do not undergo large excursions in eccentricity.  We find empirically that systems with $\theta_{\LB,0} \lesssim 65^{\circ}$ rarely reach sufficiently high eccentricities to induce tidal migration.  In the rare cases where migration occurs, the system always results in tidal disruption, rather than HJ formation.  We therefore restrict the inclination to lie in the range $65^{\circ}\leq \theta_{\LB,0} \leq 90^{\circ}$.

Of primary interest in this paper is the fraction of total systems
that result in the production of an HJ or tidal disruption, for fixed
planet mass and stellar type, and considering the full possible ranges
of ($\theta_{\LB,0},a,a_b,e_b$).  For a given combination of host star
properties and planet mass, we run $N_{\rm run}$ trials (typically
$\sim 9000$) by repeatedly sampling the inclination randomly from the
restricted range ($65^{\circ}\leq \theta_{\LB,0} \leq
90^{\circ}$)\footnote{The only exception is in Section
  \ref{sec:quadrupole}, where we explore initial inclinations in the
  range $80^{\circ}\leq \theta_{\LB,0} \leq 90^{\circ}$, since the
  parameters considered there never produce migrating planets when $\theta_{\LB,0} \lesssim 80^{\circ}$}. The fractions of HJ formation and tidal disruption can be obtained from $f_{\rm HJ}  = \cos 65^{\circ} N_{\rm HJ}/ N_{\rm run}$ and $f_{\rm dis}  = \cos 65^{\circ} N_{\rm dis}/ N_{\rm run}$, where $N_{\rm HJ}$ and $N_{\rm dis}$ are the number of systems among $N_{\rm run}$ runs that resulted in HJs and tidal disruptions.

The ultimate goals of this section are to present distributions of
final stellar spin-orbit angles, and obtain the fractions of total
systems that result in HJs and disruptions for a given planet mass and
stellar type, sampling over the entire possible ranges of $a,a_b,e_b$.
However, we begin by fixing $e_b = 0$, thereby eliminating
complications introduced by octupole terms.  Section
\ref{sec:quadrupole} shows results for fixed binary separation $a_b$
and planet semimajor axis $a$, in order to isolate and highlight the
effects of changing the planet mass and stellar mass/spin properties.
Next, Section \ref{sec:octupole} presents results for non-zero binary
eccentricity (with fixed $a_b$ and $a$), thus showing how the octupole term in the disturbing potential of the binary companion can affect the results.  Finally, in Section \ref{sec:final}, we randomly sample over a wide range in ($a,a_b, e_b$) parameter space, and present results appropriate for comparison with the observational sample of close-in giant planets.      

\subsection{Quadrupole Results}
\label{sec:quadrupole}
To start, we fix the initial planet semimajor axis $a_0 = 1.5$ AU, binary separation $a_b = 200$ AU, and binary eccentricity $e_b = 0$ (so that the octupole contributions vanish).  We consider planet masses $M_p = 0.3, 1.0, 3.0$ and $5.0M_J$, and run a fine grid of initial inclinations, selected randomly from an isotropic distribution (uniform in $\cos \theta_{\LB,0}$).  The argument of pericenter $\omega$ and orbital node $\Omega$ are randomly sampled uniformly in $[0,2 \pi]$. The results are shown in Figs.~\ref{fig:final_quantities_Gstar} (G star) and \ref{fig:final_quantities_Fstar} (F star), where we plot the final spin-orbit angle $\theta_{\SL,\rm f}$ and semimajor axis $a_{\rm f}$ versus the initial inclination $\theta_{\LB,0}$, as well as the distributions of $\theta_{\SL, \rm f}$ for the systems that resulted in HJs (with final semimajor axis $a_{\rm f} < 0.1$ AU).
\subsubsection{G Star}
\begin{figure*}
\centering
\includegraphics[width=0.8\textwidth]{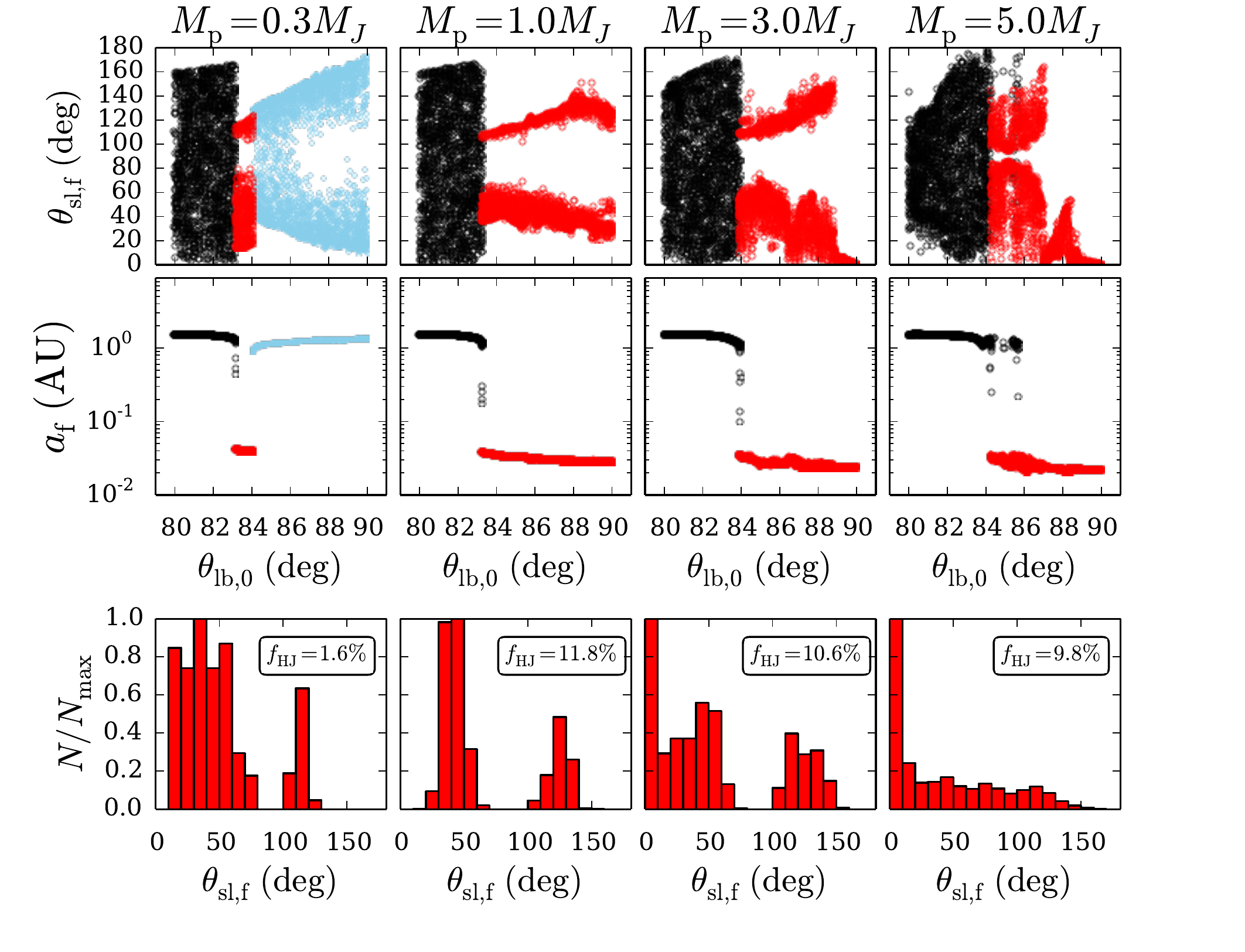}
\caption{Final spin-orbit angle $\theta_{\SL,\rm f}$ (top panels) and
  semi-major axis $a_{\rm f}$ (middle panels) as a function of
  $\theta_{\LB,0}$, for planet masses $M_p = 0.3, 1, 3$ and $5 \ M_J$
  (from left to right, as labeled).  Bottom panels show distributions
  of the final spin-orbit misalignments for the systems that
  circularized (HJs). All systems have $M_\star = 1 M_\odot$, $a =
  1.5$ AU, $a_b = 200$ AU, $e_b = 0$. Black points: non-migrating
  planets. Blue points: tidally disrupted planets. Red points: HJs.
  Note that the values of $\theta_{\SL,\rm f}$ and $a_{\rm f}$ for the
  disrupted planets are simply the values at the time-step before
  tidal disruption is achieved, and thus have no particular
  observational significance.  Tidal disruptions only occur here when $M_p = 0.3 M_J$, because the condition for disruption (Section \ref{sec:disrupt}, Eq.~[\ref{eq:disrupt}]) is not satisfied for the other planet masses.  See Table \ref{fixedbinarytable} for further information on the outcomes of the simulations.  The distribution of $\theta_{\SL,\rm f}$ is distinctly bimodal for $M_p = 1 M_J$, with a preference for prograde orbits.  As the planet mass increases, the adiabaticity parameter $\mathcal{A}_0$ increases (see Section 3), and for $M_p = 5 M_J$, the peak of the distribution occurs at low obliquities $\theta_{\SL, \rm f} = 0^\circ - 10^{\circ}$.}
\label{fig:final_quantities_Gstar}
\end{figure*}

 The dynamics considered in this section are considerably more
 complicated than the idealized analysis presented in Section
 \ref{sec:path}, since the effects of stellar spin-down ($S_\star
 \neq$ constant) and the backreaction torque from the oblate host star on the planetary orbit are now included.  Nonetheless, many of the general features remain for the G star (Fig.~\ref{fig:final_quantities_Gstar}).  The distribution of $\theta_{\SL,\rm f}$ for planets with mass $M_p = 1 M_J$ is distinctly bimodal with peaks  at $\theta_{\SL,\rm f} \sim 40^{\circ}$ and $120^{\circ}$ (compare with Figs.~\ref{typeII} and \ref{typeII_feedback} in Section \ref{sec:path}).  As $M_p$ increases, the systems with larger initial inclinations ($\theta_{\LB,0}$) show a preference for alignment due to their higher adiabaticity parameters, with $\mathcal{A}_0 \propto M_p/ \cos \theta_{\LB,0}$ (see Eq.~[\ref{adiabatic}]).  The largest mass planets ($M_p = 5 M_J$) tend to settle into low obliquity states ($\theta_{\SL,\rm f} \lesssim 10^{\circ}$), although high misalignments still remain possible.  Note that the cases with $M_p = 5 M_J$ and $\theta_{\LB,0} \sim 88^{\circ}$ (in the top, rightmost plot in Fig.~\ref{fig:final_quantities_Gstar}) have undergone adiabatic advection (see Section \ref{sec:path}).

For the lowest mass planets ($M_p = 0.3 M_J$), most systems result
either in non-migrating planets or tidal disruptions, with very few
``hot Saturns'' produced.  Tidal disruptions for low mass planets are
more common because of the larger tidal disruption radius (see
Eq.~[\ref{Rtide}]).  When $M_p = 0.3 M_J$, $r_{{\rm Tide}} \approx 4
R_{\odot}$, whereas when $M_p = 5 M_J$, $r_{{\rm Tide}} \approx 1.6
R_{\odot}$.  As a result, with $M_p = 0.3 M_J$ and the fixed values of
($a,a_b,e_b$) that we consider in this subsection, there is only a
very narrow range of initial inclinations that lead to pericenter
distances that are small enough to induce orbital decay, but large
enough to prevent tidal disruption (see
Fig.~\ref{fig:final_quantities_Gstar}, left panels).  For $a_0 = 1.5$
AU, $a_b = 200$ AU, and $e_b = 0$, systems with $M_p \geq 1 M_J$ never
result in tidal disruptions, because the condition for disruption to
be possible, derived in Section \ref{sec:disrupt} (see
Fig.~\ref{fig:migration_condition} and Eq.~[\ref{eq:disrupt}]) is
never satisfied.   However, note that these results depend on the
assumed tidal disruption radius (Eq.\ref{Rtide}).  The exact tidal
radius is somewhat uncertain, and depends on the assumed planetary
mass-radius relation, which can vary for close-in giant planets.   

\subsubsection{F Star}
\begin{figure*}
\centering       
\includegraphics[width=0.8\textwidth]{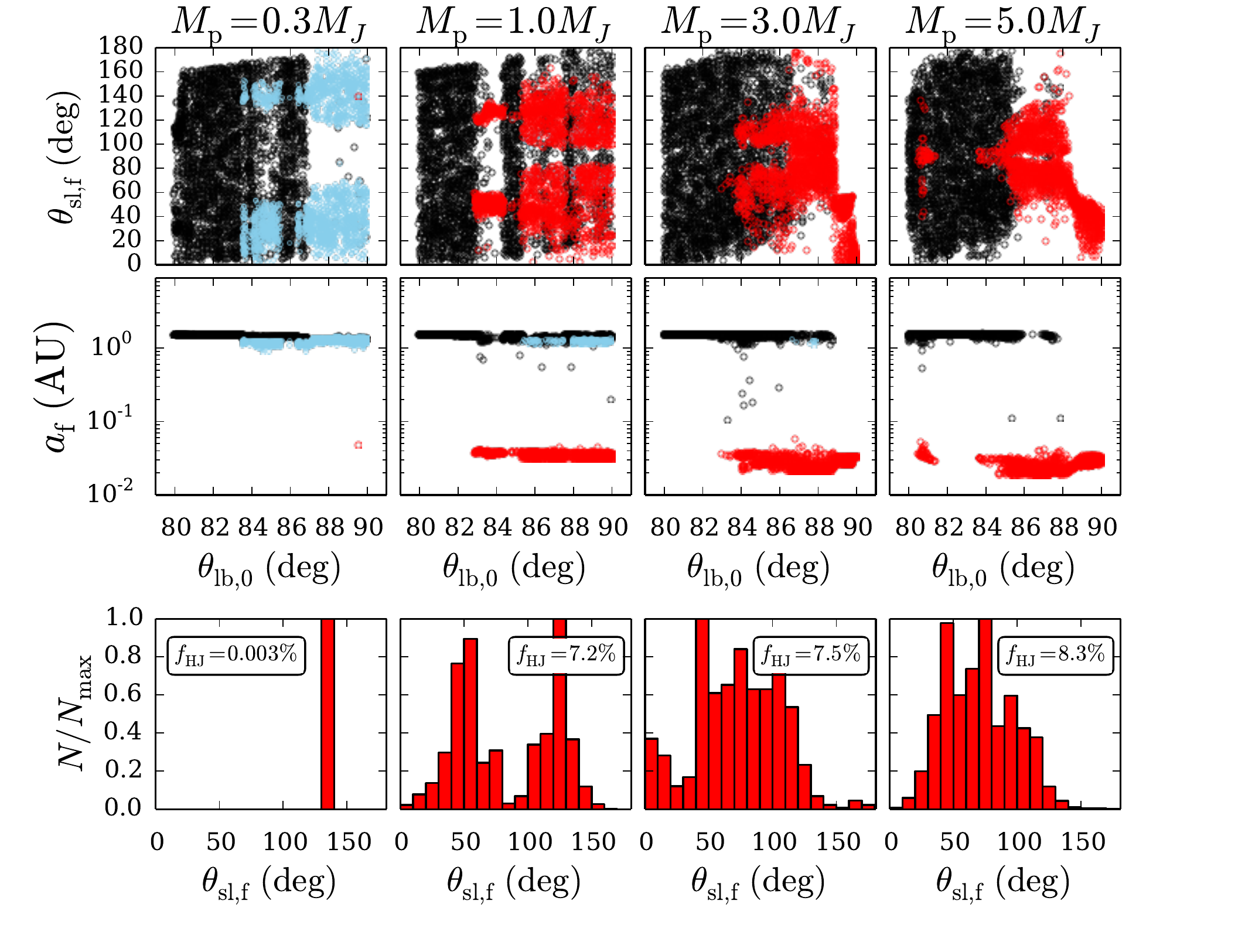}
       \caption{Same as Fig.~\ref{fig:final_quantities_Gstar}, except
         for an F-type host star, with $M_\star = 1.4 M_\odot$,
         $R_\star = 1.26 R_{\odot}$ and corresponding spin properties
         (see text).  Note that the histogram for $M_p = 0.3 M_J$ has
         only one data point.  When $M_p = 1 M_J$, the distributions of
         $\theta_{\SL,\rm f}$ are similar to those for the G star, but
         are broadened.  When $M_p = 5 M_J$, however, the strong peak
         near low obliquities ($\theta_{\SL,\rm f} = 0 - 10^{\circ}$)
        observed for planets around G stars has vanished.  We attribute these differences to the increased torque from the stellar quadrupole on the planetary orbit, as well as stronger periastron precession from SRFs.}
\label{fig:final_quantities_Fstar}
\end{figure*}

The results of identical calculations for the F star are shown in
Fig.~\ref{fig:final_quantities_Fstar}.  The HJ fractions are
consistently lower compared to the G star, for all planet masses, but
most noticeably for $M_p = 0.3 M_J$, with only a single HJ produced in
$\sim 5000$ trials.  For planet mass $M_p = 1 M_J$, the distribution
of $\theta_{\SL,\rm f}$ remains bimodal, but with larger spread.  For
$M_p = 5 M_J$, the distributions of $\theta_{\SL,\rm f}$ are
strikingly different between the F and G stars.  The peak of the
distribution occurs at $\theta_{\SL, \rm f} \approx 70^{\circ} -
80^{\circ}$, i.e. producing many HJs in near polar orbits with respect
to the stellar spin axis. This contrasts strongly with results for the
G star, where the peak occurs at $\theta_{\SL, \rm f} = 0^{\circ} -
10^{\circ}$.  These differences between the G star
(Fig.~\ref{fig:final_quantities_Gstar}) and F star
(Fig.~\ref{fig:final_quantities_Fstar}) arise for two reasons.  First,
the larger stellar mass and radius affect the net rate of pericenter
precession from SRFs, $\dot{\omega}$.  The contributions to
$\dot{\omega}$ from general relativity and the planetary tidal
deformation are higher for more massive stars, which lead to a lower
maximum achievable eccentricity and tend to reduce HJ production
fractions (however, note that the contribution to $\dot{\omega}$ from
the oblate host star has the opposite sign, and can, under come
circumstances, cancel the increases in $\dot{\omega}$ from GR and
tidal distortion).  Second, the larger stellar radius and spin frequency (compared to the G star) both lead to a more pronounced torque on the planetary orbit from the stellar quadrupole, since $(d \Lp /d t)_{\rm SL} \propto R_{\star}^5 \Omega_\star^2$; see Section \ref{sec:feedback}, Eq.~[\ref{eq:feedback}]).  The increased stellar radius alone leads to an increase in the backreaction torque of the stellar quadrupole on the orbit by a factor of $\sim 3$, with a further increase due to higher $\Omega_\star$.  

Both the wider spread in the bimodal distributions (when $M_p = 1 M_J$), and peak near $\theta_{\SL,\rm f} \sim 90^{\circ}$ (when $M_p = 5 M_J$) can be understood from the results of Section \ref{sec:path}, where we presented final spin-orbit angles for varying initial inclinations, both with and without feedback included.  Comparing the lower left panels of Figs.~\ref{typeII} and \ref{typeII_feedback} shows that in some cases, including feedback causes the bimodality to be partially preserved, but with significant broadening.  Similarly, comparing the upper left panels of Figures \ref{typeII} and \ref{typeII_feedback} shows that in other cases, including feedback completely erases the bimodality, causing $\theta_{\SL, \rm f}$ to instead cluster around $\sim 90^{\circ}$.  Thus, we attribute the qualitative differences in $\theta_{\SL,\rm f}$ between the G and F star to enhanced feedback from the oblate F star on the orbit.

\subsection{Octupole Results: Fixed Binary Eccentricity and Separation} 
\label{sec:octupole}
\begin{figure*}
\centering
\includegraphics[width=0.8\textwidth]{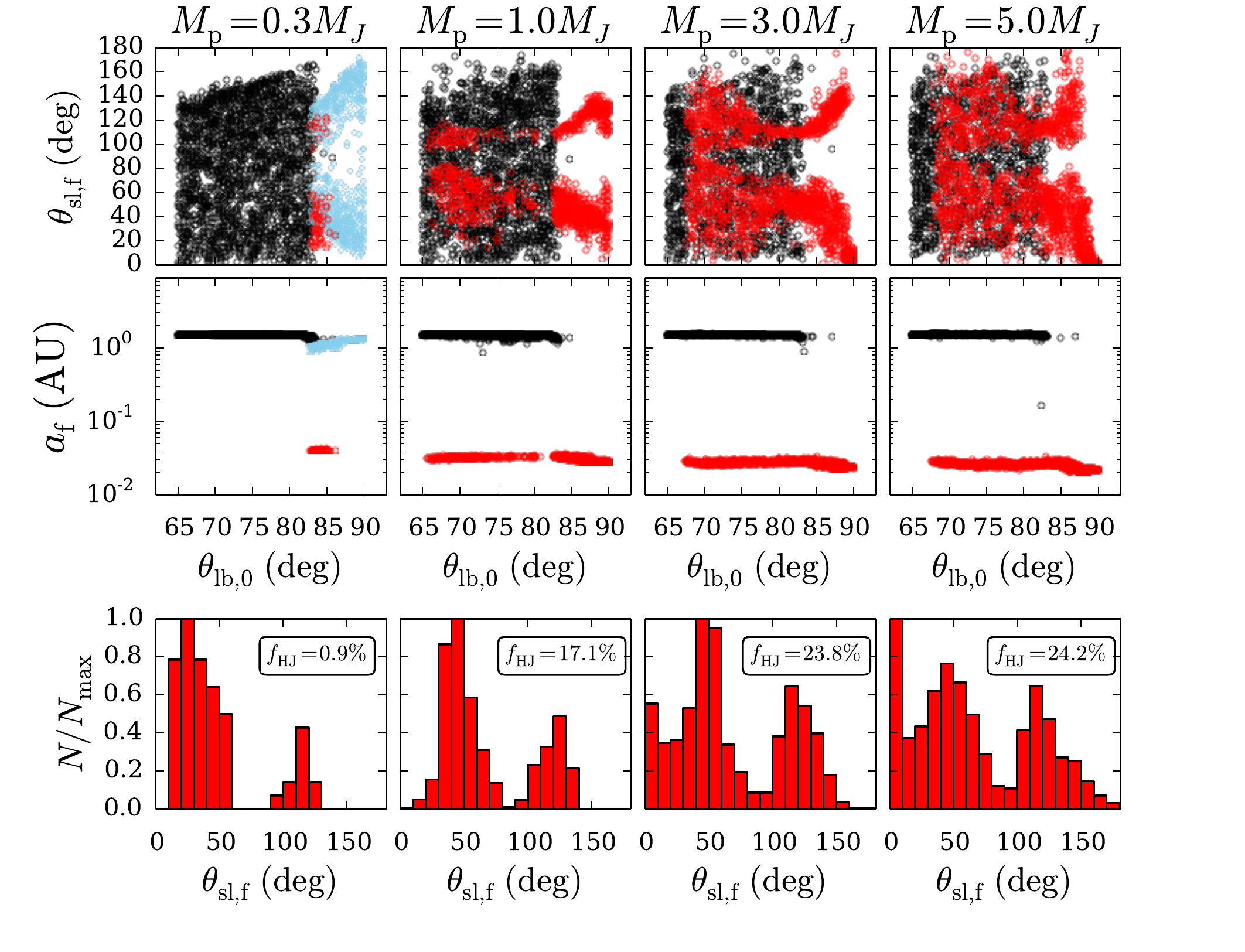}
\caption{ Same as Fig.~\ref{fig:final_quantities_Gstar}, except that
  $e_b = 0.8$, and $a_b = 333.33$ AU (so that $a_{b,\rm eff} = 200$,
  AU, and $\varepsilon_{\oct} \approx  0.01$).  For $M_p = 0.3 M_J$,
  the results are nearly unchanged (compared to Fig.~\ref{fig:final_quantities_Gstar}), because pericenter precession from
  SRFs is higher for low-mass planets (see text), and the effects of
  the octupole (e.g. extreme high eccentricities) are more easily
  suppressed.  For $M_p \geq 1 M_J$, the HJ production fraction is
  increased.  In terms of $\theta_{\SL,\rm f}$, the main effect of the octupole is to add HJs with a primarily bimodal distribution, thereby increasing the fraction of significantly misaligned planets.}
\label{fig:final_quantities_oct2}
\end{figure*}

Having demonstrated results for binary companions with zero eccentricity, we now consider binaries with non-zero eccentricity, so that the octupole terms can contribute to the dynamics.  We limit the discussion in this section to the solar-type (G) star, and present one example of fixed non-zero binary eccentricity (see Section \ref{sec:final} for general combinations of $a_b$ and $e_b$).  For a straightforward comparison with the results from Section \ref{sec:quadrupole}, and to illustrate the role of the octupole, we choose the parameters so that the quadrupole LK timescale $t_k$ (Eq.~[\ref{tk}]) is unchanged (since $t_k$ depends only on the combination $\abeff = a_b \sqrt{1 - e_b^2}$).  We thus specify the binary eccentricity $e_b$ and choose the separation $a_b$ such that the quantity $\abeff = 200$ AU.  Figure \ref{fig:final_quantities_oct2} shows results for $e_b = 0.8$, $a_b = 333$ AU, corresponding to $\varepsilon_{\oct} \approx  0.01$.  Additional results with $e_b = 0.4$, $a_b = 218$ AU, so that $\varepsilon_{\oct} \approx  0.003$ are included in Table \ref{fixedbinarytable}. Recall that $\varepsilon_{\oct}$ quantifies the ``strength'' of the octupole potential; see Eq.~(\ref{eq:epsilonOct}).

Without the octupole terms, the limiting eccentricity $e_{\rm lim}$ during an LK cycle is achieved at $\theta_{\LB,0} = 90^{\circ}$.  One effect of the octupole term is to allow this limiting eccentricity to be realized at $\theta_{\LB,0} < 90^{\circ}$ \citep{liu2015}, so that migration becomes possible for a wider range of inclinations, thereby increasing the production efficiency \citep{naoz2012}. 

Comparing Figs.~\ref{fig:final_quantities_Gstar} and \ref{fig:final_quantities_oct2} allows the role of the octupole terms to be identified, since they would produce identical results to quadrupole order.  Low mass planets are affected by the octupole potential less than high mass planets, because the rate of pericenter precession due to tidal distortion of the planet has the dependence $\dot{\omega}_\tide \propto M_p^{-1}$ (see Eq.~[\ref{eq:omega_tide_p}]).  This precession can act to suppress the extreme octupole dynamics, such as increased eccentricities and orbit flipping.  Thus for the lowest mass planets ($0.3 M_J$) the results do not differ significantly from the pure quadrupole case.  More massive planets ($M_p = 1 - 5 M_J$) are affected more strongly, with the production fraction of HJs increasing with the octupole strength $\varepsilon_{\rm oct}$ (see Section \ref{sec:migrationfrac} for further discussion of HJ and disruption fractions).  

\begin{figure*}
\centering
\includegraphics[width=0.4\textwidth]{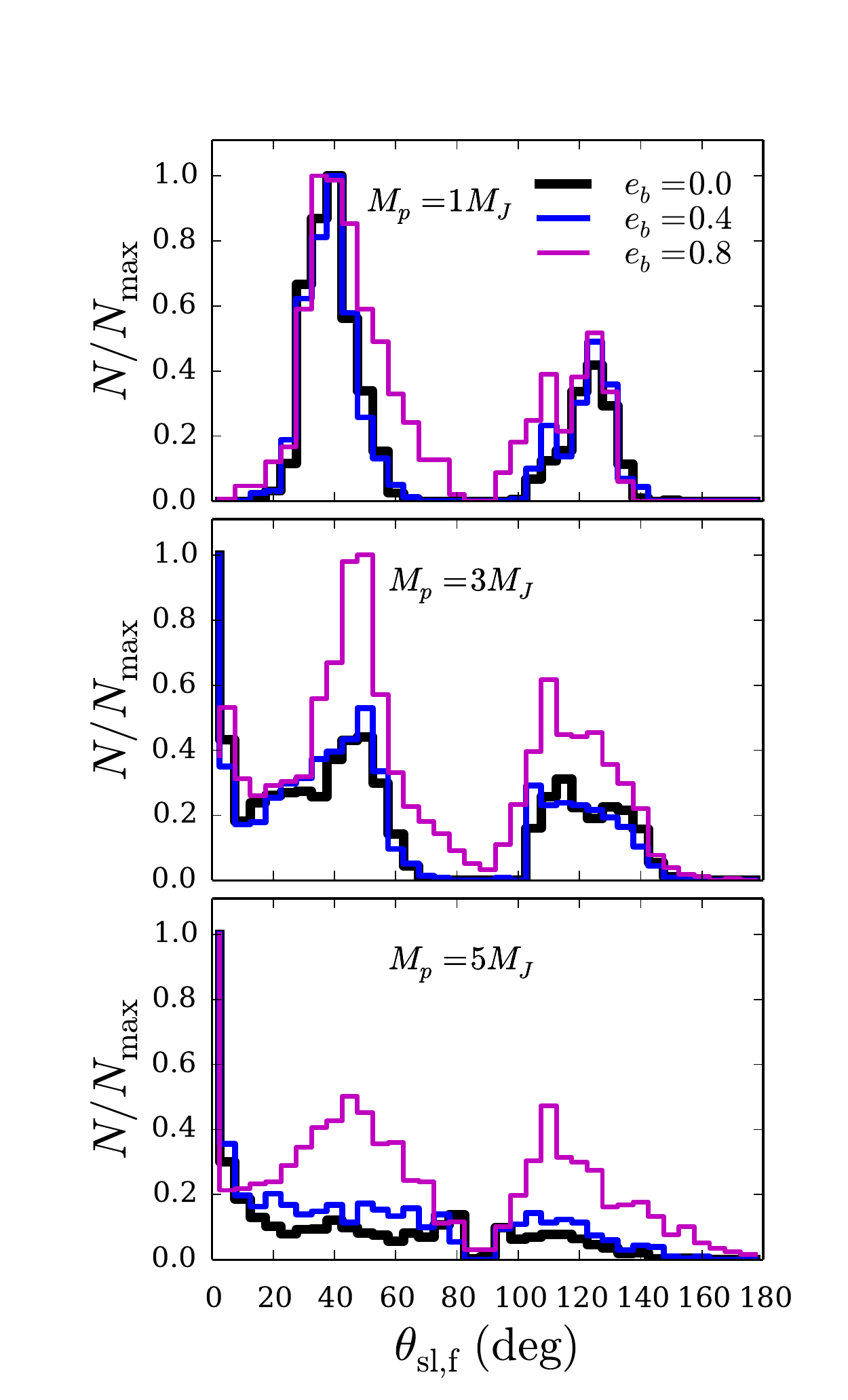}
\caption{Distributions of $\theta_{\SL,\rm f}$ for various binary eccentricities, $e_b = 0, 0.4, 0.8$, as labeled, and showing planet masses $M_p = 1, 3, 5 M_J$ (from top to bottom).  Binary separations have been chosen such that $\abeff = a_b \sqrt{1 - e_b^2} = 200$ AU.  As a result, the quadrupole LK timescale $t_k$ is identical, so that the results depicted in each panel would be identical to quadrupole order. This illustrates the role of the octupole in generating spin-orbit misalignment.}
\label{fig:oct_hist}
\end{figure*}

In terms of the final obliquity $\theta_{\SL,\rm f}$, one effect of the octupole is to increase the number of significantly misaligned $5 M_J$ planets, as demonstrated in Fig.~\ref{fig:oct_hist}.  There are two possible reasons for this. First, the octupole allows close-in planets to be produced at lower inclinations, with lower adiabaticity parameters ($\mathcal{A}_0 \propto 1 / \cos \theta_{\LB,0}$).  Since the degree of misalignment depends on $\mathcal{A}_0$, systems with low inclinations have a tendency to settle to larger obliquities, and exhibit bimodality.  Second, the chaos induced in the orbit due to the octupole terms may act to disrupt the tendency for alignment found for the pure quadrupole calculations. Despite these effects, for $5 M_J$ planets with the octupole included, the strong peak near zero obliquity observed for the pure quadrupole results ($e_b = 0$, Fig.~\ref{fig:final_quantities_Gstar}) is partially preserved.

\begin{table*}
 \centering
 \begin{minipage}{180mm}
  \caption{Input parameters and results of the calculations presented
    in Sections \ref{sec:quadrupole} and \ref{sec:octupole}.  Each
    line is the result of $N_{\rm run}$ trials with initial
    inclination $\theta_{\rm lb,0}$ randomly sampled from an isotropic
    distribution in the range $65^{\circ} - 90^{\circ}$ (the only
    exception are the first eight rows, with $e_b = 0$, where
    $\theta_{\rm lb,0}$ is sampled in $80^{\circ} - 90^{\circ}$).
    Each set of trials has a fixed $a_b$ and $e_b$, as indicated, and
    $a_0 = 1.5$ AU, and tidal enhancement factor $\chi = 10$. The initial spin-orbit angle is set to $\theta_{\LB,0} = 0^{\circ}$. We display the ``migration fraction '' $f_{\rm mig} \equiv f_{\rm HJ} + f_{\rm dis}$, as well as the ``prograde fraction'' $f_{\rm prog}$ i.e.\ the fraction of HJ systems with final obliquities $\theta_{\SL,\rm f} < 90^{\circ}$.  We also include relevant figure numbers in the rightmost column.  Note that the stellar radius is set to $R_{\star} = 1 \ R_{\odot}$ when $M_{\star} = 1 \ M_{\odot}$, and $R_{\star} = 1.26 \ R_{\odot}$ when $M_{\star} = 1.4 \ M_{\odot}$.
  }
  \begin{tabular}{@{}llllllllll@{}}
  \hline
  \hline
  $M_\star$ $(M_\odot)$ & $M_p$ $(M_J)$ & $a_b$ (AU) & $e_b$ & $N_{\rm run}$ & $f_{\rm HJ}$ (\%)& $f_{\rm dis}$ (\%) & $f_{\rm mig}$ (\%) & $f_{\rm prog}$ \% & Figure \\
\hline
{\bf Section \ref{sec:quadrupole}} \\
1.0 & 0.3 & 200.0 & 0.0 & 5000 & 1.6 & 10.3 & 12.0 & 84.3 & \ref{fig:final_quantities_Gstar}, \ref{fig:oct_hist} \\
1.0 & 1.0 & 200.0 & 0.0 & 5000 & 11.8 & 0.0 & 11.8 & 71.2 & \ref{fig:final_quantities_Gstar}, \ref{fig:oct_hist} \\
1.0 & 3.0 & 200.0 & 0.0 & 5000 & 10.6 & 0.0 & 10.6 & 72.0 & \ref{fig:final_quantities_Gstar}, \ref{fig:oct_hist} \\
1.0 & 5.0 & 200.0 & 0.0 & 5000 & 9.8 & 0.0 & 9.8 & 82.6 & \ref{fig:final_quantities_Gstar}\\
1.4 & 0.3 & 200.0 & 0.0 & 5000 & 0.003 & 7.8 & 7.8 & 0.0 & \ref{fig:final_quantities_Fstar}\\
1.4 & 1.0 & 200.0 & 0.0 & 5000 & 7.2 & 0.9 & 8.2 & 54.5 & \ref{fig:final_quantities_Fstar}\\
1.4 & 3.0 & 200.0 & 0.0 & 5000 & 7.5 & 0.0 & 7.5 & 66.8 & \ref{fig:final_quantities_Fstar}\\
1.4 & 5.0 & 200.0 & 0.0 & 5000 & 8.3 & 0.0 & 8.3 & 74.0 & \ref{fig:final_quantities_Fstar}\\
\hline
{\bf Section \ref{sec:octupole}} \\
1.0 & 0.3 & 218.22 & 0.4 & 3000 & 1.3 & 10.8 & 12.2 & 89.5 & \ref{fig:oct_hist} \\
1.0 & 1.0 & 218.22 & 0.4 & 3000 & 12.2 & 0.0 & 12.2 & 68.1 & \ref{fig:oct_hist} \\
1.0 & 3.0 & 218.22 & 0.4 & 3000 & 12.4 & 0.0 & 12.4 & 73.4 & \ref{fig:oct_hist} \\
1.0 & 5.0 & 218.22 & 0.4 & 3000 & 12.9 & 0.0 & 12.9 & 78.6 & \ref{fig:oct_hist}\\
1.0 & 0.3 & 333.33 & 0.8 & 3000 & 0.9 & 11.4 & 12.3 & 82.5 & \ref{fig:final_quantities_oct2}, \ref{fig:oct_hist} \\
1.0 & 1.0 & 333.33 & 0.8 & 3000 & 17.1 & 0.0 & 17.1 & 70.4 & \ref{fig:final_quantities_oct2}, \ref{fig:oct_hist}\\
1.0 & 3.0 & 333.33 & 0.8 & 3000 & 23.8 & 0.0 & 23.8 & 65.7 & \ref{fig:final_quantities_oct2}, \ref{fig:oct_hist}\\
1.0 & 5.0 & 333.33 & 0.8 & 3000 & 24.2 & 0.0 & 24.2 & 66.3 & \ref{fig:final_quantities_oct2}, \ref{fig:oct_hist}\\
\hline
\hline
\label{fixedbinarytable}
\end{tabular}
\end{minipage}
\end{table*}

\begin{table*}
 \centering
 \begin{minipage}{180mm}
  \caption{Same format as Table \ref{fixedbinarytable}, but showing
    results for the full population synthesis calculations in Sections
    \ref{sec:final}, \ref{sec:popsynth_dissipation}, and
    \ref{sec:primordial}. We vary $a_0$, $a_b$, and $e_b$ uniformly in
    the ranges $a_0 = (1 - 5)$ AU, $a_b = (100 - 1000)$ AU (note that $a_b$ is sampled uniformly in $\log a_b$), and $e_b = (0 - 0.8)$.  $\theta_{\LB,0}$ is sampled isotropically in the range $65^{\circ} - 90^{\circ}$.  The other parameters and notation are the same as in Table \ref{fixedbinarytable}.
  }
  \begin{tabular}{@{}llllllllll@{}}
  \hline
  \hline
  $M_\star$ $(M_\odot)$ & $M_p$ $(M_J)$ & $\theta_{\SL,0}$ $(\circ)$& $\chi$ & $N_{\rm run}$ & $f_{\rm HJ}$ (\%)& $f_{\rm dis}$ (\%) & $f_{\rm mig}$ (\%) & $f_{\rm prog}$ \% & Figure \\
\hline
{\bf Section \ref{sec:final}} \\
1.0 & 0.3 & 0.0 & 10.0 & 8988 & 0.5 & 12.3 & 12.8 & 70.4 & \ref{fig:initial_conditions_Gstar},\ref{fig:final_dist_Gstar} \\
1.0 & 1.0 & 0.0 & 10.0 & 8991 & 2.4 & 11.0 & 13.4 & 78.3 & \ref{fig:initial_conditions_Gstar},\ref{fig:final_dist_Gstar}\\
1.0 & 3.0 & 0.0 & 10.0 & 8996 & 3.8 & 9.3 & 13.1 & 72.0 & \ref{fig:initial_conditions_Gstar},\ref{fig:final_dist_Gstar}\\
1.0 & 5.0 & 0.0 & 10.0 & 8994 & 4.7 & 8.4 & 13.0 & 74.1 & \ref{fig:initial_conditions_Gstar},\ref{fig:final_dist_Gstar} \\
1.4 & 0.3 & 0.0 & 10.0 & 8993 & 0.0 & 12.3 & 12.3 & 100.0 & \ref{fig:final_dist_Fstar} \\
1.4 & 1.0 & 0.0 & 10.0 & 8994 & 1.4 & 10.9 & 12.3 & 64.9 & \ref{fig:final_dist_Fstar}\\
1.4 & 3.0 & 0.0 & 10.0 & 8998 & 3.0 & 9.8 & 12.8 & 67.7 & \ref{fig:final_dist_Fstar}\\
1.4 & 5.0 & 0.0 & 10.0 & 8997 & 3.6 & 9.1 & 12.6 & 69.4 & \ref{fig:final_dist_Fstar}\\
\hline
{\bf Section \ref{sec:popsynth_dissipation}} \\
1.0 & 0.3 & 0.0 & 1.0 & 8998 & 0.0 & 11.8 & 11.8 & 0.0 & \ref{fig:Porb_chi_Gstar}, \ref{fig:hist_chi_Gstar} \\
1.0 & 1.0 & 0.0 & 1.0 & 8991 & 0.7 & 11.1 & 11.8 & 75.6 & \ref{fig:Porb_chi_Gstar}, \ref{fig:hist_chi_Gstar}\\
1.0 & 3.0 & 0.0 & 1.0 & 8997 & 2.3 & 9.6 & 11.9 & 69.6 & \ref{fig:Porb_chi_Gstar}, \ref{fig:hist_chi_Gstar}\\
1.0 & 5.0 & 0.0 & 1.0 & 8993 & 3.1 & 9.5 & 12.5 & 70.9 & \ref{fig:Porb_chi_Gstar}, \ref{fig:hist_chi_Gstar}\\
1.4 & 0.3 & 0.0 & 1.0 & 8997 & 0.0 & 10.9 & 10.9 & 0.0 & \ref{fig:hist_chi_Fstar}\\
1.4 & 1.0 & 0.0 & 1.0 & 8995 & 0.4 & 10.6 & 10.9 & 52.0 & \ref{fig:hist_chi_Fstar}\\
1.4 & 3.0 & 0.0 & 1.0 & 8996 & 1.5 & 10.4 & 11.8 & 58.1 & \ref{fig:hist_chi_Fstar}\\
1.4 & 5.0 & 0.0 & 1.0 & 8998 & 1.9 & 9.9 & 11.8 & 61.9 & \ref{fig:hist_chi_Fstar}\\
1.0 & 0.3 & 0.0 & 100.0 & 8995 & 2.4 & 11.6 & 14.0 & 61.6 & \ref{fig:Porb_chi_Gstar}, \ref{fig:hist_chi_Gstar}\\
1.0 & 1.0 & 0.0 & 100.0 & 8997 & 4.1 & 9.7 & 13.8 & 68.7 & \ref{fig:Porb_chi_Gstar}, \ref{fig:hist_chi_Gstar}\\
1.0 & 3.0 & 0.0 & 100.0 & 8994 & 6.4 & 5.9 & 12.4 & 71.8 & \ref{fig:Porb_chi_Gstar}, \ref{fig:hist_chi_Gstar}\\
1.0 & 5.0 & 0.0 & 100.0 & 8994 & 7.8 & 4.1 & 12.0 & 71.0 & \ref{fig:Porb_chi_Gstar}, \ref{fig:hist_chi_Gstar}\\
1.4 & 0.3 & 0.0 & 100.0 & 8997 & 1.5 & 11.7 & 13.2 & 65.5 & \ref{fig:hist_chi_Fstar}\\
1.4 & 1.0 & 0.0 & 100.0 & 8996 & 3.3 & 9.9 & 13.2 & 65.0 & \ref{fig:hist_chi_Fstar}\\
1.4 & 3.0 & 0.0 & 100.0 & 8994 & 6.3 & 6.2 & 12.5 & 66.3 & \ref{fig:hist_chi_Fstar}\\
1.4 & 5.0 & 0.0 & 100.0 & 8999 & 7.6 & 4.1 & 11.6 & 66.7 & \ref{fig:hist_chi_Fstar}\\
\hline
{\bf Section \ref{sec:primordial}} \\
1.0 & 0.3 & 30.0 & 10.0 & 8995 & 0.3 & 12.8 & 13.1 & 67.2 & \ref{fig:primordial} \\
1.0 & 1.0 & 30.0 & 10.0 & 8996 & 2.6 & 10.6 & 13.1 & 62.1 & \ref{fig:primordial} \\
1.0 & 3.0 & 30.0 & 10.0 & 8986 & 4.0 & 9.5 & 13.5 & 61.1 & \ref{fig:primordial} \\
1.0 & 5.0 & 30.0 & 10.0 & 8995 & 4.8 & 8.8 & 13.6 & 70.6 & \ref{fig:primordial} \\
1.0 & 0.3 & 60.0 & 10.0 & 8993 & 0.4 & 12.8 & 13.2 & 52.4 & \ref{fig:primordial} \\
1.0 & 1.0 & 60.0 & 10.0 & 8995 & 2.6 & 11.2 & 13.8 & 47.5 & \ref{fig:primordial} \\
1.0 & 3.0 & 60.0 & 10.0 & 8993 & 4.4 & 10.0 & 14.5 & 49.3 & \ref{fig:primordial} \\
1.0 & 5.0 & 60.0 & 10.0 & 8993 & 4.9 & 9.4 & 14.3 & 54.5 & \ref{fig:primordial} \\
\hline
\hline
\label{popsynthtable}
\end{tabular}
\end{minipage}
\end{table*}

\subsection{General Results for a Range of Binary Separations, Eccentricities, and Planet Semi-major Axes}
\label{sec:final}
We now survey the parameter space in ($a_0, a_b, e_b$), sampling the
initial planet semi-major axis $a_0$ uniformly in the range $a_0 = 1 -
5$ AU, the binary separation $a_b = 100 - 1000$ AU (uniform in $\log
a_b$), and the binary eccentricity uniformly in $e_b = 0 - 0.8$.  This
choice of eccentricity distribution is highly approximate, as the
actual eccentricity distribution of wide binaries is uncertain
\citep{tokovinin2015}.  Moreover, planet formation at a few AU may be
quenched by the presence of a highly eccentric binary companion (when
$a_b [1 - e_b]$ is not sufficiently larger than $a_0$).  As in previous subsections, the initial inclination $\theta_{\LB,0}$ is sampled isotropically in the range $65^\circ - 90^{\circ}$. We fix the tidal enhancement factor at $\chi = 10$ in this section; we explore the effects of varying $\chi$ in Section \ref{sec:popsynth_dissipation}.

\subsubsection{Hot Jupiter and Disruption Fractions}
\label{sec:migrationfrac}
\begin{figure*}
\centering
\includegraphics[width=0.8\textwidth]{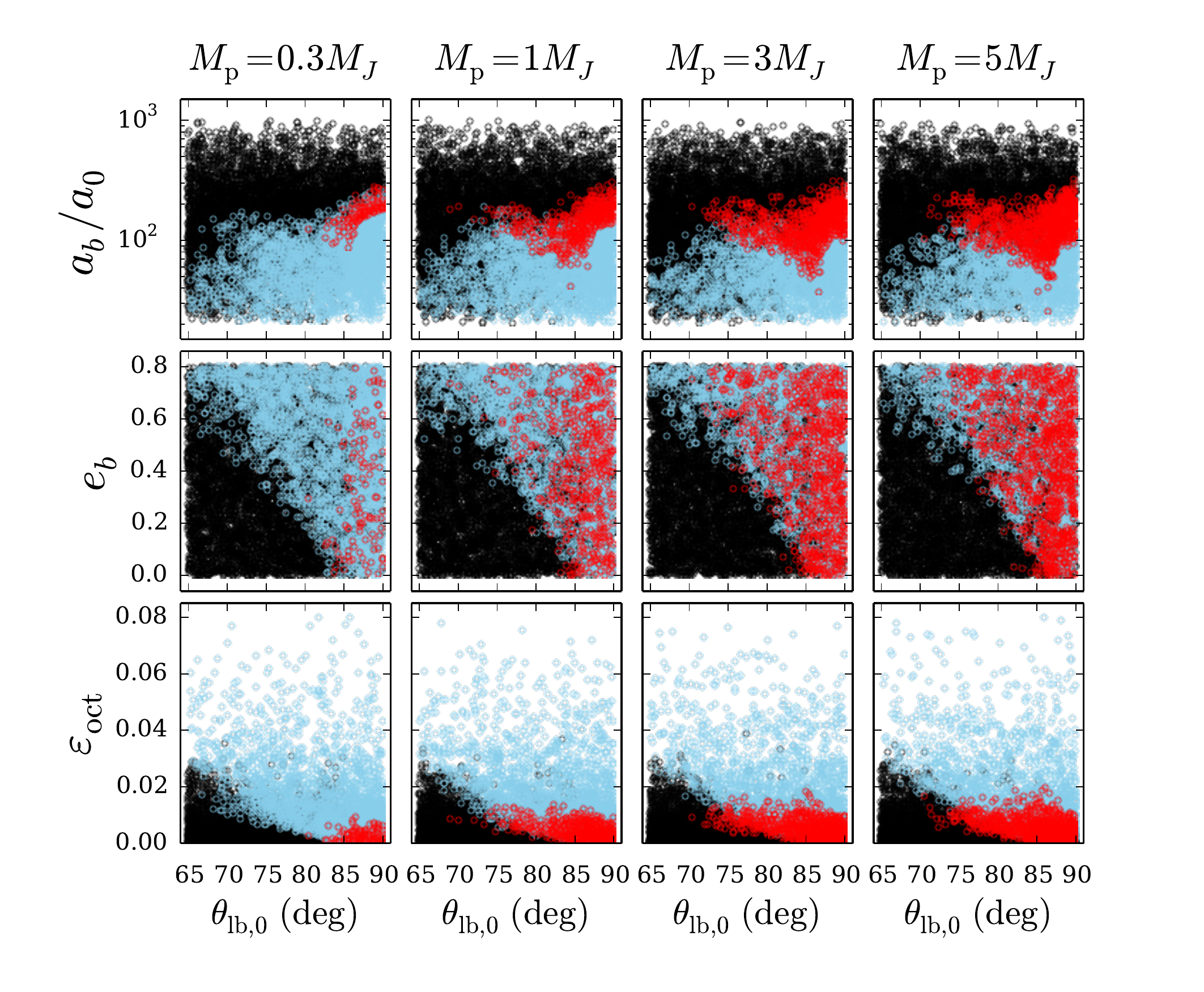}
\caption{Parameter space producing HJs (red), tidally disrupted planets (blue), and non-migrating planets (black), around G stars.  Top panels: initial binary separation ratio ($a_b/a_0$) versus the initial inclination $\theta_{\LB,0}$. Middle panels: Binary eccentricity $e_b$ .  Bottom panels: ``Octupole strength'' $\varepsilon_{\rm oct}$.  Results are separated into columns by planet mass, as labeled.  HJs are able to be produced over the full range of $e_b = [0,0.8]$, but only in a relatively narrow range of $a_b / a_0$.  As a result, the range of $\varepsilon_{\oct}$ capable of producing HJs is limited, with $\varepsilon_{\rm oct} \lesssim 0.01 - 0.02$.}
\label{fig:initial_conditions_Gstar}
\end{figure*}

Figure \ref{fig:initial_conditions_Gstar} depicts the outcomes of our simulations for planets around G stars, where we plot the initial semi-major axis ratio $a_b/a_0$ and binary eccentricity $e_b$ versus the initial inclination $\theta_{\LB,0}$.  The final outcome of each integration is indicated by the color (HJ, disrupted planet, or non-migrating).  Results for planets around F stars are qualitatively similar, and are omitted.  See Table \ref{popsynthtable} for further information, including the HJ and disruption fractions.

Figure \ref{fig:initial_conditions_Gstar} shows that HJs are produced for a relatively narrow range of the ratio $a_b / a_0$.  Planets with $a_b/a_0 \lesssim 60$ are always either tidally disrupted or non-migrating, while those with $a_b / a_0 \gtrsim 300$ never undergo migration.  This result places constraints on the requirements for stellar companions to induce migration without destroying the planet (see also Section \ref{sec:disrupt} for a discussion of the conditions that must be satisfied for migration and tidal disruption).  In the bottom panels of Fig.~\ref{fig:initial_conditions_Gstar}, we plot the values of $\varepsilon_{\rm oct}$ versus $\theta_{\LB,0}$.  We find that systems with $\varepsilon_{\rm oct} \gtrsim 0.03$ always lead to tidal disruptions, and that no HJs are produced for $\varepsilon_{\rm oct} \gtrsim 0.01 - 0.02$.  This finding can be understood by examining Fig.~\ref{fig:mig_condition_data}, where we plot the initial conditions in terms of ($a_{b,\rm eff}, a_0$) for the $1 M_J$ planets that resulted in tidal disruptions and HJs, along with the criteria for migration (disruption) to occur, shown as solid red (blue) curves (see also Fig.~\ref{fig:migration_condition}).  We see that the migration/disruption conditions derived in Section \ref{sec:disrupt} are in good agreement with our numerical calculations.  

Also plotted in Fig.~\ref{fig:mig_condition_data} are curves of constant $\varepsilon_{\rm oct} = 0.015$ (dashed black curves, with $e_b = 0.4, 0.6, 0.8$, from bottom to top).  The uppermost dashed line, with $e_b = 0.8$, nearly coincides with the tidal disruption boundary, so that $\varepsilon_{\rm oct} \gtrsim 0.015$ can only be achieved for combinations of $(a_{b,\rm eff},a_0)$ that are located in the ``disruption zone'' i.e. below the solid blue curve, where systems are likely to result in tidal disruption, rather than HJs.  Since we consider a range of binary eccentricities uniform in $e_b = [0,0.8]$, all of our systems with $\varepsilon_{\rm oct} \gtrsim 0.015$ reside in the disruption zone, thereby explaining the lack of circularized planets in our calculations with $\varepsilon_{\rm oct} \gtrsim 0.015$.

Planets with mass $M_p = 1 - 3 M_J$ around G stars have HJ production fractions $f_{\rm HJ}$ in the range $2.4 - 3.8 \%$, and $f_{\rm HJ}$ for planets around F stars is somewhat lower ($1.4 - 3 \%$).  For both stellar types, the fraction of HJs produced increases with planet mass (see also Table \ref{popsynthtable}, and the discussion in Section \ref{sec:octupole}).  This arises from our tidal disruption criterion (Eq.~[\ref{Rtide}]), with $r_{\rm Tide} \approx 4 R_{\odot}$ for the sub-Jupiter mass planet ($M_p = 0.3 M_J$), and $r_{\rm Tide} \approx 1.6 R_{\odot}$ for $M_p = 5 M_J$.  Low mass planets are therefore much more susceptible to tidal disruption, and are more readily removed from the sample of surviving planets.  We find that the fraction of ``hot Saturns'' ($M_p = 0.3 M_J$) produced is especially low, with $f_{\rm HJ} (0.3 M_J) \approx 0.5 \%$ and $0.02 \%$ for the G and F stars respectively. 

Comparing the results of Sections \ref{sec:quadrupole} and \ref{sec:octupole} (see Table \ref{fixedbinarytable}), and this subsection (Table \ref{popsynthtable}), we see that although certain combinations of ($a_0,a_b,e_b$) can lead to HJ fractions of $f_{HJ} \sim 24 \%$ (specifically when the octupole effect is included; see also \citealt{naoz2012}), when ranges of $(a_0,a_b,e_b)$ are considered, the overall HJ fraction is always less than a few percent for planets with mass $M_p = 1 M_J$.

Inspection of Table \ref{popsynthtable} reveals that the ``migration
fraction'' $f_{\rm mig} \equiv f_{\rm HJ} + f_{\rm dis} \approx 12 -
13 \%$ is nearly constant for all planet masses and stellar types,
varying by only $\sim 1 \%$.  Given the complicated interplay between
the various ingredients in our system (SRFs, octupole-level dynamics,
tidal dissipation), and the dependence of these physical processes on
planet and stellar mass, this result is not necessarily expected, but
can be qualitatively understood from the discussion in Section
\ref{sec:disrupt}.  To achieve planet migration (either HJ formation
or tidal disruption) within the lifetime of the host star, two
conditions must be satisfied: (i) The  planet must attain a
sufficiently large eccentricity ($\sim e_{\lim}$) so that the
corresponding periastron distance $a(1 - e_{\lim})$ is less than a
critical value ($\simeq 0.025$ AU).  This translates into a necessary
condition for migration as given by
Eq.~(\ref{eq:migration_condition}). (ii) For systems that satisfy this
condition, whether or not migration actually occurs depends on the
initial inclination $\theta_{\LB,0}$.  As discussed in Section
\ref{sec:disrupt}, without the octupole effect, $e_{\rm lim}$ is
achieved very close to $\theta_{\LB,0} = 90^{\circ}$. With octupole,
$e_{\rm lim}$ can be achieved for initial inclinations
$\theta_{\LB,0}$ in the range $\theta_{\LB,\rm crit} \leq
\theta_{\LB,0} \leq 90^{\circ}$, where $\theta_{\LB, \rm crit}$ (the
minimum inclination that can lead to $e_{\rm max} = e_{\rm lim}$) is
determined by $\varepsilon_{\rm oct} \simeq a e_b/a_b(1 - e_b^2)$,
with no dependence on planet or stellar mass \citep[see][]{liu2015}.
The fact that the ``window of extreme eccentricity'' ($\theta_{\LB,\rm
  crit} \leq \theta_{\LB,0} \leq 90^{\circ}$) is independent of $M_p$
and $M_{\star}$, combined with the weak dependence of
Eq.~(\ref{eq:migration_condition}) on $M_p$ and $M_{\star}$ explains
the nearly constant migration fraction observed in our calculations.
Note however that the migration fraction does depend on the assumed
distributions of the planetary and binary orbital properties ($a_0$,
$a_b$, $e_b$, $\theta_{\LB,0}$), and alternate choices for these
distributions would yield different migration fractions.  A
semi-analytic calculation of the migration/distruption fractions,
based on the idea discusssed here, is presented in \citet[submitted]{munoz2016}. 

Regardless of the reason, the fact that $f_{\rm mig} \approx$ constant
is a useful finding.  Recall that the disruption fractions quoted
herein depend on the disruption condition, which depends on the
planetary mass-radius relation, and is somewhat uncertain.  However,
noting that $f_{\rm mig} \approx$ constant allows us to estimate an
upper limit on the possible HJ fraction for any giant planet mass, by
setting $f_{\rm dis} \to 0$, so that $f_{\rm mig} \to f_{\rm HJ, max}
\sim 13 \%$. 

\subsubsection{Final HJ Orbital Periods and Spin-Orbit Misalignments}
Figures \ref{fig:final_dist_Gstar} and \ref{fig:final_dist_Fstar} show the final orbital periods and spin-orbit misalignments versus the initial inclination $\theta_{\LB,0}$ for the HJs produced in our calculations. Note that we have removed the systems that resulted in tidal disruptions and non-migrating planets for clarity.  

We see that the distribution of the final stellar obliquities are distinctly bimodal for $M_p = 1 - 3 M_J$ around both G and F host stars, with peaks around $30^\circ - 40^\circ$, and $120^\circ - 130^\circ$. As planet mass increases, greater differences emerge between the results for G and F stars.  For the G-type host star, massive planets tend to settle to lower obliquities.  When $M_p = 5 M_J$, the peak of the histogram occurs in the first bin ($\theta_{\SL,\rm f} = 0^{\circ} - 10^{\circ}$), with an underlying bimodal distribution of larger misalignments (Fig.~\ref{fig:final_dist_Gstar}).  Thus, the tendency for spin-orbit alignment for massive planets presented in Section \ref{sec:octupole} and in \cite{storch2014} is partially preserved when sampling over arbitrary binary eccentricities and separations.  By contrast, the results for massive planets ($5 M_J$) around the F-type host star (Fig.~\ref{fig:final_dist_Fstar}) show a greater degree of misalignment, with the peak of the distribution at $\theta_{\SL,\rm f} \sim 45^{\circ}$.  This is in qualitative agreement with the pure quadrupole calculations in Section \ref{sec:quadrupole} (see Fig.~\ref{fig:final_quantities_Fstar}).

We find that all combinations of stellar type and planet mass lead to a greater fraction of prograde ($\theta_{\SL,\rm f} \leq 90^{\circ}$), rather than retrograde ($\theta_{\SL,\rm f} \geq 90^{\circ}$) configurations (see Table \ref{popsynthtable}).  However, the percentage of prograde planets around F stars is consistently lower than around G stars.  For example, we find that for $M_p = 1 M_J$, the prograde percentage is $\approx 78 \%$ for the G star, and $\approx 65 \%$ for the F star.

The bimodal $\theta_{\SL,\rm f}$ distributions for Jupiter-mass planets around G stars shown in Fig.~\ref{fig:final_dist_Gstar} is quite different from those obtained by \cite{naoz2012} and \cite{petrovich2015b}.  These authors find much broader $\theta_{\SL,\rm f}$ distributions, with no apparent ``gap'' at $\theta_{\SL,\rm f} \sim 90^\circ$.  A key reason for this difference is that the previous works considered slowly-rotating host stars (and non-evolving spin rates), which have weak spin-orbit couplings.

Also depicted in Figs.~\ref{fig:final_dist_Gstar} and \ref{fig:final_dist_Fstar} are the final orbital periods $P_{\rm orb,f}$ as a function of initial inclination.  After the LK oscillations are suppressed, the tidal evolution occurs at nearly constant angular momentum, so that all planets settle to a final semi-major axis $a_{\rm f}\gtrsim 2 r_{\rm Tide}$.  Since $r_{\rm Tide}$ depends inversely on planet mass, high mass planets are able to achieve shorter final orbital periods than low mass planets.  As a result, the lowest mass planets ($M_p = 0.3 M_J$) reside farthest from their host stars, and exhibit the smallest spread in $P_{\rm orb,f}$. All calculations result in extremely close-in planets, with $P_{\rm orb,f} \lesssim 3$ days.  This lack of longer period HJs produced by the LK mechanism is in agreement with calculations by \cite{petrovich2015b}.

\subsubsection{Migration Time}
For the subset of planets that undergo migration (resulting in either HJ formation or tidal disruption), it is useful to examine the migration time $t_{\rm mig}$.  For systems that result in HJs, we define $t_{\rm mig}$ as the moment when the semi-major axis has decayed to $a < 0.1$ AU, so that the planet is classified as an HJ (this is also the time at which we stop our integrations).  For disrupted planets, $t_{\rm mig}$ is the point at which the planet crosses the tidal radius.  

Figure \ref{fig:circ_time} shows cumulative distributions of the migration time $t_{\rm mig}$ for HJs and disrupted planets obtained from our simulation with G-type host stars (as in Figures \ref{fig:initial_conditions_Gstar} and \ref{fig:final_dist_Gstar}).  Two trends are apparent: First, most tidal disruptions occur early, with more than $75 \%$ occurring within $0.1$ Gyr.  Second, the range of the HJ formation time varies with planet mass.  For $5 M_J$ planets, $2 {\rm Myr} \lesssim t_{\rm mig} \leq 5 {\rm Gyr}$.  In contrast, the HJ formation time for $0.3 M_J$ planets lies in the much more restricted range $2 {\rm Gyr} \lesssim t_{\rm mig} \leq 5 {\rm Gyr}$.  The minimum migration time for low mass planets thus differs significantly for low mass planets.  

The cause behind the lengthier HJ formation times for low mass ($M_p = 0.3 M_J$) planets is as follows.  Recall that the orbital decay rate for planets undergoing LK migration (Eq.~[\ref{eq:maxrate}]) has the dependence
\be
\left| \frac{1}{a} \frac{da}{dt} \right|_{\rm Tide, LK} \propto M_p^{-1} a_F^{-7}  \quad {\rm where} \quad a_F = a (1 - e_{\rm max}^2),
\ee
so that the tidal decay timescale $t_{\rm Tide} \propto M_p a_F^7$.  Since systems that produce surviving planets must satisfy $a_F/2 \geq r_{\rm Tide}$, for each planet mass there is a minimum tidal decay timescale
\be
t_{\rm Tide, min} \propto M_p r_{\rm Tide}^7 \propto M_p^{-4/3}.
\ee
The minimum decay time needed to produce a surviving HJ thus increases for lower mass planets, as we find in our numerical calculations.

Finally, we note that LK migration is often attributed to need a long time to operate, usually $\sim 0.1 - 1$ Gyr timescales, in contrast with disk-driven migration, which must occur before the gas dispersal time of a few Myr.  While we confirm that this is indeed the case for Jupiter and sub-Jupiter mass planets, we find that massive planets ($M_p \sim 3 - 5 M_J$) can migrate more quickly, within tens or occasionally even a few Myr, much more comparable to the timescale for disk-driven migration.

\begin{figure*}
\centering
\includegraphics[width=0.7\textwidth]{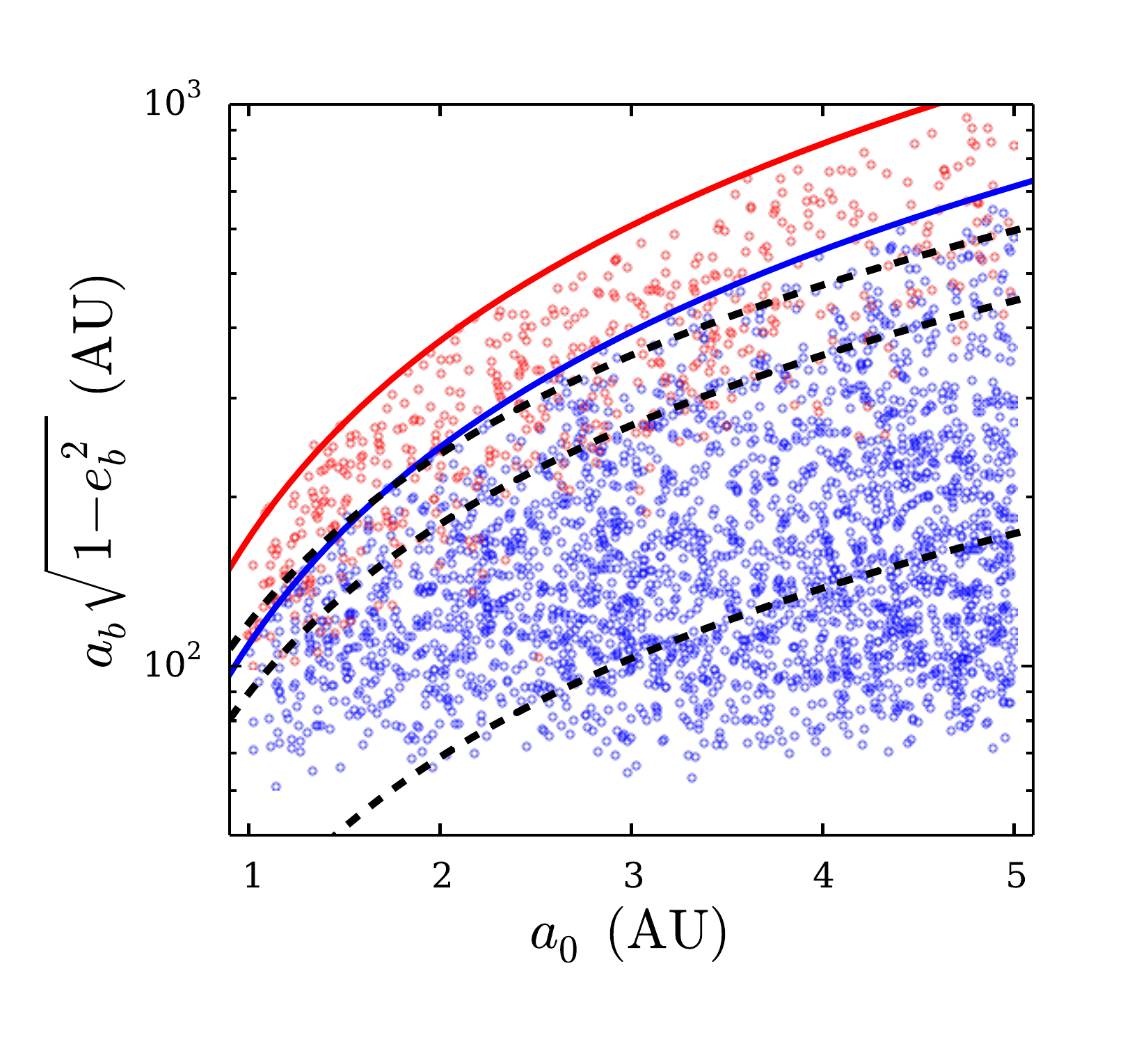}
\caption{Parameter space producing tidally disrupted planets (blue points) and HJs (red points) for the calculations presented in Fig.~\ref{fig:initial_conditions_Gstar} with $M_p = 1 M_J$.  The red solid curve shows the maximum value of $a_{\rm b, eff} = a_b \sqrt{1 - e_b^2}$ for migration to be possible, as a function of $a_0$ (Eq.~[\ref{eq:migration_condition}] with $a_{p,\rm crit} = 0.025$ AU), and the blue solid curve shows the maximum value of $a_{\rm b, eff}$ for tidal disruption to be possible (Eq.~[\ref{eq:disrupt}], with $f = 1$).  If a given combination of $(a_0, a_{\rm b, eff})$ is located below the red (blue) curve, migration (disruption) is possible, but not guaranteed.  See also Fig.~\ref{fig:migration_condition}. The dashed lines depict curves of constant $\varepsilon_{\rm oct} = 0.015$ in ($a_{\rm b, eff}, a_0$) space, with $e_b = 0.8, 0.6$ and $0.4$ (from top to bottom).  The region above the top black dashed curve cannot have $\varepsilon_{\rm oct} > 0.015$, unless $e_b > 0.8$.  Since the location of this black curve coincides with the tidal disruption limit (blue curve), there is very little parameter space with $\varepsilon_{\rm oct} > 0.015$ capable of inducing planet migration, without tidal disruption.}
\label{fig:mig_condition_data}
\end{figure*}

\begin{figure*}
\centering
\includegraphics[width=0.8\textwidth]{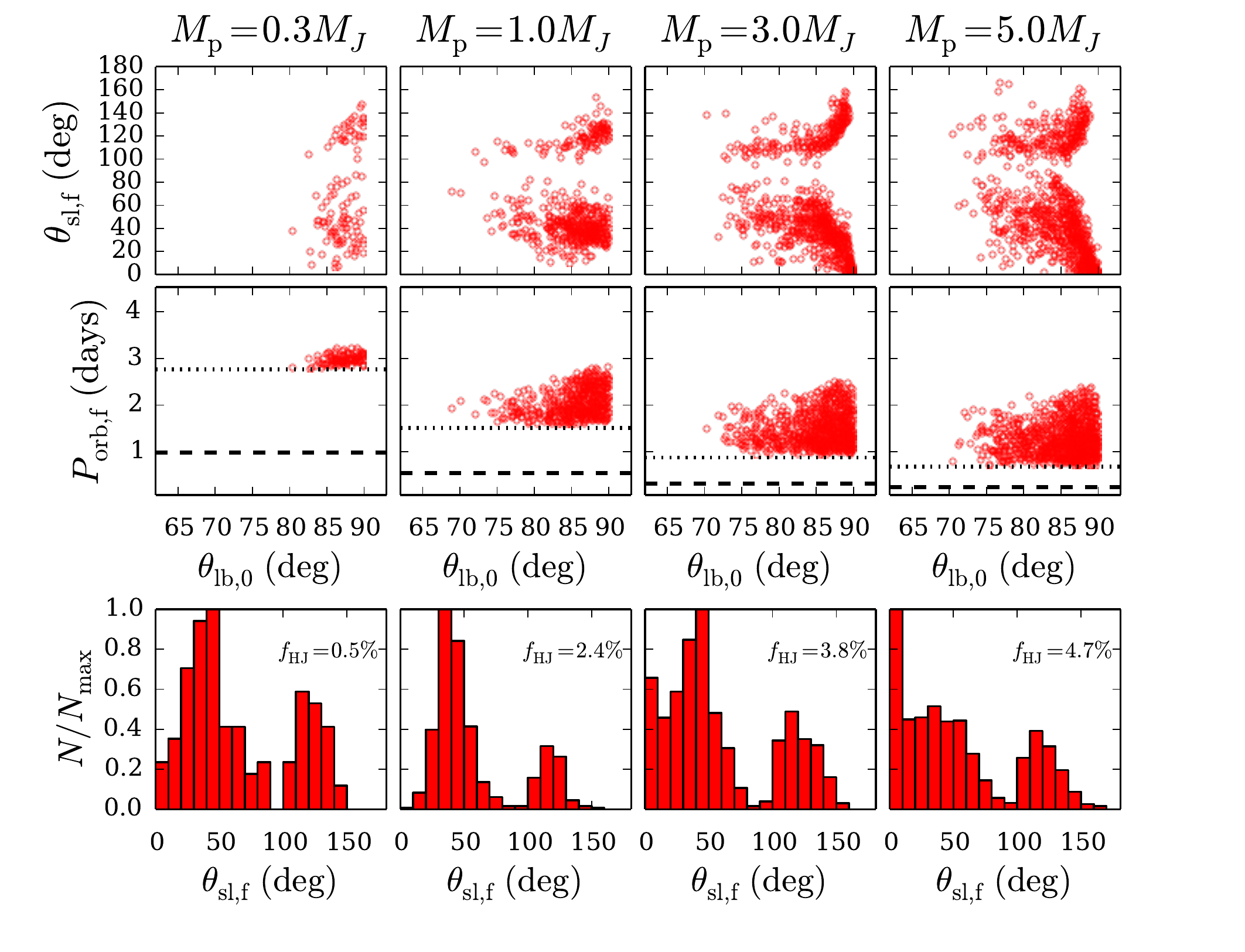}
\caption{Final stellar obliquities $\theta_{\SL,\rm f}$ and orbital periods $P_{\rm orb,f}$ for the systems shown in Figure \ref{fig:initial_conditions_Gstar} that resulted in HJs. Parameters are $M_{\star} = 1.0 M_{\odot}$ (the G-type star), and $a_0,a_b,e_b,\theta_{\LB,0}$ randomly sampled over wide ranges, as described in the text, and indicated in Table \ref{popsynthtable}. Top and middle panels depict the final spin-orbit angle $\theta_{\SL,\rm f}$ and orbital period $P_{\rm orb,f}$ versus $\theta_{\LB,0}$.  The dashed lines, included for reference, indicate the orbital period at the tidal disruption radius, and the dotted lines indicate the minimum achievable orbital period, defined by $a_{\rm f} \geq 2 R_{\rm tide}$.  Bottom panels show histograms of $\theta_{\SL,\rm f}$, with a bin width $\Delta \theta_{\SL,\rm f} = 10^{\circ}$.}
\label{fig:final_dist_Gstar}
\end{figure*}

\begin{figure*}
\centering
\includegraphics[width=0.8\textwidth]{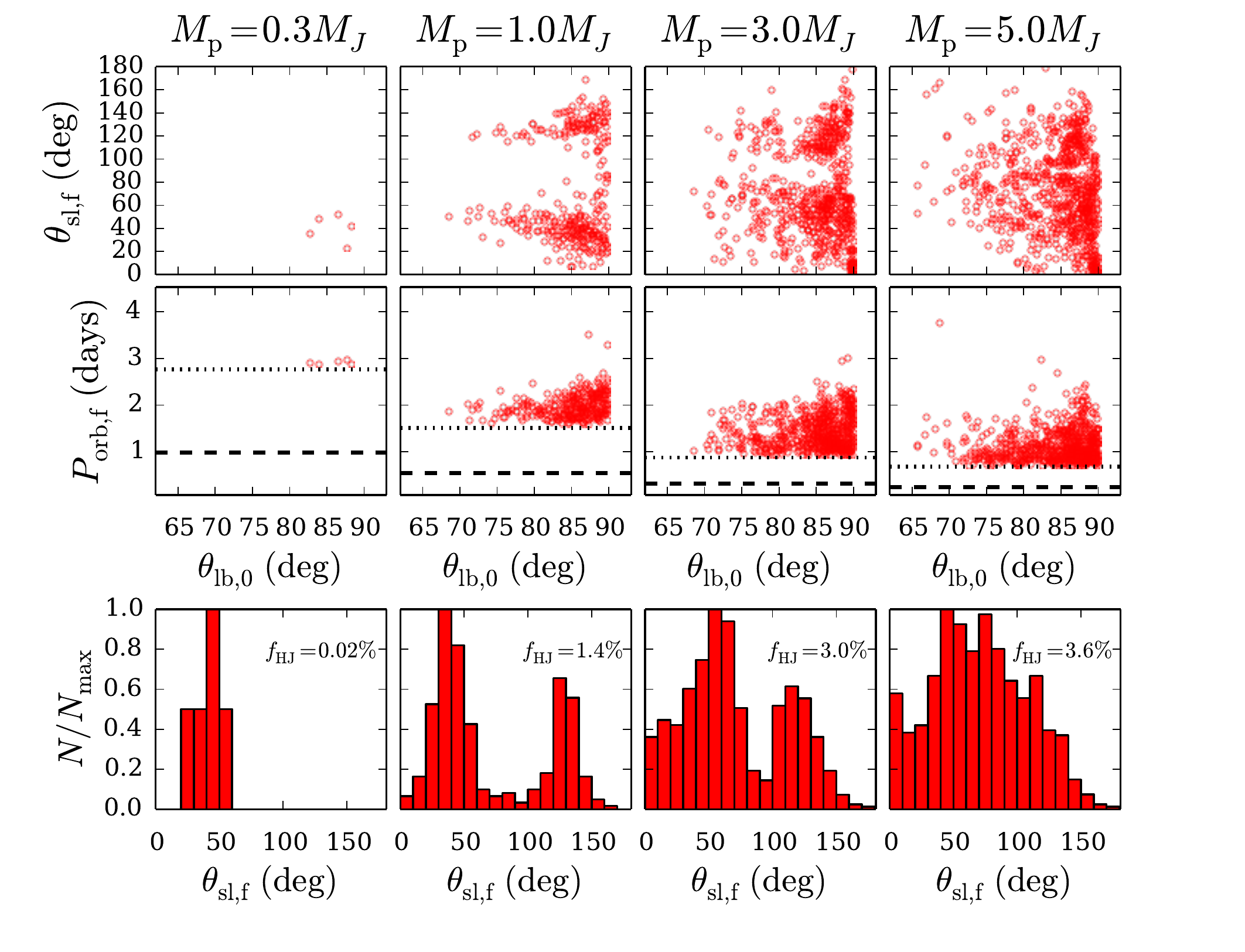}
\caption{Same as Fig.~\ref{fig:final_dist_Gstar}, but showing results for planets around F stars.}
\label{fig:final_dist_Fstar}
\end{figure*}

\begin{figure*}
\centering
\includegraphics[width=0.4\textwidth]{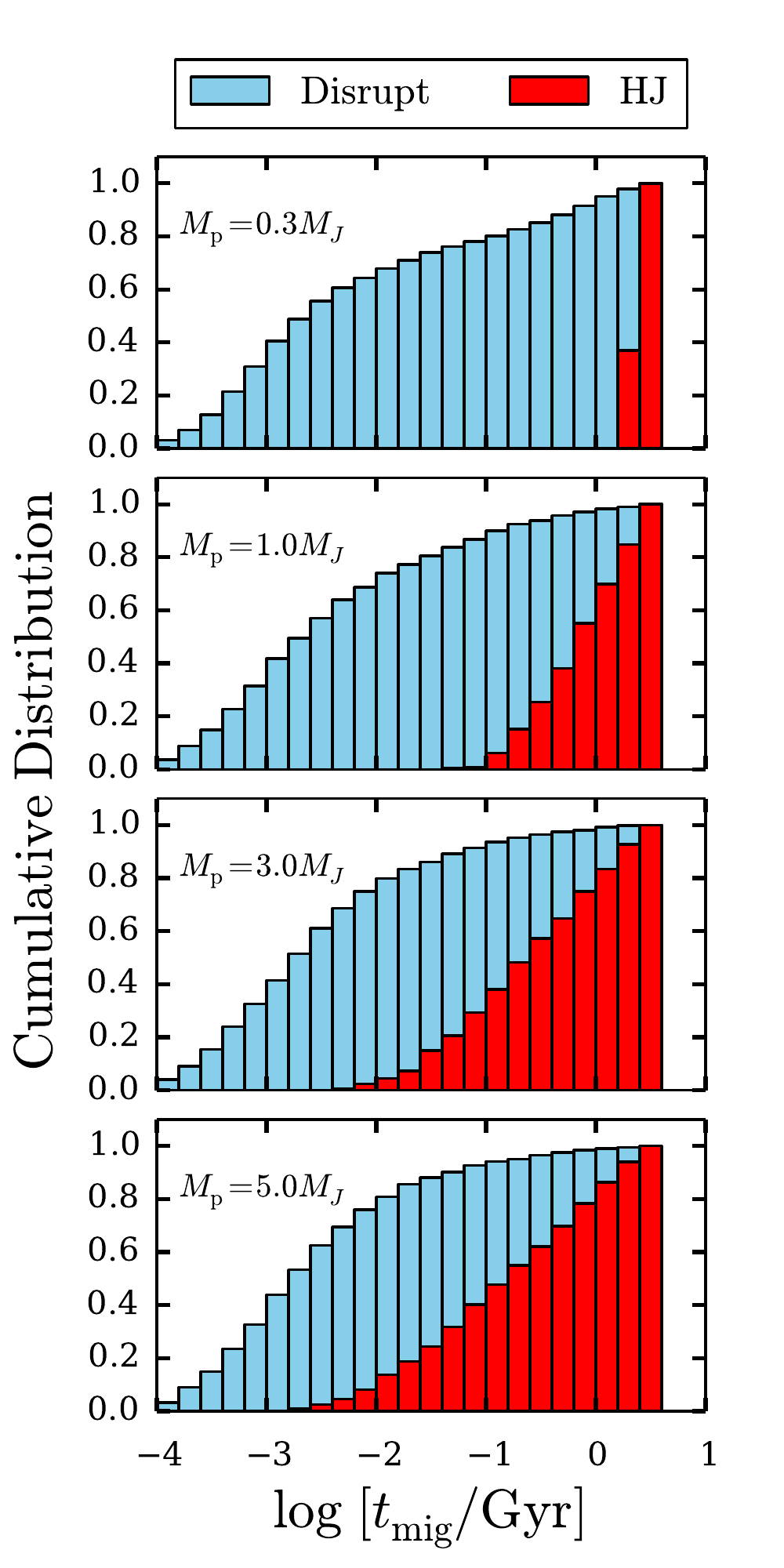}
\caption{Cumulative distributions of migration times $t_{\rm mig}$, defined as the time at which the planet crosses the tidal radius (for the disrupted planets), or the time at which the semi-major axis decreases below $0.1$ AU (for the HJs).  The results shown are the same set of simulations as depicted in Figs.~\ref{fig:initial_conditions_Gstar} and \ref{fig:final_dist_Gstar}.  Most tidal disruptions occur relatively early, with $\gtrsim 75\%$ occurring within 0.1 Gyr. The minimum time needed to produce an HJ depends on planet mass, and is $\sim 2$ Gyr for $0.3 M_J$ planets, but $\sim 2$ Myr for $5 M_J$ planets.}
\label{fig:circ_time}
\end{figure*}

\subsection{Dependence on Tidal Dissipation Strength}
\label{sec:popsynth_dissipation}
All results presented thus far adopt the tidal dissipation strength $\chi = 10$, corresponding to tidal lag time $\Delta t_L = 1$ second.  We now examine the effect of varying dissipation rate, by considering tidal enhancement factors $\chi = 1$ and $\chi = 100$, so that $\Delta t_L = 0.1$ and 10 seconds respectively.  All simulations presented in Section \ref{sec:final} were repeated with these values of $\chi$; see Table \ref{popsynthtable}.  

\begin{figure*}
\centering
\includegraphics[width=0.8\textwidth]{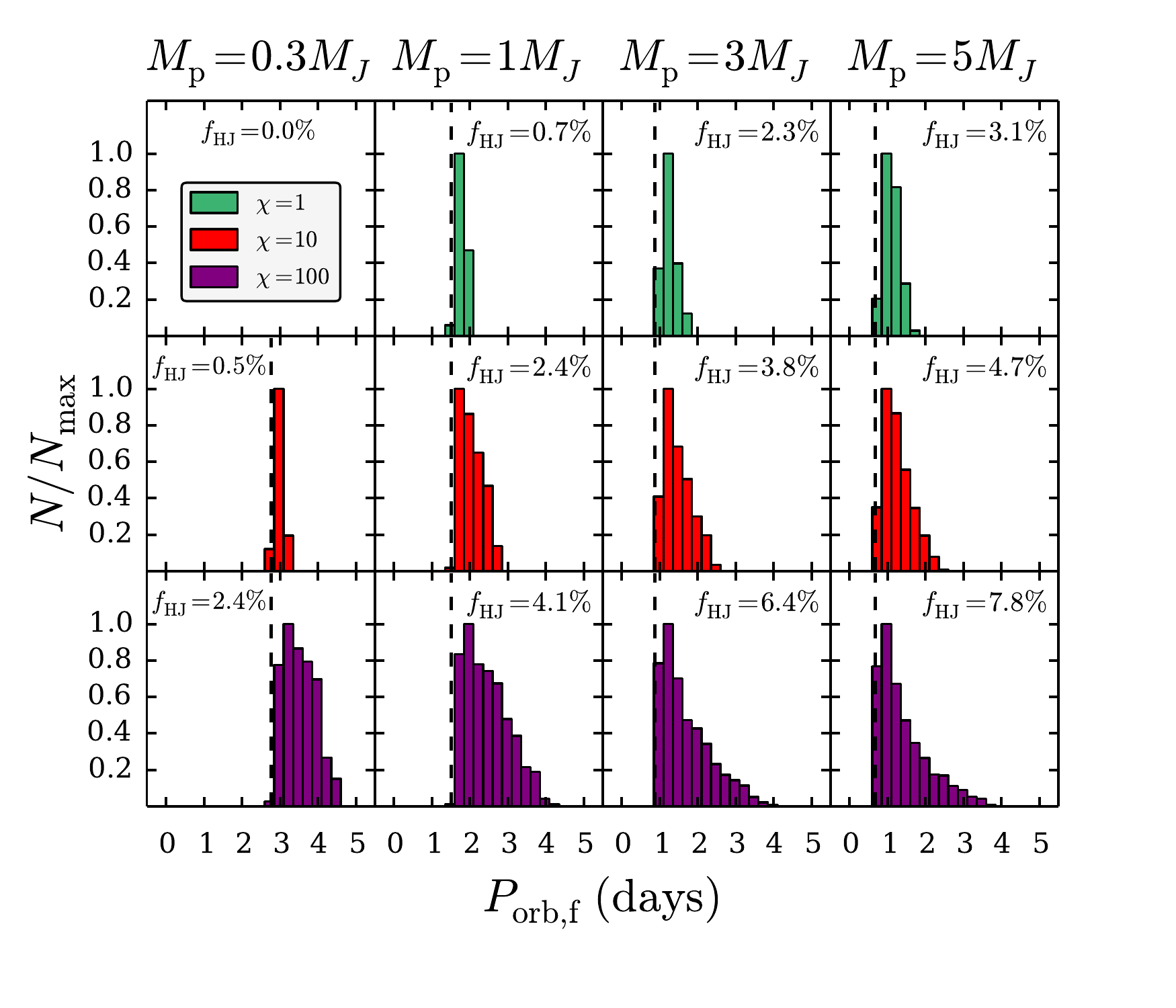}
\caption{Effects of varying tidal dissipation strength $\chi$ on the distribution of final HJ orbital periods $P_{\rm orb,f}$ for planets around G stars.  We show $\chi = 1$ (green, top row), $\chi = 100$ (purple, bottom row), along with our canonical value $\chi = 10$ (red, middle row).  The distributions shown are the result of $N_{\rm run} \sim 9000$ total trials, out of which a fraction $f_{\rm HJ}$ resulted in HJ formation (see also Table \ref{popsynthtable}). Each column shows a different planet mass, as labeled.  The vertical dashed lines, included for reference, indicate the minimum achievable orbital period, at $a_{\rm f} = 2 R_{\rm Tide}$.  For $M_p = 0.3, 1, 3, 5 M_J$ respectively, the number of data points $N_{\rm HJ}$ in each histogram are as follows: top row, $\chi = 1$, $N_{\rm HJ} = 0, 156, 490, 650$; middle row, $\chi = 10$, $N_{\rm HJ} = 108, 502, 811, 990$; bottom row, $\chi = 100$, $N_{\rm HJ} = 513, 875, 1370, 1670$.  Note that no close-in planets were produced for the combination $M_p = 0.3 M_J$, $\chi = 1$.}
\label{fig:Porb_chi_Gstar}
\end{figure*}

Figure \ref{fig:Porb_chi_Gstar} shows distributions of the HJ final orbital periods $P_{\rm orb,f}$ around the G star for each tidal dissipation strength (note that the corresponding results for the F star are nearly identical, and are not shown).  The distributions for $\chi = 1$ are narrow, and concentrated toward low orbital periods, with $P_{\rm orb,f} \lesssim 2$ days across all planet masses.  As $\chi$ increases, the distributions widen, since the enhanced tidal dissipation strength allows planets with larger pericenters to migrate inward within 5 Gyr (see Eq.~[\ref{eq:maxrate}]).  However, note that regardless of the tidal dissipation strength, no HJs with final orbital periods $P_{\rm orb,f} \gtrsim 4.6$ days were produced.  This lack of longer period HJs is consistent with previous calculations of HJ formation via the LK mechanism \citep{petrovich2015b}.

Not surprisingly, the HJ fraction $f_{\rm HJ}$ increases as $\chi$ increases.  However, the migration fraction $f_{\rm mig} = f_{\rm HJ} + f_{\rm dis}$ remains roughly constant, varying by only a few percent across all combinations of planet mass, stellar type, and dissipation strength, between $\sim 11 - 14 \%$.  This is consistent with the discussion in Section \ref{sec:migrationfrac} (see last two paragraphs of that subsection). Most of the migrating planets originate from systems where the octupole effect plays an important role, and the ``window of extreme eccentricity'' (needed for achieving migration) is independent of $M_p$, $M_\star$, and $\chi$.  On the other hand, most  HJs originate from systems with low $\varepsilon_{\rm oct}$ and high $\theta_{\LB,0}$ (see Figs.~\ref{fig:initial_conditions_Gstar} and \ref{fig:mig_condition_data}), where the octupole effect is not essential for migration.  For these systems, enhanced tidal dissipation allows planets with larger periastron distances to migrate (see Eq.~[\ref{eq:maxrate}]), leading to a larger $f_{\rm HJ}$.

\begin{figure*}
\centering
\includegraphics[width=0.8\textwidth]{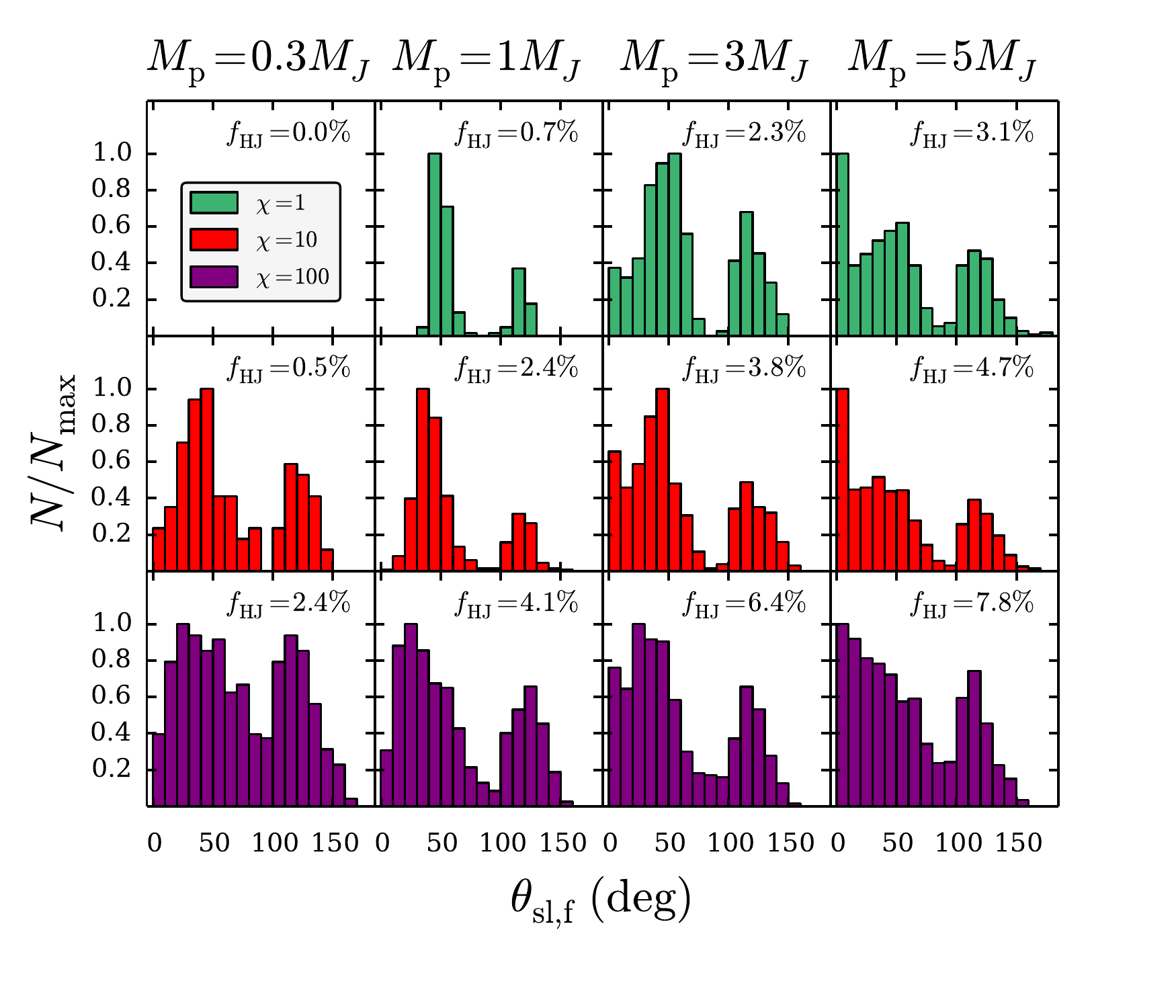}
\caption{Effects of varying tidal dissipation strength $\chi$ on the distributions of $\theta_{\SL,\rm f}$ for HJs around G stars (the same sample as in Fig.~\ref{fig:Porb_chi_Gstar}).  We show $\chi = 1$ (green, top row), $\chi = 100$ (purple, bottom row), along with our canonical value $\chi = 10$ shown previously in Fig.~\ref{fig:final_dist_Gstar} (red, middle row). For $M_p = 0.3, 1, 3, 5 M_J$ respectively, the number of data points $N_{\rm HJ}$ in each histogram are as follows: top row (from left to right), $\chi = 1$, $N_{\rm HJ} = 0, 156, 490, 650$; middle row, $\chi = 10$, $N_{\rm HJ} = 108, 502, 811, 990$; bottom row, $\chi = 100$, $N_{\rm HJ} = 513, 875, 1370, 1670$.  Note that no close-in planets were produced for $M_p = 0.3 M_J$, $\chi = 1$.  For most planet masses, increasing $\chi$ broadens the distribution of $\theta_{\SL,\rm f}$, but the overall shape (usually bimodal) remains unchanged. Increasing $\chi$ leads to more planets with low obliquities ($\theta_{\SL,\rm f} \lesssim 20^{\circ}$)}.
\label{fig:hist_chi_Gstar}
\end{figure*}

\begin{figure*}
\centering
\includegraphics[width=0.8\textwidth]{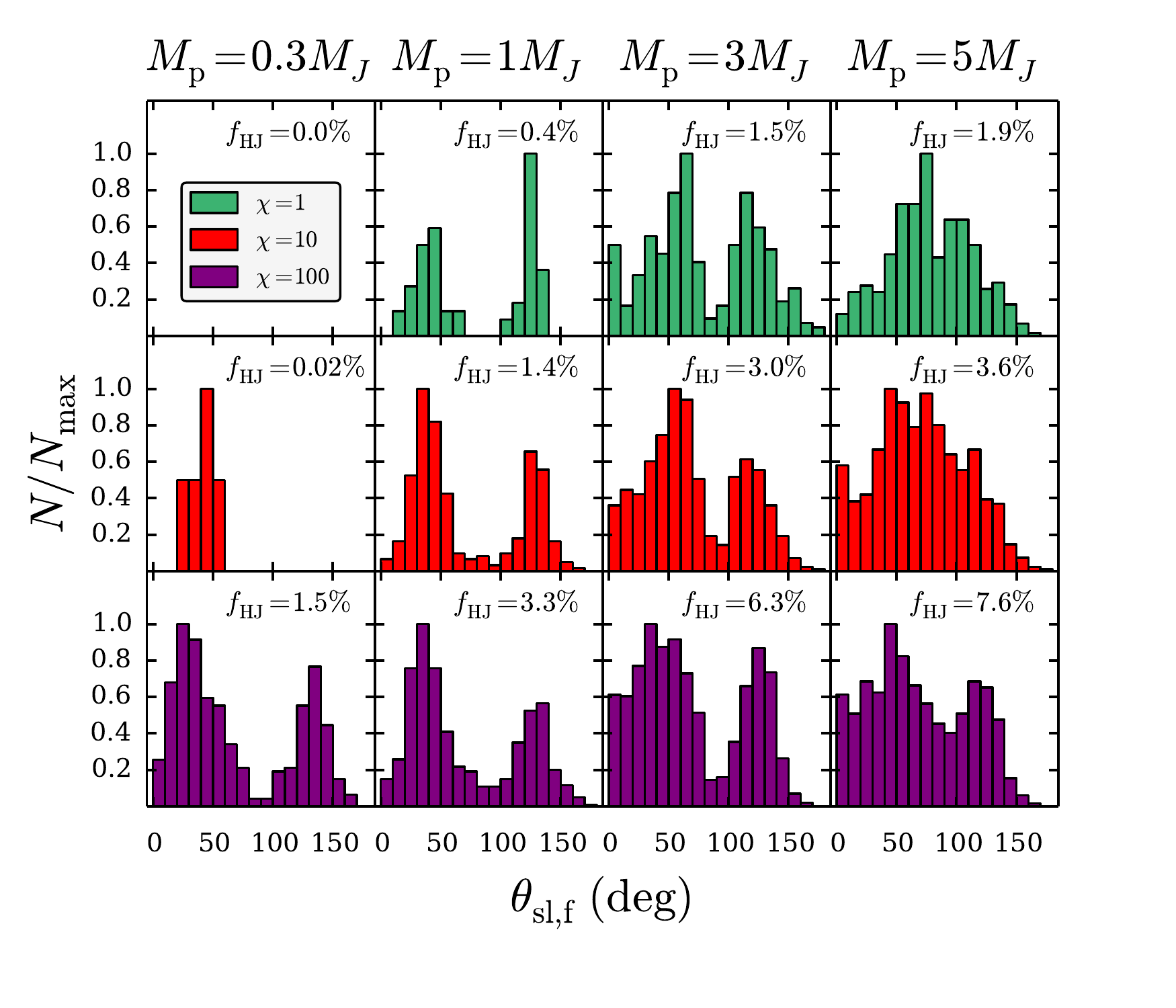}
\caption{Same as Figure \ref{fig:hist_chi_Gstar}, but showing results for planets around F stars.  The distributions shown are the result of $N_{\rm run} \sim 9000$ total trials, out of which a fraction $f_{\rm HJ}$ resulted in HJ formation (see also Table \ref{popsynthtable}).  For $M_p = 0.3, 1, 3, 5 M_J$ respectively, the number of data points $N_{\rm HJ}$ in each histogram are as follows: top row, $\chi = 1$, $N_{\rm HJ} = 0, 75, 310, 394$; middle row, $\chi = 10$, $N_{\rm HJ} = 5, 305, 640, 764$; bottom row, $\chi = 100$, $N_{\rm HJ} = 330, 711, 1339, 1609$.}
\label{fig:hist_chi_Fstar}
\end{figure*}

Figures \ref{fig:hist_chi_Gstar} and \ref{fig:hist_chi_Fstar} compare the effects of varying $\chi$ on the distribution of $\theta_{\rm sl,f}$ for planets around G and F stars.  Increasing $\chi$ generally leads to broader distributions, with a greater fraction of planets at relatively low obliquities ($\theta_{\rm sl,f} \lesssim 30^{\circ}$), but has little effect on the overall shape.  In particular, the bimodality observed previously for $(1 - 3)M_J$ planets is preserved.  

\subsection{Primordial Misalignment}
\label{sec:primordial}
Finally, we present HJ stellar obliquity distributions for systems in which the initial stellar spin-orbit angle is misaligned, i.e.\ $\theta_{\SL,0} \neq 0$.  Such initially misaligned configurations are relevant because various works \citep[e.g.][]{bate2010, lai2011, batygin2012, batygin2013, lai2014} have suggested the possibility of ``primordial misalignments'' in which the protoplanetary disk becomes tilted relative to the stellar spin axis.  We limit the discussion to planets around G stars, and the canonical tidal dissipation strength $\chi = 10$.  We fix $\theta_{\SL,0}$, and integrate a series of systems with the initial phase of $\hatS$ around $\hatLp$ (i.e. $\phi_{\SL,0}$, where $\phi_{\SL,0}$ is the azimuthal angular coordinate in the frame where $\hatLp$ is along the $z$-axis) randomly sampled uniformly in $[0, 2 \pi]$.

\begin{figure*}
\centering
\includegraphics[width=0.8\textwidth]{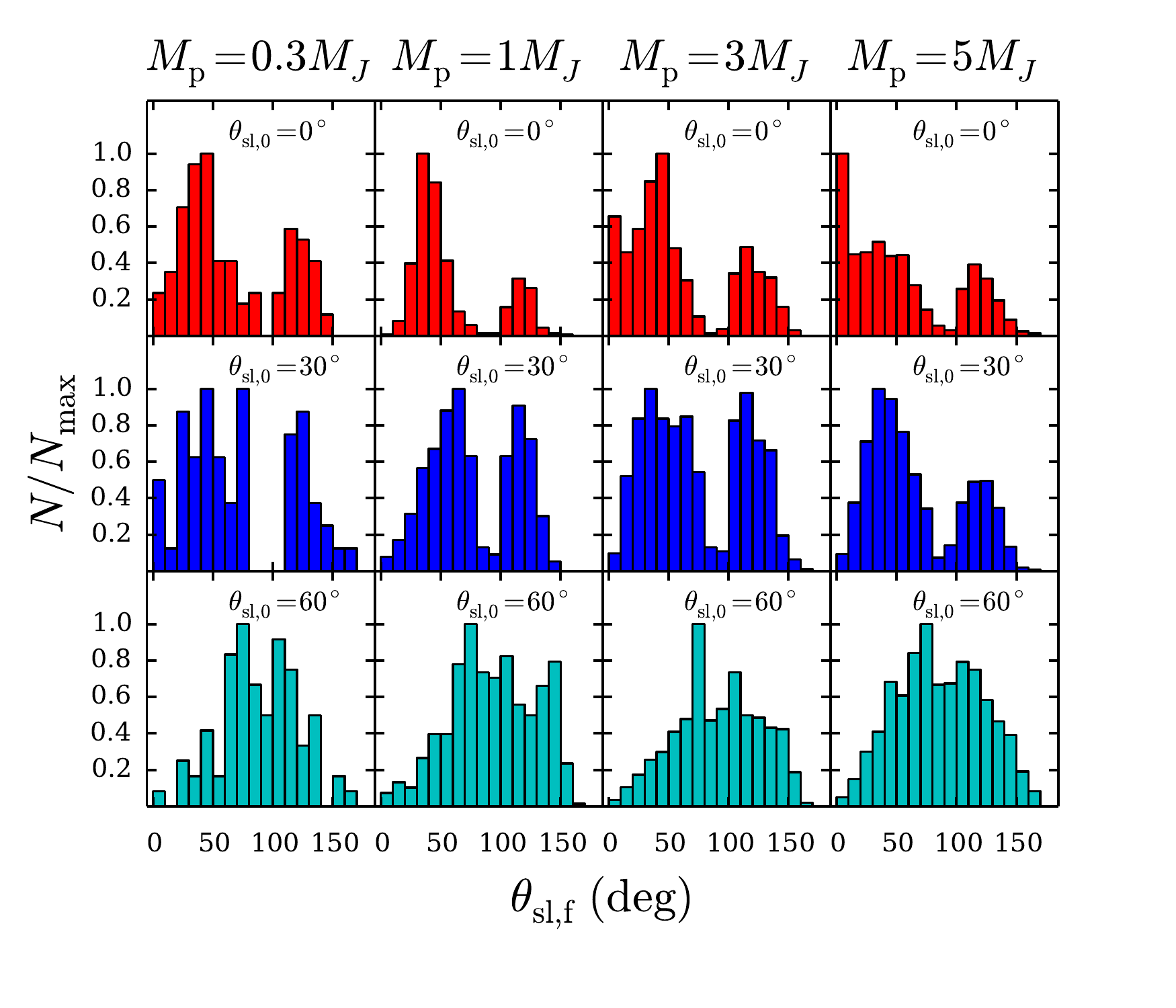}
\caption{The effect of primordial misalignment ($\theta_{\SL,0} \neq 0$) on distributions of $\theta_{\rm sl,f}$.  We show results for planets around G stars, with the canonical dissipation strength $\chi = 10$.  Top row (red): $\theta_{\rm sl,0} = 0^{\circ}$, as shown previously in Fig.~\ref{fig:final_dist_Gstar}.  Middle row (blue): $\theta_{\rm sl,0} = 30^{\circ}$.  Bottom row (cyan): $\theta_{\rm sl,0} = 60^{\circ}$.  For $M_p = 0.3, 1, 3, 5 M_J$ respectively, the number of data points $N_{\rm HJ}$ in each histogram are as follows: top row (from left to right), $\theta_{\SL,0} = 0^\circ$, $N_{\rm HJ} = 108, 502, 811, 990$.  Middle row, $\theta_{\SL,0} = 30^\circ$, $N_{\rm HJ} = 61, 544, 844, 1021$.  Bottom row, $\theta_{\SL,0} = 60^\circ$, $N_{\rm HJ} = 82, 556, 943, 1037$. See Table \ref{popsynthtable} for further information.}
\label{fig:primordial}
\end{figure*}

Figure \ref{fig:primordial} shows results for $\theta_{\SL,0} = 30^{\circ}$ and $60^{\circ}$, along with the canonical $\theta_{\SL,0} = 0^{\circ}$ case shown previously in Fig.~\ref{fig:final_dist_Gstar}.  When $\theta_{\SL,0} = 30^{\circ}$, the distributions of $\theta_{\SL,\rm f}$ are bimodal for all planet masses, including planets with $M_p = 5 M_J$.  For $\theta_{\SL,0} = 60^{\circ}$, the bimodality has vanished, and the distributions are roughly symmetric around $90^{\circ}$.  We conclude that non-zero initial obliquities can affect the final spin-orbit misalignment, such that the bimodal peaks present for $\theta_{\SL,0} = 0^{\circ}$ tend to merge as $\theta_{\SL,0}$ increases.

\section{Conclusion}

\subsection{Summary of Results}

The main goal of this paper is to conduct a thorough population
synthesis of the production of misaligned close-in giant planets (Hot Jupiters, HJs)
in stellar binaries by the mechanism of Lidov-Kozai (LK) oscillations
with tidal dissipation, examining the previously unexplored dependence
on planet mass, and stellar type and spin properties.  The complex
evolution of the stellar spin axis in systems with planets undergoing
LK oscillations poses a rich dynamical problem \citep[see
also][]{storch2014,storch2015}, and can affect the final
distributions of spin-orbit misalignments.  We have calculated the HJ production fractions and planet tidal disruption fractions
for a wide variety of systems, exploring their dependence on planet mass,
stellar properties and tidal dissipation rate. We have also presented a number of semi-analytical calculations, which
are useful in understanding the results of our population synthesis.
Our main results can be summarized as follows.

\begin{itemize}

\item Planet mass is important in determining the HJ formation and
  tidal disruption fractions (see Table \ref{popsynthtable}).  The
  fraction of systems resulting in HJs ($f_{\rm HJ}$) increases with
  planet mass, due to fewer tidal disruptions.  For Jupiter-mass
  planets, we find that $f_{\rm HJ} \approx 0.5 \% - 4 \%$ depending
  on the assumed tidal dissipation rate and host star mass.
  In general $f_{\rm HJ}$ increases with the tidal dissipation rate
  and decreases with stellar mass.  For more massive ($5 M_J$)
  planets, we find a higher fraction, with $f_{\rm HJ} \approx 3 \% -
  7.5 \%$.  The fraction of systems resulting in ``hot Saturns'' ($M_p
  \sim 0.3 M_J$) are low, especially around massive ($M_{\star} = 1.4
  M_{\odot}$, spectral type F) stars.  As a result, hot Saturns around
  massive stars are unlikely to be produced by LK migration in
  binaries, unless the tidal dissipation strength in the planet is
  high (with $\chi \gtrsim 100$, corresponding to $\Delta t_L \gtrsim
  10$ sec).

\item{We find that the ``migration fraction,'' defined as the sum of
  the HJ and disruption fractions, $f_{\rm mig} = f_{\rm HJ} + f_{\rm dis}$, 
  has a rather weak dependence on planet mass, stellar type and tidal
  dissipation rate, and is always in the range of $11-14\%$ (see Table
  3).  This behavior can be qualitatively understood from 
  analytical migration criteria (see Section 3.4 and Section 5.4.1,
  particularly Eq.~(\ref{eq:migration_condition}). Since the tidal disruption fraction for
  lower mass planets is higher (due to the increased tidal radius), a
  constant migration fraction implies that $f_{\rm HJ}$ should
  decrease with planet mass, as described above.}

\item HJs are produced only in systems when the ratio of the binary
  semi-major axis $a_b$ and the initial planet semi-major axis $a_0$
  lies in the range $60\lesssim a_b/a_0\lesssim 300$ (see
  Figs~18-19). In addition, no HJs are produced for systems with the
  dimensionless octupole parameter (see Eq.~[\ref{eq:epsilonOct}]) $\varepsilon_{\rm
    oct} \gtrsim 0.01-0.02$, where the range depends on the planet
  mass (see Figs.~\ref{fig:initial_conditions_Gstar}-\ref{fig:mig_condition_data}).  These place constraints on the types of
  binary properties and initial planet semi-major axes that are able
  to induce migration without causing tidal disruption.

\item The distribution of final spin-orbit misalignment angles depends
  on planet mass and the spin history of host stars (see
  Figs.~24-25). For $M_p = (1-3) M_J$, the distributions are always
  bimodal, with peaks near $\theta_{\SL,\rm f} \approx 40^{\circ}$ and
  $130^{\circ}$.  This bimodality is independent of stellar type.  For solar-type stars, higher-mass
  planets ($M_p = 5 M_J$) exhibit a preference for low final
  obliquities, with $\theta_{\SL,\rm f} < 10^{\circ}$ (see
  Fig.~\ref{fig:final_dist_Gstar} and Fig.~\ref{fig:hist_chi_Gstar}), although misalignment
  still remains possible.  By contrast, for F-stars, the
  $\theta_{\SL,\rm f}$ distributions for massive planets are broad,
  with no clear bimodality (see Fig.~\ref{fig:hist_chi_Fstar}).  We attribute the higher
  degree of misalignment around F stars to the stronger torque from
  the (more rapidly rotating) host star acting on the orbit, thereby
  erasing the tendency towards alignment observed for $5 M_J$ planets
  around G stars.  In general, the backreaction torques from the
  stellar quadrupole on the planet's orbit, as well as the octupole
  effect from the binary companion, give rise to a variety of
  evolutionary paths toward spin-orbit misalignments during LK
  migration (Section 4), and result in a complicated dependence of the
  $\theta_{\SL,\rm f}$-distribution on planet mass and stellar type.

\item The final stellar obliquity distribution does not depend
  significantly on tidal dissipation rate within the planet, although
  higher rates of dissipation do tend to broaden the
  distributions.

\item 
  While most of the calculations in this paper assume initial
  alignment between the stellar spin and planet's orbit axis
  ($\theta_{\SL,0}=0^\circ$), we also explore the effect of an initial
  (``primordial'') misalignment. We find that the bimodality present when $\theta_{\SL,0} = 0^\circ$ begins to merge as $\theta_{\SL,0}$ increases (see Fig.~\ref{fig:primordial}). For modest initial misalignments ($\theta_{\SL,0} = 30^\circ$), the final
  $\theta_{\SL,\rm f}$ distribution remains bimodal across all planet masses, with the peaks slightly shifted towards $90^\circ$.  For higher initial misalignment ($\theta_{\SL,\rm 0} = 60^\circ$) the bimodality has nearly vanished, and the distribution is broadly distributed and centered near $\theta_{\SL,\rm f} \sim 70^\circ - 80^\circ$.

\end{itemize}

\subsection{Discussion}

Previous studies of HJ production in stellar binaries that include the
octupole potential \citep{naoz2012,petrovich2015b} focused on a single planet mass and
initial planet semi-major axis ($M_p = 1 M_J$, $a_0 = 5$ AU), and a
single host star type ($M_\star = 1 M_{\odot}$, with constant spin
rate).  This paper has expanded upon these previous works by exploring
a range of giant planet masses and orbital separations ($M_p = 0.3 - 5
M_J$, $a_0 = 1 - 5$ AU) and two host stellar types ($M_\star = 1, 1.4
M_{\odot}$), with each stellar type governed by differing magnetic
braking laws.  We also consider systems with ``primordial
misalignment'' where the initial stellar obliquity $\theta_{\SL,0}
\neq 0$.

In terms of HJ production fractions ($f_{\rm HJ}$), our results are in
good agreement with \cite{petrovich2015b}.  We find $f_{\rm HJ} \sim$
a few percent typically, except for sub-Jupiter mass planets which can
have much lower fractions ($f_{\rm HJ} \lesssim 1 \%$).  In terms of
tidal disruptions, \cite{petrovich2015b} finds a much higher
disruption fraction, with $f_{\rm dis} \sim 25 \%$, in part because he places
all planets initially at $a_0 = 5$ AU from the host star, whereas we
vary the initial semi-major axis uniformly in the range $a_0 = 1 - 5$
AU.  Planets that begin at larger orbital separations experience 
stronger forcing from the binary and less pericenter precession due to
SRFs, and thus can achieve sufficiently high eccentricities such that
the pericenter distance $a_p = a(1 - e_{\rm max})$ is smaller,
resulting in more disruptions (see Fig.~6).  Another reason for the
higher disruption fractions quoted in \cite{petrovich2015b} lies in
the choice of binary eccentricity range (he chooses a maximum $e_b =
0.9 - 0.95$, in contrast with $0.8$ assumed in this work).  As noted
before (see the beginning of Section 5.4), the actual eccentricity
distribution of stellar binaries (especially those that allow planet
formation) is very uncertain.  Also, including binaries with $e_b
\gtrsim 0.9$ may result in over-populating systems close to the
stability limit (with small $a_b (1 - e_b)/a_0$).  Our HJ fractions (for
$M_p=1\,M_J$ around solar-type stars) are lower than those found in
\cite{naoz2012}, who give $f_{\rm HJ} \sim 15 \%$.
One major reason for the difference is that \cite{naoz2012} use the
tidal radius Eq.~(\ref{Rtide}), but set $f \simeq 0.6$, whereas we use $f = 1$. 
Note that since the migration fraction $f_{\rm mig}=f_{\rm HJ}+f_{\rm dis}$
is always in range of 11-14$\%$ regardless of planet mass and 
stellar type (see Section 5.4.1 and Table 3), 
in the extremely unlikely event that all of our tidally disrupted planets
actually survived as HJs, the maximum possible HJ production fraction 
from our simulations is $f_{\rm HJ,max} = f_{\rm mig} \sim 13 \%$.

Observations constrain the HJ occurrence rate around solar-type stars
to be $\sim 1 \%$ \citep[e.g.][]{wright2012}.  Since the observed
stellar companion fraction in HJ systems is $\lesssim 50 \%$
\citep{ngo2015}, our calculations imply that LK migration from stellar
companions can probably explain around $\sim 15 \%$ of observed HJs
(using $f_{\rm HJ} = 3 \%$, and assuming a giant planet
occurrence rate of $10 \%$).

The calculations presented in this paper never produce HJs with final
orbital periods $P_{\rm orb,f} \gtrsim 4.5$ days, with typical periods
in the range of $1-3$~days, depending on planet mass and tidal
dissipation strength (see Fig.~23).  More massive planets tend to have
shorter periods (sometimes $\lesssim 1$~day) because they can survive tidal
disruption during the high-eccentricity periastron passage. Thus, it is
clear that LK migration in stellar binaries cannot explain the
observed population of HJs with periods greater than 4 days (see also
\citealt{petrovich2015b} for an in-depth discussion of the tendency
for LK migration to produce an excess of ``Very Hot Jupiters''
compared to observations.)  In addition, for both types of stars, our
calculations yield very few planets in the process of migration.  In
particular, very few ``warm Jupiters'' are produced with $0.1 \lesssim
a \lesssim 0.5$ AU after evolving the system for 5 Gyr (see also \citealt{petrovich2015b}).

In the absence of primordial misalignment (so that $\theta_{\SL,0} = 0^{\circ}$), our calculations always
predict, for planet masses $M_p = 1 - 3 M_J$, a bimodal distribution
of final stellar spin-orbit misalignments, with peaks at
$\theta_{\SL,\rm f} \approx 40^{\circ}$ and $130^{\circ}$, and a dearth around $90^\circ$.  This
result is independent of host stellar type and tidal dissipation
strength (see Figs.~24-25).  Such bimodality results from the stellar
spin evolution transitioning from the non-adiabatic to fully adiabatic
regime \citep[submitted]{SLA15}, and thus may be interpreted as a clear signature
of HJ formation from LK oscillations with tidal dissipation.  However, for $M_p=5 M_J$ planets, the shape of the distribution of $\theta_{\SL,\rm f}$ differs substantially, and for planets around F stars, nearly polar orbits
($\theta_{\SL,\rm f} \sim 90^{\circ}$) are commonly produced (see
Fig.~\ref{fig:hist_chi_Fstar}, right panels).  

On the other hand, when
significant primordial misalignments are present, with $\theta_{\SL,0} \gtrsim 60^{\circ}$ (see Section 5.6),
the bimodality of the final misalignment distribution disappears, and
planets on polar orbits are easily produced (see
Fig.~\ref{fig:primordial}, bottom row).
Observationally, the distribution of HJ spin-orbit misalignments does not
exhibit a clear bimodal structure \citep[e.g.][]{albrecht2012a} and a handful of observed systems have nearly polar orbits, such as WASP-1b \citep{simpson2011}, WASP-7b \citep{albrecht2012b}, and WASP-79b \citep{addison2013} (these
systems mostly have $M_p\sim 1M_J$ and host star mass $M_{\star}
\approx 1.2 - 1.5 M_{\odot}$). Thus, without substantial primordial misalignments, LK migration in
stellar binaries cannot explain the observed $\theta_{\SL,\rm f}$
distribution of HJs. This again suggests that the majority ($\sim 85 \%$)
of HJs are probably formed by other mechanisms (e.g., disk-driven migration).

One physical effect not included in this paper is tidal dissipation in
the host stars.  This can in principle affect the semi-major axis of
very close-in giant planets, and change the spin-orbit misalignment
angle, as studied in numerous papers
\citep[e.g.,][]{barker2009,jackson2009,winn2010,matsumura2010,lai2012,rogers2013,xue2014,valsecchi2014}. We
neglect stellar tidal dissipation on purpose in this paper
because, compared to tidal dissipation in planets, stellar tides play
a negligible role in circularizing high-eccentricity planets
undergoing LK oscillations. 
Moreover, the stellar tidal dissipation rate is highly uncertain, and
likely depends on the stellar type and planet mass (see Ogilvie 2014
for a review); it is also possible that the tidal process and timescale
for spin-orbit alignment are different from those for orbital decay
(Lai 2012). Once an HJ has formed through high-eccentricity migration, it is
straightforward to examine the effect of stellar tides (using parameterized
tidal models) on the subsequent evolution of the system.

\section*{Acknowledgments}

We thank Diego Mu\~{n}oz and Smadar Naoz for
useful discussions, and especially Cristobal Petrovich for useful
discussion and comments on the manuscript.  This work has been supported in part by NSF grant
AST-1211061, and NASA grants NNX14AG94G and NNX14AP31G.  K.R.A. is
supported by the NSF Graduate Research Fellowship Program under Grant
No. DGE-1144153.

\appendix
\section{Equations}
\label{sec:eqns}
In this Appendix we present the secular equations of motion governing the planetary orbit and stellar spin axis.  The reader is referred to Table \ref{table:variables} for a concise summary of the notation used in this paper.
\subsection{Lidov-Kozai Oscillations}
The hierarchical triple systems studied in this paper consist of an inner binary $M_\star$ (host star) and $M_p$ (planet), with total mass $M_{\rm tot} = M_\star + M_p$, with an outer stellar mass binary companion $M_b$.  The planet has semi-major axis $a$ and eccentricity $e$, and the binary companion has semi-major axis $a_b$ and eccentricity $e_b$.  The inner binary is characterized by the unit vectors $\hatLp$ and $\hatep$, where $\hatLp$ is in the direction of the orbital angular momentum vector $\Lp$, and $\hatep$ is in the direction of the eccentricity vector $\ep$.   Similarly, the outer binary is characterized by the unit vectors $\hatLb$ and $\hat{{\bf e}}_b$. Since we are considering systems in the regime $M_p \ll M_b$, the effect of the planet on the outer binary is negligible, and $\hatLb$ and $\hat{{\bf e}}_b$ are held constant.  The inclination of the planetary orbit relative to the outer binary is specified by $\cos \theta_{\LB} = \hatLp \cdot \hatLb$. If the outer binary companion has $\theta_{\LB} \gtrsim 40^\circ$, the planet undergoes periodic variations in its orbital eccentricity and inclination \citep{lidov1962, kozai1962}, denoted in this paper as Lidov-Kozai (LK) oscillations.  The secular equations of motion for $\Lp$ and $\ep$ are, to octupole order in the disturbing potential of the binary (Liu et al.~2015, see also Petrovich 2015b),
\begin{equation} \label{eq:Lvec}
\begin{split}
\left . \frac{d{\Lp}}{dt} \right |_{{\rm LK}} &= \left . \frac{d{\Lp}}{dt} \right |_{{\rm LK,\, quad}} + \left . \frac{d{\Lp}}{dt} \right |_{{\rm LK,\, oct}}\\
 &=\frac{3}{4}\frac{L}{t_k (1 - e^2)^{1/2}}\Big[(\jvec \cdot \hatLb)~\jvec \times\hatLb
-5(\ep \cdot\hatLb)~\ep \times\hatLb \Big]\\
&-\frac{75}{64}\frac{\varepsilon_{\oct} L}{t_k (1 - e^2)^{1/2}}\Bigg\{
\bigg[2\Big[(\ep \cdot\hatLb)(\jvec \cdot\hatLb)\\
&+(\ep \cdot\hatLb)(\jvec \cdot\hateb)\Big]~\jvec +2\Big[(\jvec \cdot\hateb)(\jvec \cdot\hatLb)\\
&-7(\ep \cdot\hatLb)(\ep \cdot\hatLb)\Big]~\ep \bigg]\times\hatLb \\
&+\bigg[2(\ep \cdot\hatLb)(\jvec \cdot\hatLb)~\jvec 
+\Big[\frac{8}{5}e^2-\frac{1}{5}\\
&-7(\ep \cdot\hatLb)^2+(\jvec \cdot\hatLb)^2\Big]~\ep \bigg]
\times\hateb \Bigg\},
\end{split}
\end{equation}
and
\begin{equation} \label{eq:evec}
\begin{split}
\left . \frac{d{\ep}}{dt} \right |_{{\rm LK}} &= \left . \frac{d{\ep}}{dt} \right |_{{\rm LK, \, quad}} + \left . \frac{d{\ep}}{dt} \right |_{{\rm LK, \, oct}} \\
&=\frac{3}{4~t_k}\Big[(\jvec\cdot\hatLb)~\ep\times\hatLb
+2~\jvec\times\ep -5(\ep\cdot\hatLb)\jvec \times\hatLb\Big]\\
&-\frac{75\varepsilon_{\oct}}{64~t_k}\Bigg\{
\bigg[2(\ep\cdot\hatLb)(\jvec\cdot\hatLb)~\ep\\
&+\Big[\frac{8}{5}e^2-\frac{1}{5}-7(\ep\cdot\hatLb)^2+(\jvec \cdot\hatLb)^2\Big]~\jvec\bigg]\times\hateb\\
&+\bigg[2\Big[(\ep\cdot\hateb)(\jvec\cdot\hatLb)+(\ep \cdot\hatLb)(\jvec\cdot\hateb)\Big]~\ep \\
&+2\Big[(\jvec\cdot\hatLb)(\jvec\cdot\hateb)-7(\ep \cdot\hatLb)(\ep\cdot\hateb)\Big]~\jvec
\bigg]\times\hatLb\\
&+\frac{16}{5}(\ep \cdot\hateb)~\jvec\times\ep \Bigg\}~~,
\end{split}
\end{equation}
where we have defined $\jvec = \sqrt{1 - e^2} \hatLp$.  The terms in braces describe the octupole-level perturbation of the binary companion, where the relative ``strength'' of the octupole term is quantified through the parameter $\varepsilon_{\oct}$, defined by Eq.~(\ref{eq:epsilonOct}).
Note that in Eqs.~(\ref{eq:Lvec}) and (\ref{eq:evec}) we have introduced a characteristic (quadrupole) timescale for LK oscillations $t_k$, given by Eq.~(\ref{tk}).
Focusing only on the quadrupole terms, we note that the binary companion induces simultaneous precession and nutation of the orbital axis $\hatLp$ at a rate $\Omega_L \equiv |d \hatLp/dt_{\rm quad}| = [(\Omega_{\rm pl} \sin \theta_{\LB})^2 + \dot{\theta}_{\LB}^2]^{1/2}$, see Eq.~(\ref{OmegaPL}).
From the standard equations for LK oscillations (in terms of orbital elements) to quadrupole order \citep[e.g.][]{innanen1997},
\ba
\Omega_{\rm pl} \sin \theta_{\LB} & = & \frac{3}{8 t_k} \sin 2 \theta_{\LB} \frac{(5 e^2 \cos^2 \omega - 4 e^2 - 1)}{\sqrt{1 - e^2}} \nonumber\\
\dot{\theta}_{\LB} & = & -\frac{15}{16 t_k} e^2 \frac{\sin 2 \theta_{\LB} \sin 2 \omega }{\sqrt{1 - e^2}}.
\ea
The value of $\Omega_L$ therefore depends on the argument of pericenter $\omega$.  A good approximation to $\Omega_L$ is
\be
\Omega_L \simeq \frac{3 (1 + 4 e^2)}{8 t_k \sqrt{1 - e^2}} |\sin 2 \theta_{\LB}|.
\ee
This expression is exact at both $e = 0$ and $e = \emax$ (when $\omega = \pi/2$).

\subsection{Spin Evolution Due to the Stellar Quadrupole}
We denote the spin angular momentum of the host star as $\Ss = I_\star \Omega_\star \hatS$, where $I_\star = k_\star M_\star R_\star^2$ is the moment of inertia, $\Omega_\star$ is the spin frequency, and $\hatS$ is a unit vector along the spin axis.  Note that we have introduced a coefficient $k_\star$, describing the interior mass distribution, where $k_\star = 0.1$ is used throughout this paper.

Due to the rotational distortion of the star, the stellar spin axis $\Ss$ precesses around the orbital axis $\hatLp$ according to
\be
\left . \frac{d \Ss}{d t} \right |_{{\rm SL}} = \Omega_{\PS} \hatLp \times \Ss,
\ee
with the spin precession frequency $\Omega_{\PS}$ (see Section \ref{sec:spin}) given by Eq.~(\ref{OmegaPS}).

The effects on the planetary orbit due to the stellar
quadrupole are
\be
\left . \frac{d \Lp}{d t} \right |_{{\rm SL}} =  - \left . \frac{d \Ss}{d t} \right |_{{\rm SL}} = \Omega_{\PS}
\Ss \times \hatLp,
\ee
and 
\be
\left . \frac{d \ep}{d t} \right |_{{\rm SL}} = - \dot{\omega}_\star
 \left[ \cos
\theta_{\SL} \hatS \times \ep + \frac{1}{2}(1 - 5 \cos^2\theta_{\SL}) \hatLp
\times \ep \right],
\label{eq:evecfeedback}
\ee
where $\dot{\omega}_\star$ quantifies the rate of apsidal precession due to the oblate star, and is given by
\be
\dot{\omega}_\star = - \frac{S_\star}{L} \frac{\Omega_{\PS}}{\cos \theta_{\SL}} = \frac{3}{2} k_{q \star} \left(\frac{R_\star}{a}
\right)^2 \frac{\hat{\Omega}_\star^2}{(1 - e^2)^2} \, n.
\ee

\subsection{Pericenter Precession Due to Short Range Forces}
Besides the pericenter precession induced by the oblate host star, given in Eq.~(\ref{eq:evecfeedback}), additional short range forces (SRFs), due to general relativistic corrections, the (static) tidal bulge in the planet, and rotational distortion of the planet, induce precession of the eccentricity vector, given by \citep[e.g.][]{correia2011,liu2015}
\be
\begin{split}
  \left . \frac{d \ep}{d t} \right |_{{\rm SRF}}  & = \left . \frac{d \ep}{d t} \right |_{{\rm GR}} + \left . \frac{d \ep}{d t} \right |_{{\rm Tide}} +  \left . \frac{d \ep}{d t} \right |_{{\rm rot}}  \\
& = (\dot{\omega}_{\rm GR} + \dot{\omega}_{\rm Tide} +  \dot{\omega}_{\rm rot}) \hatLp \times \ep,
\end{split}
\ee
where the precession frequencies take the form
\be
\dot{\omega}_{\rm GR} = \frac{3 G M_{\rm tot}}{c^2 a (1 - e^2)} n,
\ee
\be
\dot{\omega}_{\rm Tide} = \frac{15}{2} k_{2p} \frac{M_\star}{M_p} \left(
  \frac{R_p}{a} \right)^5 \frac{f_4(e)}{j^{10}} n,
\label{eq:omega_tide_p}
\ee
and
\be
\dot{\omega}_{\rm rot} = \frac{3}{2} k_{qp} \left(\frac{R_p}{a}
\right)^2 \frac{\hat{\Omega}_p^2}{(1 - e^2)^2} n,
\label{eq:omega_rot_p}
\ee
where $f_4(e)$ in Eq.~(\ref{eq:omega_tide_p}) is a dimensionless function of eccentricity, given in Eq.~(\ref{eq:F4}), and in Eq.~(\ref{eq:omega_rot_p}) we have introduced a ``planetary rotational distortion coefficient'' $k_{qp} = 0.17$, analogous to the stellar rotational distortion coefficient.

\subsection{Dissipative Tides in the Planet}
The planet has spin angular momentum $\Sp = I_p \Omega_p \hatSp$, where $I_p = k_p M_p R_p^2$ is the moment of inertia, $\Omega_p$ is the rotation rate, and where $k_p = 0.25$ throughout this paper.  Averaged over an eccentricity precession timescale, the change in the planet spin due to tidal dissipation is \citep{correia2011}
\be
\frac{1}{S_p} \frac{d \Sp}{d t} = - \frac{1}{2 t_a j^{13}} \frac{L}{S_p} \left[ j^3 f_5(e) (\hatSp + \cos \theta_p\hatLp) \frac{\Omega_p}{2 n} -f_2(e) \hatLp
\right],
\label{eq:dSdt_general}
\ee
where $\cos \theta_p = \hatSp \cdot \hatLp$, and $f_2(e)$ and $f_5(e)$ are given in Eqs.~(\ref{eq:F2}) and (\ref{eq:F5}).  The timescale $t_a$ is 
\ba
\frac{1}{t_a} & = & 6 k_{2p} \Delta t_{\rm L} 
\frac{M_*}{M_p} \left( \frac{R_p}{a}\right)^5 n^2 \nonumber\\
& \approx & \frac{7.3 \times 10^{-21}}{\rm yr} \chi \bar{k}_{2p} \frac{\Msunit \Mtunit}{\Mpunit} \frac{\Rpunit^5}{\aunit^8},
\ea
where $\Delta t_L$ is the lag time, $k_{2p}$ is the tidal Love number, and where we have introduced a tidal enhancement factor $\chi$ (relative to Jupiter), defined such that $\Delta t_L = 0.1 \chi$ sec.
In this paper we assume $\Sp = S_p \hatLp$ (see Section \ref{sec:planetspin} for a justification of this approximation), so that Eq.~(\ref{eq:dSdt_general}) becomes
\be
\frac{1}{S_p} \frac{d S_p}{d t} = - \frac{1}{2 t_a j^{13}} \frac{L}{S_p} \left[ j^3 f_5(e) \frac{\Omega_p}{n} -f_2(e) \right].
\ee
The effect of tidal dissipation on the orbit is
\be
\left . \frac{d \Lp}{d t} \right |_{\rm Tide}  = - \frac{d \Sp}{d t} = - \dot{S}_p \hatLp,
\ee
The change in the eccentricity vector due to tidal dissipation takes the form
\be
\begin{split} 
\left . \frac{d \ep}{d t} \right |_{{\rm Tide}} = & - \frac{1}{2 t_a j^{13}} \left [j^{3} f_4(e) \frac{\Omega_p}{2n} (\ep \cdot \hatSp) \hatLp \right . \\
& - \left . \left( \frac{11}{2} j^3 f_4(e) \frac{\Omega_p}{n}  - 9 f_3(e)\right) \ep \right],
\end{split}
\ee
where the first term inside the brackets vanishes if $\hatSp = \hatLp$.
The dimensionless functions of eccentricity used to describe the tidal evolution take the form
\ba
&& f_1(e) = 1 + \frac{31e^2}{2} + \frac{255e^4}{8} + \frac{185 e^6}{16} + \frac{25 e^8}{64} \label{eq:F1} \\           
& & f_2(e) = 1 + \frac{15e^2}{2} + \frac{45e^4}{8} + \frac{5e^6}{16}   \label{eq:F2} \\
& & f_3(e) = 1 + \frac{15e^2}{4} + \frac{15e^4}{8} + \frac{5e^6}{64} \label{eq:F3} \\ 
& & f_4(e) = 1 + \frac{3e^2}{2} + \frac{e^4}{8} \label{eq:F4} \\
& & f_5(e) =  1 + 3e^2 + \frac{3e^4}{8}. \label{eq:F5}
\ea

\subsection{Stellar Spin-down due to Magnetic Braking}
We use the Skumanich law \citep{skumanich1972}, given by 
\be
\frac{d {\bf \Omega}_\star}{d t} = - \alpha_{\rm MB} \ \Omega_\star^2 {\bf \Omega}_\star,
\ee
where we set $\alpha_{\rm MB} = 1.5 \times 10^{-14}$ yr to model G-type stars, and $\alpha_{\rm MB} = 1.5 \times 10^{-15}$ yr to model F-type stars \citep[from][]{barker2009}.  See also Section \ref{sec:spin}.


\begin{thebibliography}{}

\bibitem[Addison et al.(2013)]{addison2013} 
Addison, B.~C., Tinney, C.~G., Wright, D.~J., et al.\ 2013, ApJL, 774, L9 

\bibitem[Albrecht et al.(2012a)]{albrecht2012a} 
Albrecht, S., Winn, J.~N., Johnson, J.~A., et al.\ 2012a, ApJ, 757, 18  

\bibitem[Albrecht et al.(2012b)]{albrecht2012b} 
Albrecht, S., Winn, J.~N., Butler, R.~P., et al.\ 2012b, ApJ, 744, 189  

\bibitem[Alexander(1973)]{alexander1973}
Alexander M.~E., 1973, ASS, 23, 459

\bibitem[Bate et al.(2010)]{bate2010} 
Bate, M.~R., Lodato, G., \& Pringle, J.~E.\ 2010, MNRAS, 401, 1505

\bibitem[Barker \& Ogilvie(2009)]{barker2009} 
Barker, A.~J., \& Ogilvie, G.~I.\ 2009, MNRAS, 395, 2268

\bibitem[Batygin(2012)]{batygin2012} 
Batygin, K.\ 2012, Nature, 491, 418

\bibitem[Batygin \& Adams(2013)]{batygin2013} 
Batygin, K., \& Adams, F.~C.\ 2013, ApJ, 778, 169

\bibitem[Beaug{\'e} \& Nesvorn{\'y}(2012)]{beauge2012} 
Beaug{\'e}, C., \& Nesvorn{\'y}, D.\ 2012, ApJ, 751, 119

\bibitem[Becker et al.(2015)]{becker2015} 
Becker, J.~C., Vanderburg, A., Adams, F.~C., Rappaport, S.~A., 
\& Schwengeler, H.~M.\ 2015, ApJL, 812, L18

\bibitem[Bouvier(2013)]{bouvier2013} 
Bouvier, J.\ 2013, EAS Publications Series, 62, 143

\bibitem[Chatterjee et al.(2008)]{chatterjee2008} 
Chatterjee, S., Ford, E.~B., Matsumura, S., \& Rasio, F.~A.\ 2008, ApJ, 686, 580

\bibitem[Claret \& Gimenez(1992)]{claret1992}
Claret, A., \& Gimenez, A.\ 1992, A\&AS, 96, 255 

\bibitem[Correia et al.(2011)]{correia2011} 
Correia, A.~C.~M., Laskar, J., Farago, F., 
\& Bou{\'e}, G.\ 2011, Celestial Mechanics and Dynamical Astronomy, 111, 105 

\bibitem[Dawson \& Murray-Clay(2013)]{dawson2013} 
Dawson, R.~I., \& Murray-Clay, R.~A.\ 2013, ApJL, 767, L24

\bibitem[Dawson et al.(2015)]{dawson2015} 
Dawson, R.~I., Murray-Clay, R.~A., \& Johnson, J.~A.\ 2015, ApJ, 798, 66

\bibitem[Eggleton \& Kiseleva-Eggleton(2001)]{eggleton2001} 
Eggleton, P.~P., \& Kiseleva-Eggleton, L.\ 2001, ApJ, 562, 1012 

\bibitem[Fabrycky \& Tremaine(2007)]{fabrycky2007} 
Fabrycky, D., \& Tremaine, S.\ 2007, ApJ, 669, 1298 

\bibitem[Feng et al.(2015)]{feng2015} 
Feng, Y.~K., Wright, J.~T., Nelson, B., et al.\ 2015, ApJ, 800, 22

\bibitem[Fielding et al.(2015)]{fielding2015} 
Fielding, D.~B., McKee, C.~F., Socrates, A., Cunningham, A.~J., 
\& Klein, R.~I.\ 2015, MNRAS, 450, 3306

\bibitem[Ford et al.(2000)]{ford2000} 
Ford, E.~B., Kozinsky, B., \& Rasio, F.~A.\ 2000, ApJ, 535, 385

\bibitem[Ford \& Rasio(2008)]{ford2008} 
Ford, E.~B., \& Rasio, F.~A.\ 2008, ApJ, 686, 621 

\bibitem[Foucart \& Lai(2011)]{foucart2011} 
Foucart, F., \& Lai, D.\ 2011, MNRAS, 412, 2799

\bibitem[Guillochon et al.(2011)]{guillochon2011} 
Guillochon, J., Ramirez-Ruiz, E., \& Lin, D.\ 2011, ApJ, 732, 74

\bibitem[H{\'e}brard et al.(2008)]{hebrard2008} 
H{\'e}brard, G., Bouchy, F., Pont, F., et al.\ 2008, A\&A, 488, 763  

\bibitem[Holman et al.(1997)]{holman1997} 
Holman, M., Touma, J., \& Tremaine, S.\ 1997, Nature, 386, 254

\bibitem[Hut(1981)]{hut1981}
Hut P., 1981, A\&A, 99, 126

\bibitem[Innanen et al.(1997)]{innanen1997}
Innanen, K.~A., Zheng, J.~Q., Mikkola, S., \& Valtonen, M.~J.\ 1997, AJ, 113, 1915

\bibitem[Jackson et al.(2009)]{jackson2009} 
Jackson, B., Barnes, R., \& Greenberg, R.\ 2009, ApJ, 698, 1357 

\bibitem[Juri{\'c} \& Tremaine(2008)]{juric2008} 
Juri{\'c}, M., \& Tremaine, S.\ 2008, ApJ, 686, 603

\bibitem[Knutson et al.(2014)]{knutson2014} 
Knutson, H.~A., Fulton, B.~J., Montet, B.~T., et al.\ 2014, ApJ, 785, 126

\bibitem[Kozai(1962)]{kozai1962}
Kozai, Y.  \ 1962, AJ, 67, 591

\bibitem[Lai et al.(2011)]{lai2011} 
Lai, D., Foucart, F., \& Lin, D.~N.~C.\ 2011, MNRAS, 412, 2790

\bibitem[Lai(2012)]{lai2012} 
Lai, D.\ 2012, MNRAS, 423, 486

\bibitem[Lai(2014)]{lai2014} 
Lai, D.\ 2014, MNRAS, 440, 3532

\bibitem[Laughlin et al.(2011)]{laughlin2011} 
Laughlin, G., Crismani, M., \& Adams, F.~C.\ 2011, ApJL, 729, LL7

\bibitem[Li et al.(2014)]{li2014} 
Li, G., Naoz, S., Holman, M., \& Loeb, A.\ 2014, ApJ, 791, 86

\bibitem[Lidov(1962)]{lidov1962} 
Lidov, M.~L.\ 1962, Planet. Space Sci., 9, 719 

\bibitem[Lithwick \& Naoz(2011)]{lithwick2011} 
Lithwick, Y., \& Naoz, S.\ 2011, ApJ, 742, 94

\bibitem[Liu et al.(2015)]{liu2015} 
Liu, B., Mu{\~n}oz, D.~J., \& Lai, D.\ 2015, MNRAS, 447, 751

\bibitem[Mardling \& Aarseth(2001)]{mardling2001} 
Mardling, R.~A., \& Aarseth, S.~J.\ 2001, MNRAS, 321, 398

\bibitem[Matsumura et al.(2010)]{matsumura2010} 
Matsumura, S., Peale, S.~J., \& Rasio, F.~A.\ 2010, ApJ, 725, 1995

\bibitem[Mazeh \& Shaham(1979)]{mazeh1979}
Mazeh, T., \& Shaham, J. 1979, A\&A, 77, 145

\bibitem[McQuillan et al.(2014)]{mcquillan2014} 
McQuillan, A., Mazeh, T., \& Aigrain, S.\ 2014, ApJS, 211, 24

\bibitem[Moutou et al.(2011)]{moutou2011} 
Moutou, C., D{\'{\i}}az, R.~F., Udry, S., et al.\ 2011, A\&A, 533,
A113 

\bibitem[Mu{\~n}oz et al.(2016)]{munoz2016}
Mu{\~n}oz, D.~J., Lai, D., \& Liu, B. 2016, submitted

\bibitem[Naoz et al.(2012)]{naoz2012} 
Naoz, S., Farr, W.~M., \& Rasio, F.~A.\ 2012, ApJL, 754, LL36

\bibitem[Naoz \& Fabrycky(2014)]{naoz2014} 
Naoz, S., \& Fabrycky, D.~C.\ 2014, ApJ, 793, 137 

\bibitem[Narita et al.(2009)]{narita2009} 
Narita, N., Hirano, T., Sato, B., et al.\ 2009, PASJ, 61, 991  

\bibitem[Nagasawa et al.(2008)]{nagasawa2008} 
Nagasawa, M., Ida, S., \& Bessho, T.\ 2008, ApJ, 678, 498

\bibitem[Neveu-VanMalle et al.(2015)]{neveu2015} 
Neveu-VanMalle, M., Queloz, D., Anderson, D.~R., et al.\ 2015, arXiv:1509.07750

\bibitem[Ngo et al.(2015)]{ngo2015} 
Ngo, H., Knutson, H.~A., Hinkley, S., et al.\ 2015, ApJ, 800, 138 

\bibitem[Petrovich(2015a)]{petrovich2015a} 
Petrovich, C.\ 2015a, ApJ, 805, 75  

\bibitem[Petrovich(2015b)]{petrovich2015b} 
Petrovich, C.\ 2015b, ApJ, 799, 27  

\bibitem[Raghavan et al.(2010)]{raghavan2010} 
Raghavan, D., McAlister, H.~A., Henry, T.~J., et al.\ 2010, ApJS, 190, 1

\bibitem[Rasio \& Ford(1996)]{rasio1996} 
Rasio, F.~A., \& Ford, E.~B.\ 1996, Science, 274, 954 

\bibitem[Rogers \& Lin(2013)]{rogers2013} 
Rogers, T.~M., \& Lin, D.~N.~C.\ 2013, ApJL, 769, L10

\bibitem[Simpson et al.(2011)]{simpson2011} 
Simpson, E.~K., Pollacco, D., Cameron, A.~C., et al.\ 2011, MNRAS, 414, 3023  

\bibitem[Skumanich(1972)]{skumanich1972}
Skumanich, A.\ 1972, ApJ, 171, 565

\bibitem[Socrates et al.(2012)]{socrates2012} 
Socrates, A., Katz, B., Dong, S., \& Tremaine, S.\ 2012, ApJ, 750, 106 

\bibitem[Spalding \& Batygin(2014)]{spalding2014} 
Spalding, C., \& Batygin, K.\ 2014, ApJ, 790, 42 

\bibitem[Storch et al.(2014)]{storch2014} 
Storch, N.~I., Anderson, K.~R., \& Lai, D.\ 2014, Science, 345, 1317 

\bibitem[Storch \& Lai(2015)]{storch2015} 
Storch, N.~I., \& Lai, D.\ 2015, MNRAS, 448, 1821

\bibitem[Storch et al.(2016)]{SLA15}
Storch, N.~I., Lai, D., \& Anderson, K.~R.\ 2016, submitted

\bibitem[Thies et al.(2011)]{thies2011} 
Thies, I., Kroupa, P., Goodwin, S.~P., Stamatellos, D., \& Whitworth,
A.~P.\ 2011, MNRAS, 417, 1817 

\bibitem[Tokovinin \& Kiyaeva(2015)]{tokovinin2015} 
Tokovinin, A., \& Kiyaeva, O.\ 2015, arXiv:1512.00278

\bibitem[Triaud et al.(2010)]{triaud2010} 
Triaud, A.~H.~M.~J., Collier Cameron, A., Queloz, D., et al.\ 2010,
A\&A, 524, A25

\bibitem[Valsecchi et al.(2014)]{valsecchi2014} 
Valsecchi, F., Rasio, F.~A., \& Steffen, J.~H.\ 2014, ApJL, 793, L3

\bibitem[Wang et al.(2015)]{wang2015} 
Wang, J., Fischer, D.~A., Horch, E.~P., \& Xie, J.-W.\ 2015, ApJ, 806, 248

\bibitem[Winn et al.(2009)]{winn2009} 
Winn, J.~N., Johnson, J.~A., Albrecht, S., et al.\ 2009, ApJL, 703, L99 

\bibitem[Winn et al.(2010)]{winn2010} 
Winn, J.~N., Fabrycky, D., Albrecht, S., \& Johnson, J.~A.\ 2010, ApJL, 718, L145

\bibitem[Wright et al.(2012)]{wright2012} 
Wright, J.~T., Marcy, G.~W., Howard, A.~W., et al.\ 2012, ApJ, 753, 160

\bibitem[Wu \& Murray(2003)]{wu2003} 
Wu, Y., \& Murray, N.\ 2003, ApJ, 589, 605 

\bibitem[Wu et al.(2007)]{wu2007} 
Wu, Y., Murray, N.~W., \& Ramsahai, J.~M.\ 2007, ApJ, 670, 820

\bibitem[Wu \& Lithwick(2011)]{wu2011} 
Wu, Y., \& Lithwick, Y.\ 2011, ApJ, 735, 109

\bibitem[Xue et al.(2014)]{xue2014} 
Xue, Y., Suto, Y., Taruya, A., et al.\ 2014, ApJ, 784, 66

\end{thebibliography}
\end{document}